\documentclass[aps, prd, amsmath, floats, floatfix, onecolumn, nofootinbib]{revtex4-2}
\pdfoutput=1
\usepackage{graphicx}
\usepackage{color}
\usepackage{amsmath}
\usepackage{latexsym}
\usepackage[makeroom]{cancel}
\usepackage{tikz}
\usetikzlibrary{tikzmark}

\usepackage{pgfplots}
\pgfplotsset{compat=1.18}

\usetikzlibrary{decorations.pathreplacing,decorations.pathmorphing,shapes.geometric,calc}

\usepackage{tikz}
\usepackage{amsmath}
\usepackage{amssymb}

\usepackage{amssymb}

\newcommand{\be}{\begin{eqnarray}}
\newcommand{\ee}{\end{eqnarray}}

\newcommand{\bfk}{{\bf k}}
\newcommand{\bfp}{{\bf p}}

\newcommand{\bea}{\begin{eqnarray}}
\newcommand{\eea}{\end{eqnarray}}

\usepackage{amsmath,amssymb,amsthm}
\usepackage{epsfig}
\usepackage{graphicx}

\def\sph{\hspace{.5em}}
\def\Eq#1{Eq.~(\ref{#1})}

\def\no{\nonumber \\}
\def\tbf#1{\textbf{#1}}

\def\tit#1{\textit{#1}}

\def\expm2piOmega{e^{-2\pi\Omega}}

\def\sp{\hspace{.25em}}

\begin{document}

\title{
%
%
%
%
%
%
%
%
%
%
%
Superconductive or superfluid condensation in curved spacetime}

\author{Leonardo Modesto}
\email{leonardo.modesto@unica.it}
	\email[]{lmodesto1905@icloud.com}
	\affiliation{Dipartimento di Fisica, Universit\`a di Cagliari, Cittadella Universitaria, 09042 Monserrato, Italy}
	\affiliation{INFN, Sezione di Cagliari, Cittadella Universitaria, 09042 Monserrato, Italy}

\begin{abstract}

We provide a rigorous proof of unitarity for quantum field theory in a general spacetime. Our argument is based on expressing the Bogoliubov transformations in terms of a unitary squeezing operator that relates the initial and final Hilbert spaces. Consequently, the $S$-matrix in curved spacetime is given by the product of the squeezing operator and the $S$-matrix in the out-Hilbert space (or, equivalently, in the in-Hilbert space, typically Minkowski spacetime). Since both factors are unitary, the product is unitary.

It follows that the Bogoliubov in-vacuum of a quantum field theory in a general spacetime is described by a BCS-like state (Bardeen--Cooper--Schrieffer): 
(i) for fermions, the Bogoliubov vacuum is exactly the BCS state, but involving electrons and positrons in place of electrons with opposite spin;
(ii) for bosons, the Bogoliubov vacuum is the Bose--Einstein condensation (BEC) superfluid ground state.

Thus, gravity, or an accelerating force, creates from the vacuum a many-particle system unstable towards the formation of a new ground state consisting of Cooper pairs. Technically, gravity (or the rocket engine) converts the vacuum for electrons and positrons into the Bogoliubov in-vacuum made of an arbitrary superposition of Cooper pairs, namely an electron-positron condensate. It is with respect to this latter state that quantum field theory evolves unitarily.

By reverse engineering, we reconstruct the effective Hamiltonian of QFT in the presence of a black hole (or in the Rindler frame), together with a very simple formula for the mass gap function, which grows approximately linearly with the temperature. Hence, electrons and positrons in the black hole background condense into a superconducting state at increasingly higher temperatures as the mass decreases. Accordingly, unitarity and the particle interpretation are clearly preserved at every stage of the evaporation process. In particular, the Hawking state is exactly a BCS state.

Finally, we compute the entanglement entropy, and derive the area law.

\end{abstract}

\maketitle

\tableofcontents

\section{Introduction}

In the half-century following Hawking's dyscovery of quantum black-hole evaporation \cite{Hawking}, the theoretical physics community has devoted considerable effort—much of it unavoidably speculative—to the so-called information-loss paradox. While we acknowledge the heuristic and educational merit of this body of work, we contend that it has largely deviated from the core physical question. For this reason, we adopt a maximally conservative framework.

Let us begin with a brief recapitulation of the history that has unfolded over the past half-century. A long-standing puzzle concerns the formal analogy between certain properties of black holes—exact solutions of Einstein's theory of gravity—and the laws of thermodynamics. In particular, Hawking's area theorem established that the total horizon area of merging black holes can never decrease, in close parallel with the second law of thermodynamics in its entropic form. Motivated by this observation, and encouraged by his advisor J. Wheeler, Bekenstein proposed that a black hole possesses an entropy proportional to the area of its event horizon \cite{Bekenstein:1972tm, Bekenstein:1973ur}. His argument, simple and illuminating, ran as follows: consider an observer safely enjoying a cup of tea outside the horizon. The warm cup carries nonzero entropy; if it is dropped into the black hole, that entropy appears to be lost. Consistently with the causal structure of spacetime, Bekenstein therefore assigned to the black hole an entropy proportional to its horizon area—and hence to the square of its mass.

For some time, this proposal remained a mere curiosity. It was Hawking's subsequent discovery that black holes are quantum-mechanically unstable and emit thermal radiation, with a temperature inversely proportional to the mass, that elevated the analogy to a full-fledged physical phenomenon \cite{Hawking, Fabbri, BD, Townsend}. So far, so good. However, when the black hole evaporates completely, a difficulty arises: the initial pure state appears to have evolved into a mixed state, in apparent conflict with the principle of unitary evolution. Technically, the vacuum initial state is described by a thermal density matrix for observers in the asymptotic region $\mathcal{I}^+$. This description is perfectly adequate as long as the black hole remains sufficiently massive (i.e., much larger than the Planck mass). But when the black hole becomes too small or disappears entirely, the entanglement between the Hawking radiation and the black hole interior is lost, and the initial pure state has effectively evolved into a thermal mixed state \cite{Fabbri}. In other words, we are confronted with the so-called {\em information loss problem}.


In order to shed light on this issue, we revisit and reformulate quantum field theory in a curved background \cite{Fabbri,BD,Townsend}, expressing the Bogoliubov transformations in terms of a unitary operator---namely, the well-known squeezing operator, which we shall often denote by $S_{\rm sq}$. This reformulation reveals that QFT in curved spacetime can be mapped onto QFT in Minkowski spacetime, provided the full $S$-matrix is identified with the product $S_{\rm sq} \, S_0$, where $S_0$ is the scattering matrix in the globally Minkowskian spacetime with metric $\eta_{\mu\nu}$ (i.e., globally Minkowski in Cartesian coordinates).

The effect of a general metric on QFT for Dirac fermions is analogous to the effect of a solid on the physics of electrons: a curved spacetime turns the vacuum, initially devoid of particles, into a superconducting condensate, in close analogy with phonons that enable the creation of an arbitrary number of Cooper pairs in a solid. Similarly, in a Bosonic QFT, the analogy extends naturally to the theory of Bose--Einstein condensation and superfluidity \cite{Guo_2017} (see also He II superfluidity \cite{Superfluidity}). There is therefore no exception: 

{\em a QFT in curved spacetime, for both fermions and bosons, is described by a superconducting or superfluid condensate.}

Records show a similar idea in the contest of loop quantum gravity \cite{Alexander:2008yg}. The authors presented a reformulation of loop quantum gravity with a cosmological constant and no matter as a Fermi-liquid theory.

Mathematically, given a non-stationary spacetime, the initial and final Hilbert spaces are mapped into each other by the squeezing operator $S_{\rm sq}$, and the Bogoliubov transformations of the creation and annihilation operators are precisely realized by the very same operator $S_{\rm sq}$. It turns out that the in-vacuum state---i.e., the vacuum for the Bogoliubons---is in fact the BCS (Bardeen--Cooper--Schrieffer) state (or the BEC ground state for a Bosonic QFT), consisting of an arbitrary number of Cooper pairs, with respect to the operators defined on the out-Hilbert space.

By reverse engineering, we can therefore reconstruct the effective interaction Hamiltonian between fermions or bosons in a general background in the mean-field approximation. In the Fermionic case, the Hamiltonian is exactly the BCS one, while for bosons it is the BEC Hamiltonian \cite{Guo_2017}.

As explicit examples, we will consider the Rindler and black hole backgrounds, which are indeed very similar. In particular, we will review how the Hawking state \cite{Fabbri} can be obtained simply by acting with the squeezing operator on the out-vacuum. The number of out-particles in the initial state is then derived explicitly by making direct use of the operator formalism. We will reconstruct the BCS Hamiltonian, the explicit form of the mass gap function, and the entanglement entropy for both fermions and bosons.

To summarize, the presence of a black hole converts the vacuum into a BCS (or BEC) state consisting of an arbitrary superposition of Cooper pairs, namely the squeezing operator acting on the out-vacuum.

Finally, we will show that the evaporation process can proceed via the emission of positive-energy particles, contrary to the heuristic picture advanced by Hawking in his original paper \cite{Hawking}. Indeed, the production of negative- and positive-energy particles from the vacuum, as claimed in \cite{Hawking}, is consistent only off-shell, whereas the Hawking state itself is an on-shell construction, as will become evident from our squeezing-operator derivation. Accordingly, the original heuristic argument would require the introduction of on-shell ghosts in the spectrum, leading to a manifest violation of unitarity.
Although in our derivation the energy is always positive definite, the instability of the black hole is confirmed. Specifically, the presence of the black hole gives rise to the creation of positive-energy particle pairs, one member of which escapes to infinity while the other is reabsorbed by the black hole. In this very simple process, the black hole mass decreases by one unit.
The above scenario is consistent provided we can avoid the singularity at $r=0$ and the dynamics is complete. In this regard, we will account for the singularity resolution by means of Weyl conformal invariance \cite{Bambi:2016wdn, Bambi:2016yne, narlikar:1977nf}.

Consequently, QFT during the entire evaporation process will be perfectly unitary for any positive value of the black hole mass. At the last stage of evaporation, however, we must modify the black hole temperature according to its Compton wavelength. This natural assumption makes the theory consistent up to the complete disappearance of the black hole. 
In particular, the entanglement entropy, for both fermions and bosons, takes a very simple form cubic in the temperature, while the area law is recovered as a feature of the black hole geometry near the event horizon. 
We also derive the evaporation time based solely on the energy conservation, without invoking the Boltzmann's law.

\section{Quantum field theory in curved spacetime revisited} \label{QFTCS}
In this section we briefly review quantum field theory in curved spacetime. In short, the entire quantization procedure for field theory in a general—flat or curved—spacetime rests on two main cornerstones: (i) a complete set of solutions to the equations of motion (EoM) with respect to a covariant scalar product, which is independent of the particular 3D hypersurface on which Dirac's quantization procedure is implemented; (ii) the comparison of the solutions to the EoM, and hence of the operators, on different 3D hypersurfaces according to the invariant scalar product.
%

 Let us start expanding a scalar field $f$, which solves the EoM $\Box f = 0$ \cite{Fabbri, BD}, 
 in terms of a complete set of mode solutions 
 $u_i^{\rm in}$,
  \be
f = \sum_i \left( a_i^{\rm in} u_i^{\rm in} + a_i^{ {\rm in} \dagger } u_i^{ {\rm in} *} 
\right)  \, . 
\label{fin}
  \ee
We can also expand $f$ choosing another set of 
solutions $u_i^{\rm out}$, 
  \be
f = \sum_i \left( a_i^{\rm out} u_i^{\rm out} + a_i^{ {\rm out} \dagger } u_i^{ {\rm out} *} 
\right)   . 
\label{fout}
  \ee
In (\ref{fin}) and (\ref{fout}), the in- and out- modes satisfy the the following orthogonal relations, 
\be
&& (u_i^{\rm in} , u_j^{\rm in} ) = \delta_{ij} \, , \quad (u_i^{{\rm in} *} , u_j^{ {\rm in} *} ) = -\delta_{ij} \, , 
\quad
(u_i^{\rm in} , u_j^{ {\rm in} *} ) = 0 \, , 
\label{OrtCi}
 \\
&& (u_i^{\rm out} , u_j^{\rm out} ) = \delta_{ij} \, , \quad (u_i^{{\rm out} *} , u_j^{ {\rm out} *} ) = -\delta_{ij} \, , 
\quad
(u_i^{\rm out} , u_j^{ {\rm out} *} ) = 0 \, , 
\label{OrtCo}
\ee
respect to the generalized scalar product:
\be
(f_1, f_2) = - i \int_{\Sigma} d \Sigma^\mu \left( f_1(x) \partial_\mu f_2(x)^{*} - f^{*}_2(x) \partial_\mu f_1(x)\right) \, . 
\label{scalarP}
\ee
In (\ref{scalarP}), $d \Sigma^\mu = n^\mu d\Sigma$, with $n^\mu$ is a future directed unit vector orthogonal to the hypersurface $\Sigma$ and $d \Sigma$ is the volume element in $\Sigma$. The hypersurface $\Sigma$ is taken to be a Cauchy surface in the globally hyperbolic spacetime. It can be shown, using the Gauss theorem, that the scalar product is independent on the choice of $\Sigma$. 
Since both the sets of modes are complete, we can expand one in terms of the other, but is not guarantee that one can expand the positive frequency solutions $u^{\rm out}_i$ in terms only of the positive frequencies $u^{\rm in}_i$. Therefore, in general:
\be
u^{\rm out}_j = \sum_i \left( \alpha_{j i} u^{\rm in}_i +  \beta_{j i} u^{ {\rm in}*}_i \right)  , 
\label{uab}
\ee
which are the Bogoliubov transformations, while $\alpha_{i j}$ and $\beta_{ij}$ are called Bogoliubov coefficients. 
Using the scalar products (\ref{scalarP}) and (\ref{uab}), we can get directly the Bogoliubov coefficients, 
\be
\alpha_{ij} = ( u^{\rm out}_i, u^{\rm in}_j) \, , \quad 
\beta_{ij} = - ( u^{\rm out}_i, u^{ {\rm in} *}_j) \, .
\ee
The condition for the transformation to be canonical, namely to preserve the commutation relations, or the orthonormality conditions (\ref{OrtCo}) and (\ref{OrtCi}), implies:
%
\begin{equation}
\boxed{
\begin{aligned}
\alpha \alpha^\dagger - \beta \beta^\dagger &=   1\!\!1   \, , \\
 \alpha \beta^T - \beta \alpha^T &= 0 \,.  
\end{aligned}
}
\label{BT1}
\end{equation}
In order to prove the above conditions, one has to replace (\ref{uab}), and its complex conjugate, in (\ref{OrtCo}),  and use the orthonormality conditions (\ref{OrtCi}) for $u^{\rm in}_i$ and $u^{{\rm in}*}_i$. 

One has also to make sure that the transformation (\ref{uab}) is invertible \cite{Townsend}. For this purpose, one defines 
\be
u^{\rm in}_j  = \sum_i \left( \alpha_{j k}^{ \prime} u^{\rm out}_k +  \beta_{j k}^{ \prime} u^{ {\rm out}*}_k \right)  , 
\label{uabI}
\ee
and, replacing (\ref{uabI}) in (\ref{uab}), shows that:
\be
\alpha^\prime = \alpha^\dagger \, , \quad  
\beta^\prime = - \beta^{\rm T} \, .
\label{apbp}
\ee 
Moreover, the Bogoliubov coefficients $\alpha^\prime$ and $\beta^\prime$ must satisfy the same conditions of 
$\alpha$ and $\beta$, namely (\ref{BT1}), i.e.  
\be
&& 
 \alpha^\prime \alpha^{\prime \dagger} - \beta^\prime \beta^{\prime \dagger} =  1\!\!1  \, ,  \nonumber \\
&& \alpha^\prime \beta^{\prime T} - \beta^\prime \alpha^{\prime T} = 0 \, .
\label{BT1prime}
\ee 
Finally, Replacing (\ref{apbp}) in (\ref{BT1prime}), we get:
\begin{equation}
\boxed{
\begin{aligned}
\alpha \alpha^\dagger - \beta^T \beta^* &=   1\!\!1   \, , \\
 \alpha^\dagger \beta - \beta^T \alpha^* &= 0 \,.  
\end{aligned}
}
\label{BT2}
\end{equation}

The covariant quantization is implemented promoting the field $f$ and the conjugate momentum $\Pi_f$ to operators and imposing the Dirac commutation relations. Hence, using the completeness of the modes $u^{\rm}_i$ (or $u^{\rm out}_i$), we end up with the commutations relations for the creation and annihilation operators,
\be
\left[ a_i^{\rm in} , a^{{\rm in} \dagger}_j \right] = \delta_{i j} \, , \quad \mbox{etc} \, . 
\ee
The construction of the vacuum state, Fock space, etc, proceed exactly like in Minkowski spacetime. However, in a general metric there is an inherent ambiguity in the formalism \cite{BD}.
Indeed, using that:
\be
a^{\rm in}_i = \left( f , u^{\rm in}_i \right) \, , \quad 
a^{\rm out}_i = \left( f , u^{\rm out}_i \right) \, , 
\label{aio}
\ee
we can expand the two sets of creation and annihilation operators in terms of the other,
\be
a_i^{\rm in} = \sum_j \left( \alpha_{j i} a^{\rm out}_j +  \beta_{j i}^{*} a^{ {\rm out} \dagger}_j \right) \, , 
\label{aouttoain}
\ee
or
\be
a_i^{\rm out} = \sum_j \left( \alpha_{i j}^{*} a^{\rm in}_j -  \beta_{i j}^{*} a^{ {\rm in} \dagger}_j \right) \, ,
\ee
and, if $\beta_{ij}$ does not vanish, the vacuum $| {\rm in} \rangle$ and the vacuum $| {\rm out} \rangle$, defined by
\be
a^{\rm in} | {\rm in} \rangle = 0 \, , \quad a^{\rm out} | {\rm out} \rangle = 0 \, , 
\ee
are not the same. One can realize the issue evaluating the expectation value of the out-number of particles in the in-vacuum state, 
\be
\langle {\rm in} | N_i^{\rm out}  | {\rm in} \rangle
=
 \langle {\rm in} | a_i^{ {\rm out} \dagger} a_i^{\rm out} | {\rm in}  \rangle = \sum_j |\beta_{ij}|^2 \, , 
\ee
where we used (\ref{aouttoain}). If the coefficients $\beta_{ij}$ differ from zero the particle content of the in vacuum, with respect to the out Fock space, is non-trivial. 
Contrary, if $\beta_{ij} =0$, the first of (\ref{BT1}) turns in $\alpha^\dagger \alpha = {\rm I}$ and the positive frequency mode basis $u_i^{\rm in}$ and $u_i^{\rm out}$ are related by a unitary transformation, and the definition of the vacuum remains unchanged. 

In the next section, we will show that the Bogoliubov transformations are realized by a unitary squeezing operator and, thus, the in- and out- Hilbert spaces are unitary equivalent. Moreover, there is no mystery about the definition of particles in a general background because the effect of the geometry is to turn the vacuum (generally the Minkowski vacuum) in a condensate (superconductive or superfluid according to the spin-statistic).

%
%
%
%
%
%


So far the construction summarized in this section is quite general, but we can consider a spacetime that admits two stationary regions, meaning there exists a timelike vector field 
$\xi^\mu$ that leaves the metric invariant both prior to and following a central region where the metric is explicitly time-dependent. In this case, the solutions $u_i^{\rm in}$ and $u_i^{\rm out}$ are positive frequency solutions in the past and future region respectively.

\section{Two modes squeezed states and Bogoliubov transformations} \label{SqBT}
In this section we review the squeezing unitary representation of the Bogoliubov transformation for a Bosonic and a Fermionic two-modes system, namely we will explicitly construct the squeezing operators that realize the Bogoliubov transformations. 
As we will see later, the construction in this section will be sufficient to get the states of our interest for a quantum field theory in the Rindler or a Black Hole spacetime 
by simply making the tensor product on all possible modes.

\subsection{The Bosonic $2$-dimensional system} \label{SqBTB}
Let us consider a two dimensional harmonic oscillator defined in terms of the creation and annihilation operators 
$a^\dagger, a$ and $b^\dagger, b$ obeying the following commutation relations:
\be
[a, a^\dagger] = 1 \, , \quad [b, b^\dagger] = 1\, , \quad [ a, a] = [b, b] = [a^\dagger, a^\dagger] = [ b^\dagger, b^\dagger] = 0 \, .
\ee
The vacuum is:
\be
|0 \rangle \equiv |0_a, 0_b \rangle \, ,
\label{vab}
\ee
and the squeezing operator for a two-dimensional system reads \cite{Stone:2023fkj}:
\be
\tilde{S}_2(\xi) & = & e^{  \xi^* a b - \xi a^\dagger b^\dagger  } \label{Sexp} \\
& = & 
e^{ - e^{i \theta}  \tanh |\xi | a^\dagger b^\dagger }
e^{ - \ln  \cosh |\xi | \left[ \left( a^\dagger a + \frac{1}{2} \right) + \left( b^\dagger b + \frac{1}{2} \right) \right] }
e^{ - e^{i \theta}  \tanh |\xi | a b } \, .
\ee
Its action on the vacuum (\ref{vab}) is:
\be
\tilde{S}_2(\xi) |0 \rangle = \frac{1}{\cosh |\xi| } e^{ - e^{i \theta}  \tanh |\xi | a^\dagger b^\dagger } | 0 \rangle = 
\frac{1}{\cosh |\xi| } \sum_{n =0 }^{+\infty} \left( - e^{i \theta}  \tanh |\xi | \right)^n |n_a , n_b \rangle \, ,
\ee
where $\xi = | \xi | e^{i \theta}$ is a complex number. 
The operator $S_2(\xi)$ realizes the Bogoliubov transformation according to:
\be
&& \tilde{S}_2^\dagger (\xi) \, a \, \tilde{S}_2(\xi) = \cosh | \xi |\, a - \frac{\xi}{| \xi |} \sinh |\xi | \, b^\dagger \, , \nonumber \\
&& \tilde{S}_2^\dagger(\xi) \, b \, \tilde{S}_2 (\xi) = \cosh | \xi |\, b - \frac{\xi}{| \xi |} \sinh |\xi | \, a^\dagger \, .
\ee
In comparison to the reference \cite{LoSollie}, 
 the minus sign in front of 
the $\sinh$ is due to the chosen opposite sign at the exponent of (\ref{Sexp}). 

In order to make contact with the usual definitions, we define:
\be
\tilde{S}_2(\xi) = S_2^\dagger(\xi)  \, , \quad S_2(\xi) = \tilde{S}_2^\dagger(\xi) \, .
\ee
Since $\tilde{S}_2^\dagger(\xi) = \tilde{S}(-\xi)$
\cite{bookO} (notice that $\xi \rightarrow - \xi$ means $\theta \rightarrow \theta + \pi$), 
%
\be
\boxed{
S_2(\xi) = e^{  - \xi^* a b + \xi a^\dagger b^\dagger  }
=  
e^{ - e^{i \pi} e^{i \theta}  \tanh |\xi | a^\dagger b^\dagger }
e^{ - \ln  \cosh |\xi | \left[ \left( a^\dagger a + \frac{1}{2} \right) + \left( b^\dagger b + \frac{1}{2} \right) \right] }
e^{ - e^{i \pi} e^{i \theta}  \tanh |\xi | a b } 
}
\, , 
\label{SexpNT} 
\ee
thus, we end up with the usual convention,
\be
&& 
\boxed{
a^\prime \equiv 
S_2 (\xi) \, a \, S^\dagger_2(\xi) = \cosh | \xi |\, a - \frac{\xi}{| \xi |} \sinh |\xi | \, b^\dagger} \, , \nonumber \\
&& \boxed{
b^\prime \equiv
 S_2 (\xi) \, b \, S^\dagger_2 (\xi) = \cosh | \xi |\, b - \frac{\xi}{| \xi |} \sinh |\xi | \, a^\dagger}
  \, , 
\label{abBose}
\ee
and the Hermitian conjugates transformations are:
\be
&& 
\boxed{
a^{\prime \dagger} \equiv 
S_2 (\xi) \, a^\dagger \, S^\dagger_2(\xi) = \cosh | \xi |\, a^\dagger - \frac{\xi^*}{| \xi |} \sinh |\xi | \, b} \, , \nonumber \\
&& \boxed{
b^{\prime \dagger} \equiv
 S_2 (\xi) \, b^\dagger \, S^\dagger_2 (\xi) = \cosh | \xi |\, b^\dagger - \frac{\xi^*}{| \xi |} \sinh |\xi | \, a}
  \, , 
\label{abBoseH}
\ee

We can write (\ref{abBose}) in the following useful short matrix notation,
\be
\begin{pmatrix}
a^\prime   \\
 b^\prime  
\end{pmatrix}	
 = 
\cosh |\xi | 
\begin{pmatrix}
1 & 0 \\
 0 & 1 
\end{pmatrix}	
\begin{pmatrix}
a   \\
 b  
\end{pmatrix}
- \frac{\xi}{| \xi |}  \sinh| \xi | 
\begin{pmatrix}
0 & 1 \\
 1 & 0 
\end{pmatrix}	
\begin{pmatrix}
a^\dagger   \\
 b^\dagger  
\end{pmatrix}	
\equiv 
\alpha^T \vec{a} + \beta^\dagger \vec{a}^\dagger \, , 
\ee
where in the last equality we made contact with the definitions of Bogoliubov transformation in section (\ref{QFTCS}). 

Finally, the action of $S_2$ on the vacuum gives:
\be
{S}_2(\xi) |0 \rangle = \frac{1}{\cosh |\xi| } e^{  e^{i \theta}  \tanh |\xi | a^\dagger b^\dagger } | 0 \rangle = 
\frac{1}{\cosh |\xi| } \sum_{n =0 }^{+\infty} \left(  e^{i \theta}  \tanh |\xi | \right)^n |0 \rangle \, .
\label{S2vac}
\ee

As a check of consistency, let us verify that the conditions (\ref{BT1}) and (\ref{BT2}) are satisfied.
Let us start with (\ref{BT1}). Since $\alpha^T = \alpha$, the first condition $\alpha \alpha^\dagger - \beta \beta^\dagger = {\rm I}$ reads:
\be
&& \cosh |\xi | 
\begin{pmatrix}
1 & 0 \\
 0 & 1 
\end{pmatrix}	
\cosh |\xi | 
\begin{pmatrix}
1 & 0 \\
 0 & 1 
\end{pmatrix}^\dagger	
- ( -1) \frac{\xi}{| \xi |}  \sinh| \xi | 
\begin{pmatrix}
0 & 1 \\
 1 & 0 
\end{pmatrix}	
 ( -1) \frac{\xi^*}{| \xi |}  \sinh| \xi | 
\begin{pmatrix}
0 & 1 \\
 1 & 0 
\end{pmatrix}^\dagger	\nonumber \\
&&
= (\cosh |\xi |)^2 \, {\rm I} - (\sinh |\xi |)^2 \, {\rm I} = {\rm I} \,  . 
\ee
The second condition in (\ref{BT1}), i.e. $\alpha \beta^T - \beta \alpha^T = 0$, turns into:
\be
&& \cosh |\xi | 
\begin{pmatrix}
1 & 0 \\
 0 & 1 
\end{pmatrix}	
 ( -1) \frac{\xi^*}{| \xi |}  \sinh| \xi | 
\begin{pmatrix}
0 & 1 \\
 1 & 0 
\end{pmatrix}^T
-
 ( -1) \frac{\xi^*}{| \xi |}  \sinh| \xi | 
\begin{pmatrix}
0 & 1 \\
 1 & 0 
\end{pmatrix}
\cosh |\xi | 
\begin{pmatrix}
1 & 0 \\
 0 & 1 
\end{pmatrix}^T	 \nonumber \\
&& = \left[  ( -1) \frac{\xi^*}{| \xi |}  \cosh |\xi |  \sinh| \xi |  -
( -1) \frac{\xi^*}{| \xi |}    \sinh| \xi | \cosh |\xi | \right] \begin{pmatrix}
0 & 1 \\
 1 & 0 
\end{pmatrix}
= 0 \, .
\ee

Similarly, from the first equation in (\ref{BT2}), namely  $\alpha \alpha^\dagger - \beta^T \beta^* = 1\!\!1$, we get:
\be
&& \cosh |\xi | 
\begin{pmatrix}
1 & 0 \\
 0 & 1 
\end{pmatrix}	
\cosh |\xi |
\begin{pmatrix}
1 & 0 \\
 0 & 1 
\end{pmatrix}^\dagger 	
- ( -1) \frac{\xi}{| \xi |}  \sinh| \xi | 
\begin{pmatrix}
0 & 1 \\
 1 & 0 
\end{pmatrix}^T	
 ( -1) \frac{\xi^*}{| \xi |}  \sinh| \xi | 
\begin{pmatrix}
0 & 1 \\
 1 & 0 
\end{pmatrix}^*	
\nonumber \\
&&
= (\cosh |\xi |)^2 \, 1\!\!1   - (\sinh |\xi |)^2 \, 1\!\!1  \, = 1\!\!1   \,  . 
\ee
From the second equation in (\ref{BT2}), i.e. $\alpha^\dagger \beta - \beta^T \alpha^* = 0$, we get:
\be
&& \cosh |\xi |
\begin{pmatrix}
1 & 0 \\
 0 & 1 
\end{pmatrix}^\dagger 	
 ( -1) \frac{\xi}{| \xi |}  \sinh| \xi | 
\begin{pmatrix}
0 & 1 \\
 1 & 0 
\end{pmatrix}
-
 ( -1) \frac{\xi}{| \xi |}  \sinh| \xi | 
\begin{pmatrix}
0 & 1 \\
 1 & 0 
\end{pmatrix}^T
\cosh |\xi |
\begin{pmatrix}
1 & 0 \\
 0 & 1 
\end{pmatrix}^*	 \nonumber \\
&& = \left[  ( -1) \frac{\xi}{| \xi |}  \cosh |\xi |  \sinh| \xi |  -
( -1) \frac{\xi}{| \xi |}    \sinh| \xi | \cosh |\xi | \right] \begin{pmatrix}
0 & 1 \\
 1 & 0 
\end{pmatrix}
= 0 \, .
\ee
Therefore, the unitary squeezing operator (\ref{Sexp}), faithfully implements the Bogoliubov transformation for a two dimensional Bosonic system.

\subsection{The Fermionic $2$-dimensional system} \label{Sq2Fermions}
For a Fermionic two-dimensional system \cite{Stone:2023fkj}, creation and annihilation operators 
$a, a^\dagger$ and $b, b^\dagger$ obey the Fermionic algebra involving anti-commutators, namely 
\be
\{ a, a^\dagger \} = 1 \, , \quad \{ b, b^\dagger \} = 1\, , \quad \{ a, a \} = \{ b, b \} 
= \{ a^\dagger, a^\dagger \} = \{ b^\dagger, b^\dagger \} = 0 \, .
\ee
The vacuum is such that:
\be
(a^2)^\dagger |0_a, 0_b \rangle =  
(b^2)^\dagger |0_a, 0_b \rangle = 0 
\, .
\label{vaF}
\ee
The two mode squeezing operator for the system reads \cite{Stone:2023fkj}:
\be
\boxed{
S_2( z)  = e^{ z \, a^\dagger b^\dagger - z^* b a    } 
\label{SexpF} 
 = 
e^{ e^{i \theta}  \tan |z | a^\dagger b^\dagger }
e^{  \ln  \cos | z | \left[ \left( a^\dagger a + \frac{1}{2} \right) + \left( b^\dagger b + \frac{1}{2} \right) \right] }
e^{ - e^{- i \theta}  \tan | z  | b a }
}
 \, , 
\ee
whose action on the vacuum (\ref{vab}) gives exactly the single mode BCS state on which we will comment further and extensively later, 
\be
S_2(z) |0 \rangle = \cos | z |  e^{  e^{i \theta}  \tan |z | a^\dagger b^\dagger } | 0 \rangle = 
\cos |z|  \left(  1 + e^{i \theta}  \tan | z  | a^\dagger b^\dagger \right) | 0  \rangle  
\equiv
| \Psi_{\rm BCS}(z) \rangle  \, ,
\label{PsiBCS}
\ee
where $z= | z  | e^{i \theta}$. Notice the normalization condition arising from (\ref{PsiBCS}), i.e 
$(\cos |z| )^2+ (\sin |z|)^2 =1$. 

Finally, the Bogoliubov-Vatalin transformations are implemented by the Squeezing operator as follows,
\be
&& 
\boxed{
a^\prime = S_2(z) a S_2^\dagger(z) =  (\cos |z|) a - (e^{i \theta }  \sin | z | ) b^\dagger}  \, , 
\nonumber 
\\
&& \boxed{
b^\prime = S_2(z) b S_2^\dagger(z) =  ( e^{i \theta }  \sin |z|) a^\dagger + (\cos | z | ) b 
}
\, , 
\label{transBosons}
\ee
which we can rewrite implementing a matrix notation similarly to the Bosonic case, i.e. 
\be
\begin{pmatrix}
a^\prime   \\
 b^\prime  
\end{pmatrix}	
 = 
\cos | z | 
\begin{pmatrix}
1 & 0 \\
 0 & 1 
\end{pmatrix}	
\begin{pmatrix}
a   \\
 b  
\end{pmatrix}
+ e^{i \theta}  \sin | z | 
\begin{pmatrix}
0 & - 1 \\
 1 & 0 
\end{pmatrix}	
\begin{pmatrix}
a^\dagger   \\
 b^\dagger  
\end{pmatrix}	
\equiv 
\alpha^T \vec{a} + \beta^\dagger \vec{a}^\dagger \, . 
\label{contFT}
\ee
The result (\ref{PsiBCS}) defines the vacuum for the operators $a'$ and $b'$, namely
\be
&& a'  \, | {\rm BCS} \rangle = a' \, S_2 (z) | 0 \rangle = S_2(z) \, a \, S_2^\dagger(z) S_2 (z) | 0 \rangle 
= S_2(z) \, a  | 0 \rangle = 0 \, , \nonumber \\
&& b'  \, | {\rm BCS} \rangle = b' \, S_2 (z) | 0 \rangle = S_2(z) \, b \, S_2^\dagger(z) S_2 (z) | 0 \rangle 
= S_2(z) \, b  | 0 \rangle = 0 \, .
\ee
It deserves to be notice that:
\be
\alpha \alpha^\dagger = \frac{\cos^2 |z|}{\sin^2 |z| } \beta \beta^\dagger
= \frac{1}{\tan^2 |z| } \beta \beta^\dagger
 \, .
\ee
Hence, if $\tan |z| = \exp ( - \pi \, \Omega)$ (see next section and (\ref{bogo_ops_a_bdag})), we get the usual relation between the Bogolubov coefficients, namely 
\be
\alpha \alpha^\dagger = e^{2 \pi \, \Omega}  \beta \beta^\dagger
\label{usuab}
 \, .
\ee

For Fermions the first condition in (\ref{BT1}) turns into:
\be
\alpha \alpha^\dagger + \beta \beta^\dagger =  1\!\!1  \, .
\ee
Therefore, according to the definitions in (\ref{contFT}), 
\be
&& 
\cos | z | 
\begin{pmatrix}
1 & 0 \\
 0 & 1 
\end{pmatrix}	
\cos | z | 
\begin{pmatrix}
1 & 0 \\
 0 & 1 
\end{pmatrix}^\dagger	
+ e^{ - i \theta} \sin | z | 
\begin{pmatrix}
0 & - 1 \\
 1 & 0 
\end{pmatrix}^\dagger	
 e^{  i \theta} \sin | z | 
\begin{pmatrix}
0 & -1 \\
 1 & 0 
\end{pmatrix}	 \nonumber \\
&& 
\cos | z | 
\begin{pmatrix}
1 & 0 \\
 0 & 1 
\end{pmatrix}	
\cos | z | 
\begin{pmatrix}
1 & 0 \\
 0 & 1 
\end{pmatrix}^\dagger	
+ e^{ - i \theta} \sin | z | 
\begin{pmatrix}
0 & 1 \\
 - 1 & 0 
\end{pmatrix}	
 e^{  i \theta} \sin | z | 
\begin{pmatrix}
0 & -1 \\
 1 & 0 
\end{pmatrix}	
= (\cosh |\xi |)^2 + (\sinh |\xi |)^2 = 1 \,  . 
\ee
For Fermions, the second condition in (\ref{BT1}) reads:
 \be
 \alpha \beta^T +  \beta \alpha^T = 0 \, , 
 \ee
 and we get:
\be
&& \cos | z | 
\begin{pmatrix}
1 & 0 \\
 0 & 1 
\end{pmatrix}	
 e^{  i \theta} \sin | z | 
\begin{pmatrix}
0 & - 1 \\
 1 & 0 
\end{pmatrix}^T
+
e^{  i \theta} \sin | z |  
\begin{pmatrix}
0 & -1 \\
 1 & 0 
\end{pmatrix}
\cosh | z | 
\begin{pmatrix}
1 & 0 \\
 0 & 1 
\end{pmatrix}^T	 \nonumber \\
&& = \left[  \cos | z | e^{  i \theta} \sin | z |  
-
 e^{  i \theta} \sin | z |  \cosh | z | \right] \begin{pmatrix}
0 & 1 \\
 - 1 & 0 
\end{pmatrix}
= 0 \, .
\ee

For later reference let us also take the dagger of the second transformation in (\ref{transBosons}), which tuns into:
\be
&& 
a^\prime = S_2(z) a S_2^\dagger(z) =  (\cos |z|) a - (e^{i \theta }  \sin | z | ) b^\dagger   \, , \nonumber \\
&& 
b^{\prime\dagger} = S_2(z) b^\dagger S_2^\dagger(z) =  ( e^{ - i \theta }  \sin |z|) a + (\cos | z | ) b^\dagger 
\, .
\label{apbpd}
\ee

Later we will need the following identifications respectively for the Rindler and the Schwarzschild spacetime, 
\be
&& a \equiv c^{\rm I}_\bfk \equiv c^{(\rm out)}_\bfk \, , \nonumber \\
&& b^\dagger \equiv  d^{\rm II \dagger}_{-\bfk} \equiv d^{(\rm int) \dagger}_{- \bfk} \,  , \nonumber  \\
&& a' \equiv a_\bfk \equiv a^{(\rm in)}_\bfk \, ,  \nonumber \\
&& b^{' \dagger} \equiv  b^{\dagger}_{-\bfk} \equiv b^{(\rm in) \dagger}_{- \bfk} \, .
\label{correspond}
\ee

 
For future reference and in order to fix the notation, we notice that 
in Fabbri-Salas \cite{Fabbri} and Townsend \cite{Townsend}, the Bogoliubov transformation relates the operators $a$ and $b$ to $a'$ and $b'$ according to (\ref{transBosons}), while later for the case of the Rindler spacetime, the Bogoliubov transformation will be understood for $a'$ and $b^{' \dagger}$ like in (\ref{apbpd}).

 \subsection{Unitarity theorem in a general background}
{\em Theorem}. The general Bogoliubov transformation derived in Section~(\ref{QFTCS}) is always realized by a unitary squeezing operator $S_{\rm sq}$.

For two-dimensional systems, the proof follows directly from the results in Section~(\ref{SqBTB}) for bosons and Section~(\ref{Sq2Fermions}) for fermions.
The generalization of the Bogoliubov transformation to the systems of interest is simply an infinite block-diagonal matrix of two-by-two blocks, indexed by ${\bf k}$ and $-{\bf k}$ for all ${\bf k}$. Hence, the full squeezing operator is the product over ${\bf k}$ of single-mode squeezing operators.
The reader can find the fully general proof for non-block-diagonal Bogoliubov transformations in \cite{Stone:2023fkj} and \cite{LoSollie}.

This theorem will be crucial in defining a unitary S-matrix in curved spacetime and in demystifying the concept of particle in curved spacetime \cite{BD}.

\begin{center}
\fbox{%
\begin{minipage}{0.9\textwidth}
\begin{center}
\textbf{Theorem.} The general Bogoliubov transformation is always realized by a\\ \textbf{unitary squeezing operator} $S_{\rm sq}$.
\end{center}
\end{minipage}}
\end{center}

\vspace{0.3cm}
For 2D systems: proof from bosons/fermions sections.

\vspace{0.3cm}
\textbf{Generalization:} infinite block-diagonal matrix of $2\times2$ blocks,
indexed by ${\bf k}$ and $-{\bf k}$ $\forall\,{\bf k}$:

\[
S_{\rm sq} = \prod_{{\bf k}} S_{\rm sq}({\bf k}), 
\qquad 
S_{\rm sq}({\bf k}) = 
\begin{pmatrix}
\ast & \ast \\
\ast & \ast
\end{pmatrix}_{{\bf k}, -{\bf k}}
\]

\vspace{0.2cm}
\begin{center}
\fbox{%
\begin{minipage}{0.9\textwidth}
\begin{center}
The full squeezing operator = \textbf{product over ${\bf k}$} of single-mode squeezing operators.
\end{center}
\end{minipage}}
\end{center}

\[
S_{\rm sq} = 
\begin{pmatrix}
\boxed{S_{\rm sq}({\bf k}_1)} & 0 & 0 & \cdots \\
0 & \boxed{S_{\rm sq}({\bf k}_2)} & 0 & \cdots \\
0 & 0 & \boxed{S_{\rm sq}({\bf k}_3)} & \cdots \\
\vdots & \vdots & \vdots & \ddots
\end{pmatrix}
\]

\vspace{0.2cm}
Each block: $2\times2$ matrix mixing modes ${\bf k}$ and $-{\bf k}$.

\[
\boxed{S_{\rm sq} = \prod_{{\bf k}} S_{\rm sq}({\bf k})}
\]

\section{Quantum field theory in the Rindler/Black hole background} 
As a master example, we review quantum field theory in Rindler spacetime and implement the squeezing representation of the Bogoliubov transformations as described in the previous Section~(\ref{SqBT}).
For the case at hand, it is sufficient to consider the two-mode squeezing operator, as will shortly become evident from the Bogoliubov transformations.

\subsection{QFT in an inertial reference frame} 

In this section we review the Dirac wave equation and its solutions in Minkowski spacetime. 
The Dirac wave equation for a free fermion in Minkowski spacetime in $D=4$ reads:
\begin{equation}
i\gamma ^{\mu }\partial _{\mu }{\psi }-m\psi =0 \, ,  \notag
\end{equation}%
where $m$ is the mass of the fermion, $\gamma^{\mu }$ are the Dirac gamma
matrices, and $\psi $ is a spinorial wavefunction. The
most suitable coordinates in Minkowski spacetime are the Cartesian coordinates 
$x^{\mu} = (t,\vec{x})$ ($\mu= 0,1,2,3$).
We can expand the field in terms of positive- and negative-energy
solutions of the Dirac equation, i.e.\  
$\psi _{k}^{+}$ and $\psi _{k}^{-}$, 
which form a complete orthonormal basis.
Therefore, 
\begin{equation}
\label{Minkowski_field}
\psi =\int d k
\, ( a_{k}\psi_{k}^{+}+b_{k}^{\dagger }\psi_{k}^{-}) \, , 
\end{equation}
where $k$ stands for the three-wavevector $\vec{k}$, which labels the modes.
The positive- and negative-energy modes have the following form in Minkowski spacetime,
\be
\left(\psi_{k}^{\pm}\right)_{s} = \frac{1}{\sqrt{2\pi\omega_k}}
\,\phi^{\pm}_{s} \, e^{\pm i(\vec{k}\cdot \vec{{x}} - \omega_k t)} \, , 
\ee
where $\omega_k = (m^2 + \vec{k}^2)^{1/2}$, and
$\phi_{s}=\phi_{s}(\vec{k})$ is a spinor with $s=\{\uparrow,\downarrow\}$ indicating
spin-up or spin-down along the quantization axis, satisfying the normalization relations 
\be
\pm \bar{\phi}^{\pm}_{s}\,\phi^{\pm}_{s'} = (\omega_k/m)\,\delta_{ss'} \, ,  \quad \bar{\phi}^{\pm}_{s'}\,\phi^{\mp}_{s'}=0 \, ,
\ee
with the adjoint spinor given by $\bar{\phi}^{\pm}_{s} = \phi^{\pm\dagger}_{s}\,\gamma^0$.
The above positive- and negative-energy solutions satisfy the orthonormality relations:
\be
( \psi_{k}^{+}, \psi_{k'}^{+} ) = -(\psi_{k}^{-}, \psi_{k'}^{-} ) = \delta(k-k'), \quad
( \psi_{k}^{\pm}, \psi_{k'}^{\mp} ) = 0 \, ,
\label{scalarproM}
\ee
where the Dirac inner product for two mode functions is given by
\be
\big( \phi(\vec{x},t), \varphi(\vec{x},t) \big)
= \int \, d^3{x} \, \phi^\dagger(\vec{x},t)\,\varphi(\vec{x},t).
\ee
The modes $\psi_{k}^{\pm}$ are classified as positive- and negative-frequency with respect to
(the future-directed Minkowski Killing vector) $\partial_t$  for $\omega_k > 0$, i.e.
\be
\partial_t\, \psi_{k}^{\pm} = \mp\, i\,\omega_k \,\psi_{k}^{\pm}, \qquad \omega_k > 0 \, .
\ee

The operators $a_{k}^{\dagger },b_{k}^{\dagger}$ and $a_{k},b_{k}$
are the creation and annihilation operators for the positive- and negative-energy
solutions of momentum $k$ that satisfy the anticommutation relations:
\begin{equation}
\{a_{i},a_{j}^{\dagger }\}=\{b_{i},b_{j}^{\dagger }\}=\delta _{ij},  \notag
\end{equation}%
while all the other anticommutators vanish. 
The Minkowski vacuum is
defined by the absence of excitations, namely: 
\be
|0\rangle =\prod_{kk'}|0_{k}\rangle^{+}|0_{k'}\rangle^{-} \quad \mbox{such that}: \quad 
a_{k}|0_{k}\rangle^{+}=b_{k}|0_{k}\rangle^{-}=0 \, , 
\label{MinkVa}
\ee
where the $\{+,-\}$ refer to the
particle and anti-particle vacua.

\subsection{QFT in the Rindler reference frame}
Let us consider a uniformly accelerated observer in Cartesian coordinates 
$(t,z)$ \cite{Alsing:2006cj}. The Rindler coordinates $(\tau,\zeta)$ are appropriate for
describing an observer at rest with respect to the moving reference frame. 
However, in order to cover Minkowski spacetime, we need two sets of Rindler coordinates 
that differ from each other by an overall change of sign. 
The two sets of coordinates define the two Rindler regions that are causally disconnected from each other---the observer can choose to accelerate in two opposite directions with respect to the $z$ direction---and are given by:
\begin{eqnarray}\label{Rindler_coords}
a t &=& \sph e^{a\zeta}\sinh(a\tau), \quad a z = \sph e^{a\zeta}\cosh(a\tau), \sph \text{in region I} \, , 
\nonumber \\
&&\\ \notag a t &=& -e^{a\zeta}\sinh(a\tau), \quad a z =
-e^{a\zeta}\cosh(a\tau), \sp \text{in region II} \, , 
\end{eqnarray}
where $a$ denotes the acceleration. 
The above
set of coordinates gives rise to the same \tit{Rindler} metric, i.e.\ 
\be
ds^2 = dt^2 - dz^2 - d^2\mathbf{x}_{\perp} = e^{2a\zeta}\left(d^2\tau - d^2\zeta\right)
 - d^2\mathbf{x}_{\perp},
\ee
where $\mathbf{x}_{\perp} = (x,y)$ are the two extra directions orthogonal to the acceleration.


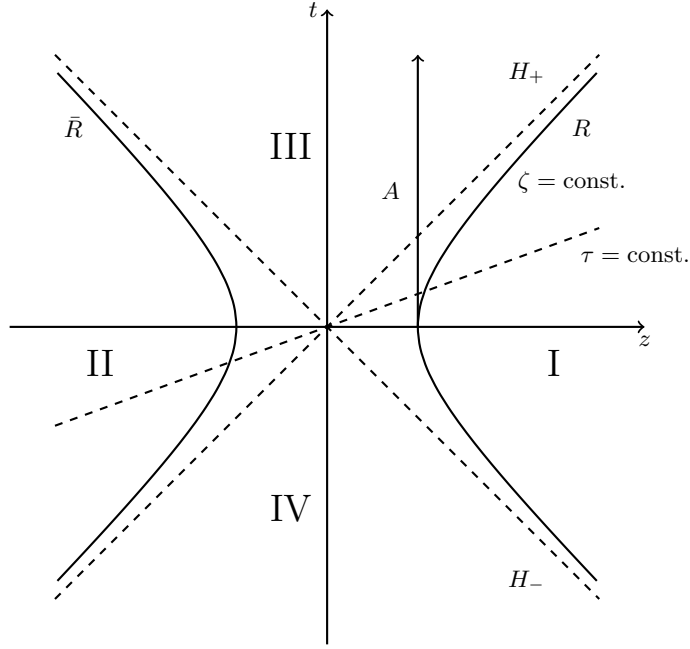
\begin{figure}[h]
\centering
\begin{tikzpicture}[scale=1.2]
    \draw[->, thick] (-3.5,0) -- (3.5,0) node[below] {$z$};
    \draw[->, thick] (0,-3.5) -- (0,3.5) node[left] {$t$};
    
    \draw[thick, dashed] (-3,-3) -- (3,3);
    \draw[thick, dashed] (-3,3) -- (3,-3);
    
    \draw[thick, dashed] (-3,-1.09) -- (3,1.09);
    
    \draw[thick, ->] (1,0) -- (1,3);
    
    \node at (0.7,1.5) {$A$};
    
    \node at (3.4,0.8) {$\tau = {\rm const.}$};
    
    \node at (2.2,2.8) {$H_+$};
    \node at (2.2,-2.8) {$H_-$};
    
    \draw[thick, domain=-2.8:2.8, smooth, variable=\x] 
        plot ({sqrt(\x*\x + 1)}, {\x});
    
    \draw[thick, domain=-2.8:2.8, smooth, variable=\x] 
        plot ({-sqrt(\x*\x + 1)}, {\x});
    
    \node at (2.7,1.6) {$\zeta = {\rm const.}$};
    
    \node at (2.8,2.2) {$R$};
    \node at (-2.8,2.2) {$\bar{R}$};
    
    \node at (-2.5,-0.4) {\Large II};
    \node at (2.5,-0.4) {\Large I};
    \node at (-0.4,2.0) {\Large III};
    \node at (-0.4,-2.0) {\Large IV};
    
\end{tikzpicture}
\caption{
The hyperbolic spacetime trajectories represent the motion of an observer at rest, $\zeta$ constant, in the accelerated reference frame. 
The lines of constant proper time $\tau$ for the accelerated observer are straight lines through the origin. 
Notice that $\tau$ flows in the direction of $t$ in region I, but flows in the direction of $-t$ in region II. 
A uniformly accelerated observer Rob ($\rm R$) with
acceleration $a$ travels on a hyperbola constrained to region I, while a fictitious observer anti-Rob
$(\bar{\rm R})$ travels on the corresponding hyperbola in region II, given by the negative of Rob's coordinates, i.e.\ it moves in the opposite direction.
The observer Alice ($\rm A$), at rest in the $z$ coordinate, crosses the horizons $H_\pm$ at a finite value of the Minkowski time $t$ that corresponds  
to $\tau=\pm\infty$ for R.
}
\label{fig:rindler}
\end{figure}


In Fig.~\ref{fig:rindler} the two curves R and $\left(\rm{\bar{R}}\right)$ correspond to two observers moving in opposite directions with uniform acceleration, namely towards positive and negative values of the coordinate $z$. Therefore, the two observers will never meet and regions I and II are causally disconnected, as is evident from drawing the light cones in the coordinates $(t,z)$.

In both regions I and II, 
the coordinates $(\tau,\zeta)$ take values from minus infinity to plus infinity, namely $\tau \in [-\infty, +\infty]$ 
and $\zeta \in [-\infty, +\infty]$. 
Hence, both regions I and II admit a separate
quantization procedure with positive- and negative-energy solutions, namely 
\be
\left( \psi^{{\rm I}+}_{ \bf k}, \psi^{{\rm I} -}_{ \bf k} \right)  \quad  \
\mbox{and} \quad \left( \psi^{{\rm II} +}_{\bf k}, \psi^{{\rm II} -}_{ \bf k} \right) . 
\ee
The Rindler metric
is static, i.e.\ it is independent of $\tau$, thus it admits solutions of the form:
\be
e^{-i\omega\tau} \, \phi_{\alpha}(\zeta,\mathbf{x}_{\perp}) \, ,
\ee
where $\phi_{\alpha}(\zeta,\mathbf{x}_{\perp})$ is a spinor depending only on the spatial coordinates.  
Therefore, particles and anti-particles will be classified with
respect to the future-directed timelike Killing vector in each region. In region I
this is given by $\partial_{\tau}$, which can be expressed in $(t,z)$ coordinates (\ref{Rindler_coords}) as follows, 
\be
\partial_{\tau} = \frac{\partial t}{\partial\tau}\,\partial_{t}
+ \frac{\partial z}{\partial\tau}\,\partial_{z}
=a\,(z\partial_{t} + t \partial_{z}) \, , 
\ee
which is a boost into the comoving frame.
 Hence, the mode solutions $\psi^{{\rm I}+}_{\bf k} \sim e^{-i\omega\tau}$ with $\omega>0$  
are 
positive-frequency solutions in region I, namely 
\be \partial_{\tau} \psi^{{\rm I} +}_{\bf k} = -i\omega \psi^{{\rm I}+}_{\bf k} \, , \quad \omega > 0 \, .
\ee
On the other hand, in region II $\partial_{\tau}$ points in the opposite direction of $\partial_{t}$
($\tau$ increases in the direction of $-t$, see Fig.~\ref{fig:rindler}).
Therefore, in region II the future-directed timelike Killing vector is given by $\partial_{-\tau} = -\partial_{\tau}$ \cite{BD,carroll}. 
Thus, a solution in region II with time dependence $e^{-i\omega\tau}$ with $\omega>0$ is actually a negative-frequency
mode, i.e.\ 
\be
\partial_{-\tau}\,e^{-i\omega\tau} = i\omega\,e^{-i\omega\tau}. 
\ee
Thus, in region II the
positive-frequency mode is given by 
\be
\psi^{{\rm II} +}_{\bf k} \sim e^{i\omega\tau}
\quad 
 \mbox{with} \quad \omega>0 \quad \Longrightarrow \quad 
 \partial_{-\tau} \psi^{{\rm II} +}_{\bf k} = -i\omega \psi^{{\rm II}+}_{\bf k} \, .
 \ee
Since regions I and II are causally disconnected, 
the modes $\psi^{{\rm I} \pm}_{\bf k}$ have support only in region I and vanish in region II, while the
opposite is true for the modes $\psi^{{\rm II} \pm}_{\bf k}$ in region II.
The Rindler modes satisfy orthonormality relations
similar to those in Minkowski spacetime (\ref{scalarproM}), i.e.\ 
\be 
(\psi^{\sigma\pm}_{\bf k},\psi^{\sigma'\mp}_{\bf k'})=0 
\quad  \mbox{and} \quad 
(\psi^{\sigma\pm}_{\bf k},\psi^{\sigma'\pm}_{\bf k'}) = \pm\,\delta_{\sigma,\sigma'}\,\delta( {\bf k}- {\bf k'} ) 
\label{RinOrt}
\ee
where we introduced the label $\sigma\in\{{\rm I,II} \}$.

In region I, we denote by $(c_{\bf k}^{\rm I},c^{{\rm I} \dagger}_{\bf k})$ the annihilation
and creation operators for fermions and by 
$(d_{\bf k}^{\rm I},d^{{\rm I} \dagger}_{\bf k})$ the annihilation and creation operators
for anti-fermions, while in region II the corresponding particle and
anti-particle operators are 
$(c_{\bf k}^{\rm II},c^{{\rm II} \dagger}_{\bf k})$ and $(d_{\bf k}^{\rm II},d^{{\rm II} \dagger}_{\bf k})$,
which obey 
the usual Dirac anti-commutation relations, i.e.\ 
\be
\{c_{\bf k}^{\sigma},c^{\sigma'\dagger}_{\bf k'}\} = \{d_{\bf k}^{\sigma},d^{\sigma'\dagger}_{\bf k'}\} =
\delta_{\sigma\sigma'}\,\delta_{\bf k k'} \, .
\ee
All other
anti-commutators, including those between operators in region I and
region II, vanish.  
Given the two sets of modes in each Rindler region, the Dirac
field can be expanded, in analogy to \Eq{Minkowski_field} in Minkowski spacetime, as 
\begin{equation}\label{Rindler_field}
\psi = \int d{\bf k} \left( c_{\bf k}^{\rm I}\psi _{\bf k}^{{\rm I}+}+d_{\bf k}^{{\rm I} \dagger }\psi _{\bf k}^{{\rm I} -}+c_{{\bf k}}^{\rm II}\psi
_{\bf k}^{{\rm II} +}+d_{\bf k}^{{\rm II} \dagger }\psi _{\bf k}^{{\rm II} -} \right).
\end{equation}
\Eq{Minkowski_field} and \Eq{Rindler_field} represent, respectively, the
decomposition of the Dirac field in Minkowski and Rindler spacetime. 
Hence, we are able to relate the Minkowski and Rindler
creation and annihilation operators by making use of the 
inner product. 
Using the Rindler orthogonality relations (\ref{RinOrt}) and
\Eq{Rindler_field} we have $c^{\sigma}_{\bf k} =
(\psi^{\sigma+}_{\bf k},\psi)$. 
The Bogoliubov coefficients are given by the
inner product between the Rindler mode wavefunctions and the Minkowski
positive- and negative-frequency modes 
\be
\label{bogo_coeffs}
\alpha^{\sigma}_{\bf kk'} = \left( \psi^{\sigma+}_{\bf k},
\psi^{+}_{\bf k'}\right), \qquad \beta^{\sigma}_{\bf kk'}  = \left(
\psi^{\sigma+}_{\bf k}, \psi^{-}_{ \bf k'}\right). 
\ee 
Therefore, replacing \Eq{Minkowski_field} for
$\psi$ in $c^{\sigma}_{\bf k} =
(\psi^{\sigma+}_{\bf k},\psi)$, we get: 
\be
\label{bogo_c} 
c^{\sigma}_{\bf k} =
\int dk' \,\left( \alpha^{\sigma}_{\bf kk'} \, a_{\bf k'} +
\beta^{\sigma}_{\bf kk'} \, b^{\dagger}_{\bf k'}\right), \quad \sigma \in
\{{\rm I}, {\rm II} \} \, .
\ee 
A similar calculation
for $d^{\sigma}_{\bf k}$ yields the corresponding expression 
\be
\label{bogo_d}
d^{\sigma}_{\bf k} = \int d{\bf k}' \,\left( \alpha^{\sigma}_{\bf kk'} \, b_{\bf k'} +
\beta^{\sigma}_{\bf kk'} \, a^{\dagger}_{\bf k'}\right), \quad \sigma \in
\{{\rm I} , {\rm II} \}  \,  , 
\ee 
with the Bogoliubov coefficients given in
\Eq{bogo_coeffs}. 
In deriving \Eq{bogo_d} use has been made of the
following properties of the Dirac inner product: $(\phi_1,\phi_2)^*
= (\phi_1^*,\phi_2^* ) = (\phi_2,\phi_1)$. 
The details of the calculation of the Bogoliubov coefficients can be found in 
\cite{takagi,rocio,mcmahon_alsing_embid} and the outcome reads:
\be
\left[
\begin{array}{c}
  a_{\bf k} \\
  \\
  b_{-{\bf k}}^\dagger \\
\end{array}
\right] =
\left[
\begin{array}{cc}
  \cos r & -e^{-i\phi} \,\sin r \\
  \\
  e^{i\phi} \,\sin r & \cos r \\
\end{array}
\right] \,
\left[
\begin{array}{c}
  c^{\rm I}_{\bf k} \\
  \\
  d^{\rm II\dagger}_{-{\bf k}} \\
\end{array}
\right],
\label{bogo_ops_a_bdag}
\quad \mbox{where} \quad
\tan r = \exp(-\pi\Omega) 
\quad \mbox{with} \quad \Omega \equiv
\frac{\omega}{a} \, . 
\label{PIAPII}
\ee
From \Eq{bogo_ops_a_bdag} and its
adjoint, one can check that given the anti-commutation relations of the Rindler
operators, the Minkowski anti-commutation relations are preserved.

We can explicitly write \Eq{bogo_ops_a_bdag} as:
\be
&& a_{\bf k} = \cos r \, c^{\rm I}_{\bf k} - e^{- i \phi} \sin r \, d^{\rm II \dagger}_{-{\bf k}}  \, , \label{riscrivo1} \\
&& b^\dagger_{-{\bf k}} = e^{i \phi} \sin r \, c^{\rm I}_{\bf k} + \cos r \, d^{\rm II \dagger}_{-{\bf k}}  \, ,
\label{riscrivo2}
\ee
which involve only the operators $c_k^{\rm I}, \, d_{-k}^{\rm II \dagger}$ on the right-hand side. 
Taking the adjoint of the second equation (\ref{riscrivo2}) and displaying again the first equation, we get the following two equations,  
\be
&& a_{\bf k} = \cos r \, c^{\rm I}_{\bf k} - e^{- i \phi} \sin r \, d^{\rm II \dagger}_{-{\bf k}}  \, , \label{riscrivo1A} \\
&& b_{-{\bf k}} = e^{- i \phi} \sin r \, c^{\rm I \dagger}_{\bf k} + \cos r \, d^{\rm II}_{-{\bf k}}  \, .
\label{riscrivo2A}
\ee
which now involve all the operators in Rindler frame, i.e.\ 
$c_{\bf k}^{\rm I}, \, c_{\bf k}^{\rm I \dagger}, \, d_{-{\bf k}}^{\rm II}, \, d_{-{\bf k}}^{\rm II \dagger}$.
Given (\ref{riscrivo1A}) and (\ref{riscrivo2A}), we can now define the BT in terms of the matrices $\alpha$ and $\beta$
according to the definitions (\ref{contFT}) and identifying
$a \equiv c^{\rm I}_{\bf k}$ and $b \equiv d^{\rm II}_{- {\bf k} }$.

Note that the Bogoliubov transformation (\ref{PIAPII}) 
mixes a particle in region I and an anti-particle in region II.
Correspondingly, the Bogoliubov transformation that mixes an
anti-particle mode in region I and a particle in region II is given
by 
\be
\label{bogo_ops_b_adag}
 \left[
\begin{array}{c}
  b_{\bf k} \\
  \\
  a_{-{\bf k}}^\dagger \\
\end{array}
\right] =
\left[
\begin{array}{cc}
  \cos r & e^{-i\phi} \,\sin r \\
  \\
  -e^{-i\phi} \,\sin r & \cos r \\
\end{array}
\right] \,
\left[
\begin{array}{c}
  d^{{\rm I}}_{\bf k} \\
  \\
  c^{ {\rm II} \dagger}_{-{\bf k}} \\
\end{array}
\right]. 
\ee

Since the anti-commutators between particle and
anti-particle operators vanish, and since anti-commutators of operators in region I and region II are zero, it is easy to show that the Minkowski operators
in \Eq{bogo_ops_a_bdag} anti-commute with the Minkowski operators in
\Eq{bogo_ops_b_adag} (as they should, because ${\bf k}$ and $-{\bf k}$ represent
two separate modes).

Let us now come to the squeezing realization of the Bogoliubov transformation. 
In particular, after having related Minkowski and Rindler creation and annihilation operators,
we now wish to relate the Minkowski vacuum to
the corresponding Rindler vacuum.
We notice that (\ref{bogo_ops_a_bdag}) can be written
as a two-mode squeezing transformation \cite{book2}, 
\be
\left[
\begin{array}{c}
  a_{\bf k} \\
  \\
  b_{-{\bf k}}^\dagger \\
\end{array}
\right] =
S\,
\left[
\begin{array}{c}
  c^{{\rm I}}_{\bf k} \\
  \\
  d^{{\rm II} \dagger}_{-{\bf k}} \\
\end{array}
\right]\,
S^\dagger
\label{primaBTF}
\ee
for a single mode ${\bf k}$, with $S$ given by
\be
S = \exp \left[  r\,\left( c^{{\rm I} \dagger}_{\bf k} \,d^{{\rm II} \dagger}_{-{\bf k}} \,e^{-i\phi}
+  c^{{\rm I}}_{\bf k} \,d^{{\rm II} }_{-{\bf k}} \,e^{i\phi} \right)  \right].
\ee
Moreover, the adjoint of (\ref{primaBTF}) gives:
\be
\left[
\begin{array}{c}
  a^\dagger_{\bf k} \\
  \\
  b_{-{\bf k}} \\
\end{array}
\right] =
S\,
\left[
\begin{array}{c}
  c^{ {\rm I} \dagger }_{\bf k} \\
  \\
  d^{ {\rm II} }_{-{\bf k}} \\
\end{array}
\right]\,
S^\dagger \, .
\ee

Using the Campbell identity:
\be e^X \,Y \,e^{- X} = Y + [X , Y ] + \frac{1}{2 !} [X,[X,Y]] +  \frac{1}{3!} [X,[X,[X,Y]]] + \cdots \, , 
\ee
we can directly verify by brute force the correctness of (\ref{primaBTF}): 
\be
&& \label{a_k}
a_{\bf k} = S \, c^{ \rm I}_{\bf k} \, S^\dagger = c^{\rm I}_k - r e^{-i\phi}\,d^{{\rm II} \dagger}_{- {\bf k}}
- \frac{r^2}{2!}\,c^{\bf I}_{\rm k} + \frac{r^3}{3!}\,e^{-i\phi}\,d^{{\rm II} \dagger}_{-{\bf k}} + \cdots 
 = \cos r \,c^{\rm I}_{\bf k} - e^{-i\phi} \sin r \,d^{{\rm II} \dagger}_{-{\bf k}} \, ,
\\
&&
\label{b_negk_dag}
b^\dagger_{- {\bf k}} = S d^{{\rm II} \dagger}_{- {\bf k} } S^\dagger = d^{{\rm II} \dagger}_{- {\bf k} } + r e^{i\phi}\,c^{ \rm I}_{\bf k}
- \frac{r^2}{2!}\,d^{{\rm II} \dagger}_{-{\bf k}} - \frac{r^3}{3!}\,e^{i\phi}\,c^{\rm I}_{{\bf k}} + \cdots 
 = \cos r \,d^{{\rm II} \dagger}_{-{\bf k}} + e^{ {\color{red}+ } i\phi} \sin r \,c^{ \rm I}_{\bf k} \, .
\ee

According to (\ref{MinkVa}), the operators $a_{\bf k}$ and $b_{- {\bf k}}$ respectively annihilate the single-mode particle and anti-particle Minkowski vacua, namely 
\be
 a_{\bf k} |0_{\bf k}\rangle^+ = 0 \, , \quad 
b_{-{\bf k}} |0_{-k}\rangle^- = 0 \, . 
\ee
Since, by (\ref{a_k}), $a_k$ mixes particles in
region I and anti-particles in region II, we can postulate that the
Minkowski particle vacuum for the mode ${\bf k}$ in terms of Rindler Fock
states is given by 
\be
\label{vacM_1} |0_k\rangle^+ = \sum_{n=0}^1 A_n
\,|n_{\bf k}\rangle^+_{\rm I} \,|n_{-{\bf k} }\rangle^-_{\rm II} \, , 
\ee 
where the Hilbert spaces in I and II are defined by
\be
&&  c^{\rm I}_{\bf k} |0_{\bf k} \rangle^+_{\rm I} = 0 \, , \quad 
  d^{\rm II}_{-{\bf k}} |0_{- {\bf k}}\rangle^-_{\rm II} = 0 \, , 
   \nonumber \\
&& c^{{\rm I}\dagger}_k |0_{\bf k}\rangle^+_{\rm I} = |1_k\rangle^+_{\rm I} \, ,  \quad 
  d^{{\rm II} \dagger}_{-{\bf k}} |0_{-k}\rangle^-_{\rm II} = |1_{-{\bf k}}\rangle^-_{\rm II} \, . 
  \label{c_d_ops}
\ee 
Notice that the Fock states in the Rindler regions I and II carry a subscript I or II, while the 
Minkowski Fock states are indicated without any 
subscript on the kets. 
Moreover, the $\{+,-\}$ ket superscript indicates a
particle or an anti-particle state, respectively. 
Applying $a_{\bf k}$ from (\ref{a_k}) to the vacuum (\ref{vacM_1}), we get:
\be 
0 &=& a_{\bf k} |0_{\bf k}\rangle^+   \nonumber \\
&=& \left( \cos r \,c^{\rm I}_{\bf k} - e^{-i\phi} \sin r \,d^{{\rm II} \dagger}_{-{\bf k}}
\right) \, \sum_{n=0}^1 A_n \,|n_{\bf k} \rangle^+_{\rm I} \, |n_{- {\bf k} }\rangle^-_{\rm II} 
\no 
&=& (A_1 \,\cos r - A_0 \, e^{-i\phi}\,\sin
r)\,|0_{\bf k}\rangle^+_{\rm I} \,|1_{-{\bf k}}\rangle^-_{\rm II}
\quad \Longrightarrow
\quad 
 A_1 =
A_0 \,e^{-i\phi}\,\tan r \, . 
\label{azero}
\ee
The normalization condition 
$^+\langle 0_{\bf k} | 0_{\bf k} \rangle^+ = |A_0|^2 + |A_1|^2 = 1$ yields $A_0 = \cos r$, so that
we finally arrive at
\be
\label{vacM} 
|0_{\bf k}\rangle^+ &=& \cos r
\,| 0_{\bf k} \rangle^+_{\rm I} \,| 0_{ - {\bf k} } \rangle^-_{\rm II} + e^{-i\phi}\,\sin r
\, |1_{\bf k} \rangle^+_{\rm I} \, |1_{- {\bf k} }\rangle^-_{\rm II}, \no
&=&\left(
\cos r \, + e^{-i \phi}\,\sin r \,c^{{\rm I} \dagger}_{\bf k}\,d^{{\rm II} \dagger}_{-{\bf k} }
\right) \,|0_{\bf k} \rangle^+_{\rm I} \, |0_{-{\bf k} }\rangle^-_{\rm II}
= 
\cos r \left( 1 + e^{-i\phi}\,\tan r \, c^{{\rm I} \dagger}_{\bf k} \,d^{\rm II\dagger}_{-{\bf k} }
\right) \,|0_{\bf k} \rangle^+_{\rm I} \,| 0_{- {\bf k} } \rangle^-_{\rm II} \, , 
\ee
which exactly coincides with (\ref{PsiBCS}) for $\phi = - \theta$, 
$a^\dagger \equiv  c^{{\rm I} \dagger}_{\bf k}$ and $b^\dagger \equiv  d^{{\rm II} \dagger}_{-{\bf k}}$,
namely
\be
|0_{\bf k}\rangle^+ = S_2(z) \,|0_{\bf k} \rangle^+_{\rm I} \,| 0_{-{\bf k} } \rangle^-_{\rm II} \, . 
\label{works} 
\ee

Moreover, the operator $b_{-{\bf k}}$ also annihilates the vacuum (\ref{vacM}), i.e.\ 
$| 0_{\bf k} \rangle^+$ is implicitly the tensor product $| 0_k\rangle^+ | 0_{-{\bf k}}\rangle^-$, and 
\be
b_{-{\bf k} } | 0_{\bf k} \rangle^+ | 0_{-{\bf k} }\rangle^- = 0 \, . 
\ee
The full state reads:
\be
\label{vacM4} 
|0_{\bf k} \rangle^+ |0_{-{\bf k}}\rangle^- 
= \left(
\cos r \, + e^{-i\phi}\,\sin r \,c^{{\rm I}\dagger}_{\bf k}\,d^{{\rm II}\dagger}_{-{\bf k}}
\right) \,|0_{\bf k}\rangle^+_{\rm I}\,|0_{-{\bf k}}\rangle^-_{\rm II} 
= S_2(z) \,|0_{\bf k} \rangle^+_{\rm I} \,| 0_{-{\bf k}} \rangle^-_{\rm II} \, . 
\ee

As an explicit example, we calculate
$a_{\bf k}^\dagger \,|0_{\bf k}\rangle^+$ using the adjoint of (\ref{a_k}), which reads:
\be
\label{a_k_dag}
a_{\bf k}^\dagger = \cos r \,c^{{\rm I}\dagger}_{\bf k} - e^{+i\phi} \sin r \,d^{\rm II}_{-{\bf k}},
\ee
and the vacuum 
$|0_{\bf k}\rangle^+$ in (\ref{vacM}), yielding
\be
\label{1M}
a_{\bf k}^\dagger \,|0_{\bf k}\rangle^+ &=&
\left(
\cos^2 r \,c^{{\rm I} \dagger}_{\bf k} - \sin^2 r \,d^{\rm II}_{-{\bf k}}\,c^{{\rm I}\dagger}_{\bf k}\,d^{{\rm II}\dagger}_{-{\bf k}}
\right)\,\,|0_{\bf k}\rangle^+_{\rm I}\,|0_{-{\bf k}}\rangle^-_{\rm II}, \no
&=& \left(
\cos^2 r \,c^{I\dagger}_{k} + \sin^2 r \,c^{{\rm I} \dagger}_{\bf k}\,d^{\rm II}_{- {\bf k}}\,d^{{\rm II} \dagger}_{-{\bf k}}
\right)\,\,|0_k\rangle^+_{\rm I} \, | 0_{-k}\rangle^-_{\rm II}
 = c^{{\rm I} \dagger}_{\bf k} \, |0_{\bf k} \rangle^+_{\rm I} \,| 0_{-{\bf k} }\rangle^-_{\rm II}
 = |1_{\bf k} \rangle^+_{\rm I} \,| 0_{-{\bf k} }\rangle^-_{\rm II} \, , 
\ee
where in the second equality the minus sign comes from the anti-commutation relations between
$c^{{\rm I}\dagger}_{\bf k}$ and $d^{\rm II}_{-{\bf k} }$,
and in the third equality we have used the anti-commutation relations to write
$d^{\rm II}_{-{\bf k} }\,d^{ {\rm II}\dagger}_{-{\bf k}} = 1 - d^{{\rm II} \dagger}_{-{\bf k}}\,d^{\rm II}_{-{\bf k}}$, where 
the second term annihilates the anti-particle vacuum. 
One can also verify that the double action of 
$a_{\bf k}^\dagger$ on the vacuum yields zero,
\be
\label{a_k_dag_on_1_k}
a_{\bf k}^\dagger |1_{\bf k}\rangle^+ = \left( \cos r \,c^{ {\rm I} \dagger}_{\bf k} - e^{+i \phi} \sin r \,d^{\rm II}_{-{\bf k}} \right) \,
|1_{\bf k}\rangle^+_{\rm I}\,|0_{-{\bf k}}\rangle^-_{\rm II} = 0 \, ,
\ee
in accordance with Fermi statistics.

Finally, (\ref{bogo_ops_b_adag}) is also realized by a squeezing operator, i.e.\ 
\be
\left[
\begin{array}{c}
  b_{\bf k} \\
  \\
  a_{-{\bf k}}^\dagger \\
\end{array}
\right] =
\bar{S}\,
\left[
\begin{array}{c}
  d^{{\rm I}}_{\bf k} \\
  \\
  c^{ {\rm II} \dagger}_{-{\bf k}} \\
\end{array}
\right]\,
\bar{S}^\dagger
\label{altri}
\ee
and for the Hermitian conjugate operators one has:
\be
\left[
\begin{array}{c}
  b^\dagger_{\bf k} \\
  \\
  a_{- {\bf k}} \\
\end{array}
\right] =
\bar{S}\,
\left[
\begin{array}{c}
  d^{{\rm I} \dagger}_{\bf k} \\
  \\
  c^{ {\rm II} }_{- {\bf k}} \\
\end{array}
\right]\,
\bar{S}^\dagger
\label{altriH}
\ee
where
\be
\bar{S} = \exp \left[ {  -}  r\,\left( d^{{\rm I} \dagger}_{\bf k} \, c^{{\rm II} \dagger}_{-{\bf k}} \,e^{-i\phi}
+  d^{{\rm I}}_{\bf k} \, c^{{\rm II} }_{- {\bf k}} \,e^{i\phi} \right) \right] \, .
\label{BarS}
\ee
Notice that $S$ and $\bar{S}$ commute because the first involves the operators $c^{{\rm I}}$ and $d^{\rm II}$ while the second involves $c^{{\rm II}}$ and $d^{\rm I}$; moreover, they are both exponentials of quadratic anti-commuting operators, i.e.
\be
[ S, \, \bar{S} ] = 0 \, .
\ee
Therefore, the state (\ref{vacM}) was not yet the final one. Indeed, since 
\be
a_{- {\bf k}} | 0_{- {\bf k}} \rangle^+ = 0 \, , \quad b_{\bf k} | 0_{\bf k} \rangle^- = 0 \, ,
\ee
following the very same steps that led to (\ref{vacM}), we get:
\be
\label{vacM2} 
|0_{ - {\bf k}}\rangle^+ |0_{\bf k}\rangle^- 
= \bar{S}_2(z) \,|0_{ - {\bf k}} \rangle^+_{\rm II}\,|0_{\bf k}\rangle^-_{\rm I} \, .
\ee
Finally, the complete vacuum state is:
\be
|0_{\bf k} \rangle^+ |0_{- {\bf k} }\rangle^- 
|0_{ - {\bf  k} }\rangle^+ |0_{\bf k}\rangle^- 
=  S_2(z) \bar{S}_2(z) 
\, |0_{\bf k} \rangle^+_{\rm I} \, | 0_{-{\bf k} } \rangle^-_{\rm II} 
\,|0_{ - {\bf k} } \rangle^+_{\rm II}\,|0_{\bf k}\rangle^-_{\rm I} 
\, .
\label{finalS2S2}
\ee

Alice has access to both Minkowski states, $|0_{\bf k}\rangle^+$ and  $|1_{\bf k}\rangle^+$, which
correspond to the particle field of mode $k$. 
On the other hand, an observer moving with uniform
acceleration $a$ in one of the regions I or II has no access to field modes
in the causally disconnected region. 
Therefore, the observer has to trace over the inaccessible region, constituting an unavoidable loss
of information about the state, which essentially results in the
detection of a mixed state. 

Thus, when a Minkowski observer detects
a vacuum state $|0_{\bf k}\rangle^+ \langle 0_{\bf k}|$ for the mode ${\bf k}$, an
accelerated observer in region I sees a distribution of particles
according to 
\be
\label{eq:modes}
\rho_{\bf k}^{\rm I} &=& {\rm Tr}_{\rm II} \left( |0_{\bf k} \rangle^+ \langle 0_{\bf k} |\right) 
=
\cos^2 r\,|0_{\bf k} \rangle^+_{\rm I}\,^+_{\,\,\, \rm I} \langle 0_{\bf k} | + 
\sin^2 r \, |1_{\bf k} \rangle^+_{\rm I} \,^+_{\,\,\, \rm I} \langle 1_{\bf k} | \, . 
\ee
When the observer in region I
accelerates through the Minkowski particle vacuum $|0_{\bf k}\rangle^+$ of
mode ${\bf k}$, his detector registers a number of particles given by
\be
\label{thermalization}
^+\langle 0_{\bf k} | \,c^{{\rm I} \dagger}_{\bf k} \,c^{\rm I}_{\bf k} |0_{\bf k}\rangle^+
& = &{\rm Tr}_{{\rm I} , {\rm II}}
\left[c^{{\rm I} \dagger}_{\bf k} \,c^{\rm I}_{\bf k}\,|0_{\bf k} \rangle^+ \langle 0_{\bf k} |\right] 
= {\rm Tr}_{\rm I} \left[c^{{\rm I} \dagger}_{\bf k}\,c^{\rm I}_{\bf k}\,\rho_{\bf k}^{\rm I} \right] = \sin^2 r \;\; ^+_{\,\, {\rm I}} \langle 1_{\bf k} | c^{{\rm I} \dagger}_{\bf k}  \,c^{\rm I}_{\bf k} | 1_{\bf k}\rangle^+_{\rm I} \no
&=& \sin^2 r = \frac{1}{e^{2 \pi \Omega}+1} \equiv \frac{1}{e^{\hbar\omega/T}+1} \, , 
\ee
where we used 
$\tan r = \exp(-\pi\Omega)$ with $\Omega = \omega/a$, and we have
introduced the Unruh temperature 
\be
\label{T_U}
 T =\frac{a}{2\pi}.
\ee

Let us \underline{summarize} the main results of this section.
We have found that there are two Hilbert spaces, $\mathcal{H}_{\rm M}$ and $\mathcal{H}_{\rm R}$, one for an observer at rest and the other for an observer moving together with an accelerating reference frame. The two Hilbert spaces are related by a unitary squeezing operator. 
In particular, the vacuum in Minkowski spacetime is actually a superconducting condensate for the Rindler observer (to better understand the analogy with superconductivity in a solid, the reader will find all the details in Section~(\ref{BCSBEC})), namely (\ref{vacM4}), 
\be
\label{vacM3} 
|0_{\bf k}\rangle^+ |0_{- {\bf k}}\rangle^- 
= \left(
\cos r \, + e^{-i\phi}\,\sin r \,c^{{\rm I}\dagger}_{\bf k}\,d^{{\rm II}\dagger}_{-{\bf k}}
\right) \,|0_{\bf k}\rangle^+_{\rm I}\,|0_{- {\bf k} }\rangle^-_{\rm II} 
= S_{\rm sq}
 \,|0_{\bf k} \rangle^+_{\rm I}\,|0_{-{\bf k} }\rangle^-_{\rm II}
\, ,
\ee
(see also the complete state in (\ref{finalS2S2})). 
Taking the product over ${\bf k}$, we obtain the BCS state on which we will expand later in Section~(\ref{BCSBEC}), 
\be
\label{vacM2BCS} 
|0 \rangle^+
|0 \rangle^- 
&  \equiv   & 
|0_{\bf k_1} \rangle^+ 
\dots
 |0_{{\bf k}_n} \rangle^+
|0_{ {\bf - k}_{1}}\rangle^- 
\dots 
|0_{\bf -k_{n}}\rangle^- 
\nonumber \\
& = & \prod_{\bf k} \left(
\cos r_{\bf k} \, + e^{-i\phi}\,\sin r_{\bf k} \,c^{{\rm I}\dagger}_{\bf k}\,d^{{\rm II}\dagger}_{- \bf k}
\right) \,|0_{\bf k} \rangle^+_{\rm I}\,|0_{- \bf k}\rangle^-_{\rm II} 
\nonumber \\
& = &\prod_{\bf k} S_{\rm sq} \,|0_{\bf k}\rangle^+_{\rm I}\,|0_{- \bf k}\rangle^-_{\rm II} \equiv | \Psi_{\rm BCS} \rangle
.
\ee

The creation and annihilation operators are also related to each other by $S_{\rm sq}$, 
\be
&&   a_{\bf k} = S_{\rm sq} \, c^{{\rm I}}_{\bf k} \, S^\dagger_{\rm sq} \, , \quad 
a^\dagger_{\bf k} = S_{\rm sq} \, c^{{\rm I} \dagger}_{\bf k} \, S^\dagger_{\rm sq} \, 
\nonumber \\
&&
  b_{-{\bf k}}^\dagger = S_{\rm sq} \, d^{{\rm II} \dagger}_{-{\bf k}} \, S^\dagger_{\rm sq} \, , \quad 
   b_{-{\bf k}} = S_{\rm sq} \, d^{{\rm II} }_{-{\bf k}} \, S^\dagger_{\rm sq} 
  \, .
\label{primaBTF2}
\ee
Notice that in (\ref{vacM2BCS}) and (\ref{primaBTF2}) we have redefined $S \rightarrow S_{\rm sq}$ to emphasize the meaning of the operator.

\subsection{QFT in the black hole background}

In order to appreciate the squeezing interpretation of the Hawking state \cite{Adami:2023hvo, Akil:2025coj}, 
we need to briefly recap the original derivation, whose details can be found in the excellent and exhaustive book by Fabbri and Salas \cite{Fabbri} (the details of the derivation that follows are in the first part of Chapter~3 of \cite{Fabbri}).  

We consider a real Klein-Gordon scalar field $\Phi(x)$ in the black hole background. 
We assume a Cauchy foliation of the spacetime and identify two hypersurfaces: $\Sigma_{\rm in}$ (before the black hole formation) and $\Sigma_{\rm final}$ (after the black hole formation). 
The equation of motion for the scalar field in curved spacetime reads:
\be
\Box \Phi(x) = 0 .
\label{EoMphi}
\ee
Moreover, we assume the spacetime to be stationary---i.e.\ the metric does not depend on the time-like coordinate---before and after the black hole formation. However, in the transient period the metric has to evolve in order to have particle production. Such dynamics will be fully provided by the boundary condition.

As in Minkowski spacetime, in order to quantize the field, we split a generic solution of the Klein-Gordon equation into positive-frequency $u_i(t, \vec{x})$ and negative-frequency $u_i^*(t, \vec{x})$ modes relative to the Killing time $t$ in the two asymptotic stationary spacetime regions. 
The modes are chosen to form a complete orthonormal basis with respect to the covariant generalization of the Klein-Gordon scalar product \cite{Fabbri}. 
Since Poincar\'e symmetry holds only in the two asymptotic regions of the spacetime, 
a different choice of time does not lead to the same characterization of positive-frequency modes. Therefore, the vacuum state as well as the whole Fock space are not invariant.

Hence, we express the solution in different bases each one of them relative to the selected $3$-dimensional hypersurface $\Sigma_{\rm in}$ or $\Sigma_{\rm final}$. Indeed, $\Phi(x)$ has support on the spacetime, but the base respect to which it is expanded is defined on the selected $3d$ sub-manifold as explained in section (\ref{QFTCS}).

On $\Sigma_{\rm in}$, the solution of the above EoM in terms of ``in"  modes $u_{\bf k}^{\rm in}$ is:
\be
\Phi(x) =  \int_0^{+\infty} d {\bf k}
\left( u_{\bf k}^{\rm in} a_{\bf k}^{\rm in} + u_{\bf k}^{{\rm in}^*} a_{\bf k}^{{\rm in} \dagger} 
\right)  .
\label{Sigmain}
\ee

On $\Sigma_{\rm final}$, we have to take into account that there are modes outside the event horizon and modes crossing the horizon. Hence, we expand the same solution $\Phi$
in terms of ``int" modes  and ``out" modes, 
\be
%
\Phi(x) =  \int_0^{+\infty}  d {\bf k} 
 \left( u_\bfk^{\rm out} a_\bfk^{\rm out} 
+ u_\bfk^{{\rm out}^*} a_\bfk^{{\rm out} \dagger} 
 + 
u_\bfk^{\rm int} a_\bfk^{\rm int} 
+ u_\bfk^{{\rm int}^*} a_\bfk^{{\rm int} \dagger} 
\right) ,
\label{SigmainFull}
\ee
%
where $a_\bfk^{\rm int}$ and $a_\bfk^{{\rm int} \dagger}$ are the creation and annihilation operators of incoming particles at the future horizon $H^+$, $a_\bfk^{\rm out}$ and $a_\bfk^{ {\rm out} \dagger}$ are those at $\mathcal{I}^+$, 
while in (\ref{Sigmain}) $a_\bfk^{\rm in}$ and $a_\bfk^{{\rm in} \dagger}$ are the creation and annihilation operators at $\mathcal{I}^-$.

Given the above definitions, the most important result for our discussion can be found in Section~3.3.6 of \cite{Fabbri}, namely the complete Hawking state including the int-modes,
\be
| {\rm in} \rangle  & = &  \mathcal{N} \, 
{\rm e}^{  \sum_{k \geqslant 0} {\rm e}^{- 4 \pi M \omega(k) } a^{ {\rm out} \dagger}_k 
a^{  {\rm int} \dagger  }_{-k} }  | {\rm out} \rangle \otimes | {\rm int} \rangle 
\label{Sq0} \\
 & = & 
\prod_k \mathcal{N}(\bfk) \, 
\sum_N  {\rm e}^{- 4 \pi M \omega(\bfk) } \frac{1}{ N !} 
(a^{  {\rm out} \dagger}_\bfk )^N
(a^{  {\rm int} \dagger}_{-\bfk} )^N  | {\rm out} \rangle \otimes | {\rm int} \rangle 
\nonumber \\
 & = & 
\prod_\bfk \mathcal{N}(\bfk) \, \sum_N  {\rm e}^{- 4 \pi M \omega(\bfk) } 
| N^{\rm out}_\bfk \rangle \otimes  | N^{\rm int}_{-\bfk} \rangle \, , \quad 
\mathcal{N}(\bfk) = \sqrt{1 - {\rm e}^{- 8 \pi M \omega(\bfk)} } \, ,
\label{HS}
\ee
where $| N^{\rm out}_\bfk \rangle$, $| N^{\rm int}_{-\bfk} \rangle$ are the $N$-particle states with momentum 
$\vec{\bfk}$ and $-\vec{\bfk}$ respectively at $\mathcal{I}^+$ and $H^+$, and $\mathcal{N}$ is the normalization. 
Notice that in comparison to \cite{Fabbri} (see Section~3.3.6), we have labelled the modes with the $3$-momentum, as is usual. However, the outcome is unchanged because $\omega(\bfk) \equiv \sqrt{\bfk^2 + m^2}$.

Let us now focus on the first expression for the Hawking state (\ref{Sq0}) and make manifest the unitarity of the squeezing operator that is somehow hidden in it.

{\em Theorem}. 
The relation (\ref{Sq0}) between the initial and final vacuum states is equivalent to:
\be
 && | {\rm in} \rangle =  \prod_\bfk   {\rm e}^{ \zeta(\omega(\bfk))
\left( a_\bfk^{ {(\rm o)}  \dagger} a_{-\bfk}^{ {(\rm i)}  \dagger} 
- a_\bfk^{ {(\rm o)} } a_{-\bfk}^{ {(\rm i)} } \right)  }
| {\rm out} \rangle \otimes | {\rm int} \rangle 
\equiv {S_{\rm sq}} | {\rm out} \rangle \otimes | {\rm int} \rangle
\equiv{\prod_\bfk } S_{\rm sq}(\bfk) | {\rm out} \rangle \otimes | {\rm int} \rangle 
  ,  \quad 
\nonumber \\
&&  
{\rm tanh} (\zeta(\omega)) = {\rm e}^{ - 4 \pi M \omega(\bfk)}   = 
{\rm e}^{ - \frac{ \omega(\bf k)}{2 T}}
 , 
\label{Sq1}
\ee
where again $S_{\rm sq}$ stays for {\em squeezing operator}, 
and for the sake of shortness we have also introduced the notation:
 \be
{\rm  i} := \mbox{int \,\, and \,\, o} := {\rm out. }
 \ee
Notice that (also for future reference):
\be
\mbox{ for } \quad T\rightarrow 0 \, ,  \quad {\rm tanh} (\zeta(\omega)) \rightarrow 0 \quad 
\Longrightarrow \quad  \zeta(\omega) \rightarrow 0 
\quad \mbox{ and} 
\quad 
S_{\rm sq} \rightarrow 1\!\!1 \, .
\label{limitsB}
 \ee

{\em Proof.} The thesis follows from the general results recapped in Section~(\ref{SqBTB}). 
Indeed, 
by temporarily introducing the short-hand notation 
$a_\bfk^{ {(\rm o)} } \equiv a$ and $a_{-\bfk}^{ {(\rm i)} } \equiv b$, it is straightforward to apply equality (\ref{SexpNT}) to each mode $\bfk$ in (\ref{Sq1}), namely:
\be
S_{\rm sq}(\bfk)& = &{\rm e}^{\zeta \left( a^\dagger b^\dagger - a  b \right)} 
= {\rm e}^{{\rm tanh} ( \zeta ) a^\dagger b^\dagger} 
\, 
 {\rm e}^{ - {\rm ln} [{\rm cosh} ( \zeta ) ] \left( a^\dagger a+ b \,  b^\dagger  \right)}
 \,
 {\rm e}^{ - {\rm tanh} ( \zeta ) b  a} 
  \nonumber \\
 & = & {\rm e}^{{\rm tanh} ( \zeta ) a^\dagger b^\dagger} 
\, 
 {\rm e}^{ - {\rm ln} [{\rm cosh} ( \zeta ) ] \left( a^\dagger a+ b^\dagger  b + [ b , b^\dagger] \right)}
 \,
 {\rm e}^{ - {\rm tanh} ( \zeta ) b  a}
 \, . 
 \label{FullS}
 \ee
When acting on the vacuum $|0_a,0_b \rangle \equiv |0,0 \rangle$ (\ref{vab}) with the squeezing operator (\ref{FullS}) according to (\ref{S2vac}) we get:
\be
S_{\rm sq}(k) |0,0 \rangle & = &
{\rm e}^{{\rm tanh} ( \zeta ) a^\dagger b^\dagger} 
\, 
 {\rm e}^{ - {\rm ln} [{\rm cosh} ( \zeta ) ] \left( a^\dagger a+ b^\dagger  b + [ b , b^\dagger] \right)}
 \,
 {\rm e}^{ - {\rm tanh} ( \zeta ) b  a}
 |0,0 \rangle
 \nonumber \\
 & = & 
 {\rm e}^{{\rm tanh} ( \zeta ) a^\dagger b^\dagger} 
\, 
 {\rm e}^{ - {\rm ln} [{\rm cosh} ( \zeta ) ] \left( 0 + 0  + 1 \right)}
 \,
 {\rm e}^{ - {\rm tanh} ( \zeta )0}
 |0,0 \rangle
 \nonumber \\
 & = & 
 {\rm e}^{{\rm tanh} ( \zeta ) a^\dagger b^\dagger} 
\, 
 {\rm e}^{ - {\rm ln} [{\rm cosh} ( \zeta ) ]}
 |0,0 \rangle
 \nonumber \\
 & = & 
 \frac{1}{{\rm cosh} ( \zeta )} {\rm e}^{{\rm tanh} ( \zeta ) a^\dagger b^\dagger} 
 |0,0 \rangle 
 =
 \frac{1}{{\rm cosh} ( \zeta )} \sum_{n=0}^{+ \infty} ({\rm tanh} ( \zeta ))^n 
 |{\rm n}_a , {\rm n}_b \rangle 
 \label{proofF}
 \, .
\ee
One can notice the strict similarity with the ground state of Helium~II \cite{Superfluidity}.

If we look at the Hawking state (\ref{Sq0}) or (\ref{HS}), the above result is really not trivial at all and is in strict connection with unitarity in a general background. Indeed, acting on the vacuum state with the squeezing operator (see (\ref{proofF})), we lose any information about the unitarity of the original operator if one naively uses the operator
\be
\$_{\rm H} \equiv \frac{1}{{\rm cosh} ( \zeta )} {\rm e}^{{\rm tanh} ( \zeta ) a^\dagger b^\dagger} 
\label{bullsh}
\ee
as a non-unitary evolution from the initial to the final Hilbert space.

Coming back to the black hole background, we have to implement the result of above for each mode and make the following identification, 
\be
{\rm tanh}(\zeta) = {\rm e}^{- 4 \pi M \omega(\bfk)}  \quad \Longrightarrow \quad 
\frac{1}{{\rm cosh} ( \zeta )} = \sqrt{1 - {\rm e}^{- 8 \pi M \omega(\bfk)} } 
\equiv 
  \sqrt{1 - {\rm e}^{- \frac{ \omega(\bfk)}{T}} } 
\, . 
\label{ident}
\ee
Indeed,  making use of the result (\ref{proofF}) and the identification (\ref{ident}), we can turn (\ref{Sq1}) into:
\be
 | {\rm in} \rangle & =  & \prod_\bfk   {\rm e}^{ \zeta(\omega(\bfk))
\left( a_\bfk^{ {(\rm o)}  \dagger} a_{-\bfk}^{ {(\rm i)}  \dagger} 
- a_\bfk^{ {(\rm o)} } a_{-\bfk}^{ {(\rm i)} } \right)  }
| {\rm out} \rangle \otimes | {\rm int} \rangle 
\nonumber \\
& = & 
\prod_\bfk 
\frac{1}{{\rm cosh} ( \zeta(\omega(\bfk)) )} {\rm e}^{{\rm tanh} ( \zeta(\omega) ) \, a^{ ({\rm o}) \dagger}_\bfk a^{ ({\rm i}) \dagger}_{-\bfk}  } 
 |{\rm out}, {\rm int}  \rangle 
 \nonumber \\
& = & 
\prod_\bfk 
\sqrt{1 - {\rm e}^{- 8 \pi M \omega(\bfk)} } \,  {\rm e}^{{\rm e}^{- 4 \pi M \omega(\bfk)}  \, a^{ ({\rm o}) \dagger}_\bfk a^{ ({\rm i}) \dagger}_{-\bfk}  } 
 |{\rm out}, {\rm int}  \rangle \nonumber \\
&=&
 \prod_k 
\sqrt{1 - {\rm e}^{- \frac{ \omega(\bfk)}{T} } }\,  {\rm e}^{{\rm e}^{- 4 \pi M \omega(\bfk)}  \, a^{ ({\rm o}) \dagger}_\bfk a^{ ({\rm i}) \dagger}_{-\bfk}  } 
 |{\rm out}, {\rm int}  \rangle
 \equiv (\ref{Sq0})  \quad \qed \, .
 \label{proofabio}
\ee


Since the single-mode squeezing operator is unitary---because $S_{\rm sq}^\dagger(z) = S(-z)$---it is easy to show that the full squeezing operator is also unitary. Indeed:
\be
{S_{\rm sq}^\dagger}{S_{\rm sq}} & = &
\left( \prod_\bfk S_{\rm sq}(\bfk) \right)^\dagger \left( \prod_{\bfk'} S_{\rm sq}(\bfk') \right)
= \left( S_{\rm sq}(\bfk_1) S_{\rm sq}(\bfk_2) \dots S_{\rm sq}(\bfk_n)) \right)^\dagger
\left( S_{\rm sq}(\bfk_1) S_{\rm sq}(\bfk_2) \dots S_{\rm sq}(\bfk_n)) \right) \nonumber \\
& = & 
 \underbrace{S_{\rm sq}(\bfk_n)^\dagger \dots \underbrace{S_{\rm sq}(\bfk_2)^\dagger \underbrace{S_{\rm sq}(k_1)^\dagger
 S_{\rm sq}(\bfk_1)}_{1\!\!1} S_{\rm sq}(\bfk_2)}_{1\!\!1} \dots S_{\rm sq}(\bfk_n)) }_{1\!\!1} = 1\!\!1 \,  \qed \, .
\ee
Therefore, {\em the initial and final vacuum states are related by a unitary operator}. 

Hence, in short we can write (\ref{Sq1}) as:
 \be
 \boxed{ | {\rm in} \rangle = 
{ S_{\rm sq}} | {\rm out} \rangle  \otimes  | {\rm int} \rangle 
} 
\quad {\rm or} \quad 
| {\rm out} \rangle  \otimes  | {\rm int} \rangle = 
{S^\dagger_{\rm sq}} | {\rm in} \rangle
\quad \Longrightarrow \quad 
\langle {\rm out} | \otimes  \langle {\rm int} | = 
 \langle {\rm in} | S_{\rm sq}
\, . 
\label{iSoi}
\ee

As a check, we can evaluate the number of particles created by the black hole. What we need is the following expectation value, derived in \cite{bookO}:
\be
\langle {\rm in}, {\bf k} | \hat{\mathcal{O}}^{( {\rm o} ) } | {\rm in}, {\bf k} \rangle 
= \frac{1}{\rm cosh^2 \zeta} \sum_{n = 0}^{+ \infty} \left( {\rm tanh} \zeta \right)^{2 n} 
 \langle {\rm n}_{\rm out},  {\rm n}_{\rm int}  | \hat{\mathcal{O}}^{(\rm o)} |  {\rm n}_{\rm out},  {\rm n}_{\rm int} \rangle \, ,
 \label{avera}
\ee
which follows from the last equality in (\ref{proofF}), where, for the sake of simplicity, we have omitted the explicit dependence on ${\bf k}$ in $\zeta$ and in the eigenstates of the number operators.

Therefore, the number of particles of energy $\omega$ and moving towards infinity is:
\be
\langle {\rm in} | \hat{N}^{(\rm o)}_{\omega} | {\rm in} \rangle 
& =  & \langle {\rm out} | {S_{\rm sq}^\dagger  } \hat{N}^{(\rm o)}_{\omega} 
{S_{\rm sq} }
| {\rm out}  \rangle \\
& = & \langle {\rm out} |  \left( \dots { S_{\rm sq}^\dagger (\bfk)} \dots { S_{\rm sq}^\dagger (\bfk_2) } {S_{\rm sq}^\dagger (\bfk_1) } \dots 
 \right) 
\, \hat{N}^{(\rm o)}_{\omega} 
\,\left( \dots {S_{\rm sq}(\bfk_1) }  {S_{\rm sq}(\bfk_2) } \dots { S_{\rm sq}(\bfk) }\dots \right) 
| {\rm out} \rangle 
\nonumber \\
& = & 
\langle {\rm out} |  { S_{\rm sq}^\dagger (\bfk)}  
\, a^{ (\rm o) \dagger}_\bfk a^{ (\rm o)}_\bfk \, { S_{\rm sq} (\bfk)}  | {\rm out} \rangle
\nonumber \\
& = & 
\langle {\rm in}, {\bf k} |  
\, a^{ (\rm o) \dagger}_\bfk a^{ (\rm o)}_\bfk   | {\rm in}, {\bf k} \rangle
\nonumber \\
& = & \frac{1}{\rm cosh^2 \zeta} \sum_{{\rm n} = 0}^{+ \infty} \left( {\rm tanh} \zeta \right)^{2 {\rm n} } 
 \langle {\rm n}_{\rm out},  {\rm n}_{\rm int}  | \hat{N}^{(\rm o)}_{\omega} |  {\rm n}_{\rm out},  {\rm n}_{\rm int} \rangle
 \nonumber \\
& = & \frac{1}{\rm cosh^2 \zeta} \sum_{ {\rm n } = 0}^{+ \infty} \left( {\rm tanh} \zeta \right)^{2 {\rm n} } 
 {\rm n}  
= \frac{1}{\rm cosh^2 \zeta} \sum_{{\rm n} = 0}^{+ \infty} {\rm e}^{ - 8 \pi {\rm n}  M \omega} 
 {\rm n}  
 =  \frac{1}{\rm cosh^2 \zeta} 
 \frac{ {\rm e}^{ 8 \pi   M \omega} }{ \left( {\rm e}^{  8 \pi   M \omega}  - 1 \right)^2 }
= \frac{1}{{\rm e}^{8 \pi M \, \omega} -1 } \, , 
\label{Ninfinity}
\ee
where at the third step each operator ${S_{\rm sq}^\dagger (\bfk_i) }$ with $\bfk_i \neq \bfk$ commutes with  
$a^{ (\rm o) \dagger}_\bfk a^{ (\rm o)}_\bfk$ and yields the identity when it meets ${S_{\rm sq}(\bfk_i) }$. 
In the last three steps of (\ref{Ninfinity}) we made repeated use of (\ref{avera}). 
The number of particles with positive energy that cross the horizon is also exactly equal to the last expression in (\ref{Ninfinity}), i.e.\ 
$\langle {\rm in} | \hat{N}^{(\rm o)}_{\omega} | {\rm in} \rangle = \langle {\rm in} | \hat{N}^{(\rm i)}_{\omega} | {\rm in} \rangle$. 

Accordingly, we infer from (\ref{Ninfinity}) that the emitted particles are in a thermal state at the Hawking temperature,
\be
T_{\rm H} = \frac{1}{8 \pi M} . 
\label{T_H}
\ee

\subsection{Bosons in the black hole background} 
In order to make closer contact with the general results in Section~(\ref{SqBTB}), we make the following identifications:
\be 
&& a \equiv a^{\rm I}_\bfk \equiv a^{(\rm out)}_\bfk \, , \nonumber \\
&& b^\dagger \equiv  a^{\rm II \dagger}_{-\bfk} \equiv a^{(\rm int) \dagger}_{- \bfk} \,  , \nonumber  \\
&& a' \equiv a^{({\rm in})}_\bfk \equiv a^{(\rm in)}_\bfk \, ,  \nonumber \\
&& b^{' \dagger} \equiv  a^{({\rm in}) \dagger}_{-\bfk} \equiv a^{(\rm in) \dagger}_{- \bfk} \, .
\label{correspondB}
\ee
The above identifications allow us to figure out the relation between initial and final operators through the squeezing operator in the Rindler, or equivalently in the black hole, background,
\be
\left[
\begin{array}{c}
  a^{({\rm in})}_\bfk \\
  \\
  a_{-\bfk}^{({\rm in})\dagger} \\
\end{array}
\right] =
S\,
\left[
\begin{array}{c}
  a^{{\rm (out) }}_\bfk \\
  \\
  a^{{\rm (int) } \dagger}_{-\bfk} \\
\end{array}
\right]\,
S^\dagger \, .
\label{iSoS}
\ee
For the Hermitian conjugate operators, we have:
\be
\left[
\begin{array}{c}
  a^{({\rm in}) \dagger}_{\bfk } \\
  \\
  a_{-\bfk}^{({\rm in})} \\
\end{array}
\right] =
S\,
\left[
\begin{array}{c}
  a^{{\rm (out) \dagger}}_\bfk \\
  \\
  a^{{\rm (int) } }_{-\bfk} \\
\end{array}
\right]\,
S^\dagger \, .
\label{iSoSd}
\ee

\subsection{Fermions in the black hole background} 
In order to make closer contact with the general results in Section~(\ref{Sq2Fermions}) and with the case of Rindler spacetime, we set out the notation as follows (see also (\ref{correspond})). 
According to the identifications (\ref{correspond}) and (\ref{primaBTF}):
\be
\left[
\begin{array}{c}
  a^{({\rm in})}_\bfk \\
  \\
  b_{-\bfk}^{({\rm in})\dagger} \\
\end{array}
\right] =
S\,
\left[
\begin{array}{c}
  c^{{\rm (out) }}_\bfk \\
  \\
  d^{{\rm (int) } \dagger}_{-\bfk} \\
\end{array}
\right]\,
S^\dagger \, , 
\ee
and according to (\ref{altri}):
\be
\left[
\begin{array}{c}
  b^{{\rm (in})}_\bfk \\
  \\
  a_{-\bfk}^{({\rm in})\dagger} \\
\end{array}
\right] =
\bar{S} \,
\left[
\begin{array}{c}
  d^{({\rm out })}_\bfk \\
  \\
  c^{ ({\rm int} ) \dagger}_{-\bfk} \\
\end{array}
\right]\,
\bar{S}^\dagger \, .
\ee
%
%
%
%
%
%
%


\section{S-matrix unitarity in a general background}\label{S-matrixS unitarity} \label{Smatrix}
In this section, we show that the $S$-matrix for the evaporation process is a unitary operator.
Let us consider a general initial state obtained by acting on the vacuum $| {\rm in} \rangle$ with the creation operators $a^{(\rm in) \dagger}_{\bfk_i}$ ($i=1, \dots , n$) and  $a^{(\rm in) \dagger}_{-\bfp_i}$ ($i=1, \dots , m$), i.e.
\be
| \Phi(- \infty) \rangle & = & 
a^{(\rm in) \dagger}_{\bfk_1} \dots  a^{ (\rm in) \dagger}_{\bfk_{n}}   
 \, 
 a^{(\rm in) \dagger}_{-\bfp_1} \dots  a^{ (\rm in) \dagger}_{-\bfp_{m}} 
| {\rm in} \rangle
\nonumber 
\\
& = &  
a^{(\rm in) \dagger}_{\bfk_1} \, \dots  \, a^{ (\rm in) \dagger}_{\bfk_{n}}  
\, 
a^{(\rm in) \dagger}_{-\bfp_1} \dots  a^{ (\rm in) \dagger}_{-\bfp_{m}} 
 {\color{red} S_{\rm sq}} | {\rm out} , {\rm int} \rangle
\nonumber \\
& = & {\color{red} S_{\rm sq}} 
\, 
a^{(\rm out) \dagger}_{\bfk_1} 
{\color{red} S_{\rm sq}^\dagger }
\, 
\dots  
\, 
{\color{red} S_{\rm sq}} \, 
a^{ (\rm out) \dagger}_{\bfk_{n}}   
{\color{red} S_{\rm sq}^\dagger }
\, 
{\color{red} S_{\rm sq}} \,
a^{(\rm int) \dagger}_{-\bfp_1} 
{\color{red} S_{\rm sq}^\dagger }
\, 
\dots  
\, 
{\color{red} S_{\rm sq}} \, 
a^{ (\rm int) \dagger}_{-\bfp_{m}}   
{\color{red} S_{\rm sq}^\dagger }
{\color{red} S_{\rm sq}} | {\rm out} , {\rm int} \rangle
\nonumber \\
&=  & {\color{red} S_{\rm sq}} 
\, 
a^{(\rm out) \dagger}_{\bfk_1} 
\, 
\dots  
\, 
a^{ (\rm out) \dagger}_{\bfk_{n}}   
a^{(\rm int) \dagger}_{-\bfp_1} \dots  a^{ (\rm int) \dagger}_{-\bfp_{m}} 
 | {\rm out} , {\rm int} \rangle \equiv {\color{red} S_{\rm sq}}  | i \rangle 
 \, , 
 \label{menusInfinity}
\ee
where we used (\ref{iSoi}), the first of (\ref{iSoSd}), and the second of (\ref{iSoS}). 
For the sake of simplicity, in what follows we will omit the int-operators.
Notice that the state $| i \rangle$ and $S_{\rm sq}$ are defined in and on the Hilbert space 
$\mathcal{H}_{\rm out}$, respectively. 

On the other hand, the final state is obtained by evolving with the unitary $S$-matrix of the theory (which could simply be that of a two-derivative scalar field theory, the Standard Model, etc.):
\be
| \Phi(+ \infty) \rangle = S 
| \Phi(- \infty) \rangle \equiv S \, {\color{red} S_{\rm sq}}  | i \rangle \, ,
\label{S-matrix}
\ee
where $S$ is defined on the out-Hilbert space $\mathcal{H}_{\rm out}$, consistently with (\ref{menusInfinity}), 
which only involves out operators and out states. 
The transition probability from the state $| \Phi(+ \infty) \rangle$ to a general state in the Hilbert space 
$\mathcal{H}_{\rm out}$ reads:
\be
P = |\langle f |  \Phi(+ \infty) \rangle |^2 \, .
\ee
We now introduce the following standard definition, 
\be
\langle f |  \Phi(+ \infty) \rangle = \langle f |  S \, {\color{red} S_{\rm sq}}  | i \rangle \equiv 
\left( S \, {\color{red} S_{\rm sq}} \right)_{f i} \,, 
\ee 
and we expand the state at infinity by introducing the completeness relation for $\mathcal{H}_{\rm out}$ 
in (\ref{S-matrix}), 
\be
|  \Phi(+ \infty) \rangle = \sum_f | f  \rangle \langle f |  S \, {\color{red} S_{\rm sq}}  | i \rangle 
= 
\sum_f | f  \rangle \langle f |
\left( S \, {\color{red} S_{\rm sq}} \right)_{f i} \,, 
\ee 
Since the state $|  \Phi(+ \infty) \rangle$ is normalized to one,
\be
1 = \langle  \Phi(+ \infty) |  \Phi(+ \infty) \rangle \quad \Longrightarrow 
\quad 
\sum_f  \left| S \, {\color{red} S_{\rm sq}} \right|^2_{f i} = 1 \, , 
\ee
which states the {\em unitarity of an interacting QFT in a general background}. 
Indeed, if $S S^\dagger = 1\!\!1$,
\be
 \left( S \, {\color{red} S_{\rm sq}} \right)  \left( S \, {\color{red} S_{\rm sq}} \right)^\dagger 
 = S \,\underbrace{ {\color{red} S_{\rm sq}} \,  {\color{red} S_{\rm sq}^\dagger}}_{1 \!\! 1}  \, S^\dagger = 1\!\!1\, . 
\ee
Therefore, the unitarity of QFT in the black hole background, or in a general spacetime, is encoded in the operator:
\be
\$ \equiv  S {\color{red}S_{\rm sq} } \, , \quad \$^\dagger \$ =  1\!\!1 \, .
\label{dollaro}
\ee
Let us remark that all the above steps involve only states and the scalar product in $\mathcal{H}_{\rm out}$. Moreover, the $S$-matrix is defined on the Hilbert space $\mathcal{H}_{\rm out}$ as well.


The squeezing operator takes into account the presence of a general background and how it affects the matter within. Therefore, 
the {\em probability amplitude for a scalar field theory in a general curved spacetime} reads:
\be
\langle {\rm out} | a^{ (\rm out)}_{\bfk'_{n}}    \dots a^{ (\rm out)}_{\bfk'_{1}} 
\, 
S 
 \, 
  a^{(\rm in) \dagger}_{\bfk_1} \dots  a^{ (\rm in) \dagger}_{\bfk_{n}}   
| {\rm in} \rangle
= 
\langle {\rm out} | a^{ (\rm out)}_{\bfk'_{n}}    \dots a^{ (\rm out) }_{\bfk'_{1}} \, 
S  {\color{red} S_{\rm sq}} 
\, 
a^{(\rm out) \dagger}_{\bfk_1} 
\, 
\dots  
\, 
a^{ (\rm out) \dagger}_{\bfk_{n}}   
 | {\rm out} \rangle 
 \, .
 \label{SSsqG}
\ee

We can equivalently express the above amplitude in terms of states and operators on the initial hypersurface. Indeed, making use of the last implication in (\ref{iSoi}) and introducing the identity in the form 
${\color{red} S^\dagger_{\rm sq}}  {\color{red} S_{\rm sq}}$, we get:
\be
\langle {\rm out} | a^{ (\rm out)}_{\bfk'_{n}}    \dots a^{ (\rm out)}_{\bfk'_{1}} 
\, 
S 
 \, 
  a^{(\rm in) \dagger}_{\bfk_1} \dots  a^{ (\rm in) \dagger}_{\bfk_{n}}   
| {\rm in} \rangle 
& = &
\langle {\rm in} | {\color{red} S_{\rm sq}} \, a^{ (\rm out) }_{\bfk'_{n}}    \dots a^{ (\rm out) }_{\bfk'_{1}} \, 
S 
 \, 
  a^{(\rm in) \dagger}_{\bfk_1} \dots  a^{ (\rm in) \dagger}_{\bfk_{n}}   
| {\rm in} \rangle \nonumber \\
& = &
\langle {\rm in} | {\color{red} S_{\rm sq}} \, a^{ (\rm out) }_{\bfk'_{n}} \,  
{\color{red} S^\dagger_{\rm sq}}  {\color{red} S_{\rm sq}}    \dots a^{ (\rm out)}_{\bfk'_{1}} \, 
{\color{red} S^\dagger_{\rm sq}}  {\color{red} S_{\rm sq}}
\, 
S 
 \, 
  a^{(\rm in) \dagger}_{\bfk_1} \dots  a^{ (\rm in) \dagger}_{\bfk_{n}}   
| {\rm in} \rangle 
\nonumber \\
& = &
\langle {\rm in} |  a^{ (\rm in) }_{\bfk'_{n}} \,  
    \dots a^{ (\rm in) }_{\bfk'_{1}} \, 
 {\color{red} S_{\rm sq}}
\, 
S 
 \, 
  a^{(\rm in) \dagger}_{\bfk_1} \dots  a^{ (\rm in) \dagger}_{\bfk_{n}}   
| {\rm in} \rangle
 \, .
 \label{SSsqGin}
\ee 
Since everything has to be written in terms of operators in $\mathcal{H}_{\rm in}$, we have to express ${\color{red} S_{\rm sq}}$ in terms of in-operators as well. 
Let us expand on this point. The squeezing operator for the $\bfk$-mode is:
\be
S_{\rm sq}(\bfk) =   {\rm e}^{ \zeta(\omega(\bfk))
\left( a_\bfk^{ {(\rm o)}  \dagger} a_{-\bfk}^{ {(\rm i)}  \dagger} 
- a_\bfk^{ {(\rm o)} } a_{-\bfk}^{ {(\rm i)} } \right)  } \, ,
\ee
and the out- and int-operators are related to the in-operators by (\ref{iSoS}) and (\ref{iSoSd}). Hence, we can equivalently express the squeezing operator as: 
\be
S_{\rm sq}(\bfk) & = &   {\rm e}^{ \zeta(\omega(\bfk))
\left( S_{\rm sq}(\bfk)^\dagger a_\bfk^{ {(\rm in)}  \dagger} S_{\rm sq}(\bfk) \,  S_{\rm sq}(\bfk)^\dagger a_{-\bfk}^{ {(\rm in)}  \dagger} S_{\rm sq}(\bfk)
- S_{\rm sq}(\bfk)^\dagger a_\bfk^{ {(\rm in)} } S_{\rm sq}(\bfk) \, S_{\rm sq}(\bfk)^\dagger a_{-\bfk}^{ {(\rm in)} } S_{\rm sq}(\bfk)  \right)  } 
\nonumber \\
& = &
 {\rm e}^{ \zeta(\omega(\bfk))
 S_{\rm sq}(\bfk)^\dagger \left( a_\bfk^{ {(\rm in)}  \dagger}  a_{-\bfk}^{ {(\rm in)}  \dagger} 
-  a_\bfk^{ {(\rm in)} }  a_{-\bfk}^{ {(\rm in)} } \right) S_{\rm sq}(\bfk)   } 
\equiv 
 {\rm e}^{
 S_{\rm sq}(\bfk)^\dagger \mathcal{O}  S_{\rm sq}(\bfk)   } \, .
\ee
Therefore, we have to find the solution for $S_{\rm sq}(\bfk)$ of the following algebraic equation:
\be
S_{\rm sq}(\bfk) =  {\rm e}^{
 S_{\rm sq}(\bfk)^\dagger \mathcal{O}  S_{\rm sq}(\bfk)   } \, .
 \label{eqforSq}
\ee
One can verify that the solution is:
\be
S_{\rm sq}(\bfk) = e^{\mathcal{O} } \, .
\label{Ssqin} 
\ee
Indeed, substituting (\ref{Ssqin}) into the right-hand side of (\ref{eqforSq}), we get:
\be
 {\rm e}^{
 e^{\mathcal{O}^\dagger } \mathcal{O}  e^{\mathcal{O} }  } 
 = e^{e^{ - \mathcal{O} } \mathcal{O}  e^{  \mathcal{O} }  }
= e^{\mathcal{O} } 
 \label{eqforSq2}
\ee
because $\exp \mathcal{O}$ commutes with $\mathcal{O}$. Therefore, the right-hand side of (\ref{eqforSq}) is identically equal to the squeezing operator, but now expressed in terms of in-operators. 
Finally,
\be
S_{\rm sq}(\bfk) = {\rm e}^{ \zeta(\omega(\bfk))
 \left( a_\bfk^{ {(\rm in)}  \dagger}  a_{-\bfk}^{ {(\rm in)}  \dagger} 
-  a_\bfk^{ {(\rm in)} }  a_{-\bfk}^{ {(\rm in)} } \right)    } 
 \, .
 \label{Sqink}
\ee

{\em According to the amplitude (\ref{SSsqGin}) and the squeezing operator (\ref{Sqink}), 
the unitary operator $\$$ (\ref{dollaro}) and all the states are defined in Minkowski spacetime; thus, the effect of a general background metric is to create a superconducting-like condensate (see the next section on the analogy with superconductivity and superfluidity) 
in which QFT takes place. 
Therefore, QFT in a general spacetime is perfectly unitary as long as the squeezing operator is well defined.}

In Rindler spacetime, QFT is unitary as long as the acceleration takes finite values. Of course, for $a = \infty$ the result no longer makes sense. The same issue shows up in the Schwarzschild background because the temperature diverges when the black hole mass goes to zero, i.e.\ $M\rightarrow 0$. 
The parallelism with the Rindler case should be extended to the full evaporation as follows. 
We assume the astronaut to be able to change the acceleration of the rocket (notice that if $a$ changes with the time-like coordinate, the spacetime is no longer Riemann-flat) by burning more or less fuel. Hence, he can first accelerate and afterwards decelerate. Correspondingly, the temperature, initially zero, will grow and finally decrease again to zero. The different values of $a$ define different superconducting phases 
interpolating between an initial and a final Minkowski vacuum. 
In the black hole case, we have a very similar picture as long as we identify $a \propto 1/M$. Hence, small $a$ corresponds to large $M$ and vice versa. In particular, for $M=0$ the temperature blows up and we have to devise a mechanism 
to make the whole analogy with the Rindler case work.
Indeed, at any stage of the evaporation process---namely for any strictly positive value of the black hole mass---QFT is well defined and unitary, but when the mass of the black hole is such that its Compton wavelength is larger than $2M$, the horizon disappears and we can assume the temperature to be zero, consistently with the end of the evaporation process. A natural way to implement this condition is to modify the temperature according to \cite{Alesci:2011wn}:
\be
\text{SMOOTH}: \quad  T = \frac{G M}{8 \pi [ (GM)^2 + \lambda^2_{\rm C} ] } \qquad \text{or} \qquad \text{SHARP}: \quad
\quad  T = \frac{1}{8 \pi M} \Theta( 2 G M - \lambda_{\rm C} ) \, .
\label{Temperatures}
\ee
Since the Compton wavelength is related to the mass by $\lambda_{\rm C} = 1/M$, the temperatures in (\ref{Temperatures}) take the following closed form in terms of the black hole mass,
\be
T = \frac{G M^3}{8 \pi [ G^2 M^4 + 1 ] } \quad \text{or} 
\quad  T = \frac{1}{8 \pi M} \Theta \left( 2 G M - \frac{1}{M} \right) 
= \frac{1}{8 \pi M} \Theta \left( M - \frac{M_{\rm p}}{\sqrt{2}} \right)
\, .
\label{Temperatures2}
\ee
In the second case, in order to have a black hole one needs $M \gtrsim 0.7 \, M_{\rm p}$. 

The above kind of modification by hand is enough to make QFT unitary at any stage of the evaporation process, including the limit $M\rightarrow 0$. 
However, as we will expand further in the next section, it assumes the geodesic completion of the spacetime.

According to (\ref{proofabio}) or (\ref{proofF}), the probability amplitude that no particle is produced in the final state reads:
 \be
 \langle 0_{\bf k} , {\rm out} ; 0_{- \bfk} , {\rm int} | 0_{\bf k}, {\rm in} \rangle 
 =  
  \langle 0_{\bf k} , {\rm out} ; 0_{- \bfk} , {\rm int} |
 S_{\rm sq}({\bf k}) | 0_{\bf k} , {\rm out} \rangle 
 =  
 \sqrt{1 - {\rm e}^{- \frac{ \omega({\bf k})}{T} } }.
 \label{Norm}
 \ee
 The second equality in (\ref{Norm}) is due to the projection onto the mode ${\bf k}$ in the out state. 
 
 We can go beyond (\ref{Norm}) and evaluate the probability amplitude that $n$ pairs of particles in the ${\bf k}$- and $-{\bf k}$-modes are produced,
\be
| \langle n_{{\bf k} } , {\rm out} ; 
 n_{ - {\bf k} } , {\rm int}
 | 0_{\bf k}, {\rm in}  \rangle |^2 =  
| \langle n_{{\bf k} } , {\rm out} ;
 n_{ - {\bf k} } , {\rm int}
| S_{\rm sq}({\bf k}) 
| 0_{\bf k}, {\rm out} ;
0_{- {\bf k}}, {\rm int} 
 \rangle |^2 . 
\label{Zero-1}
 \ee
 In order to evaluate the amplitude we make use of (\ref{proofF}), 
 \be
&&
\hspace{-2cm}
| \langle n_{{\bf k} } , {\rm out} ;
 n_{ - {\bf k} } , {\rm int}| 
S_{\rm sq}({\bf k}) 
| 0_{\bf k}, {\rm out} ;
0_{- {\bf k}}, {\rm int} 
 \rangle |^2  =
 \nonumber \\
 && = 
| \langle n_{{\bf k} } , {\rm out} ;
 n_{ - {\bf k} } , {\rm int}| 
 \left(
   \frac{1}{{\rm cosh} ( \zeta({\bf k}) )} {\rm e}^{{\rm tanh} ( \zeta({\bf k}) ) a^{(o) \dagger} a^{(i) \dagger} }
   \right) 
  | 0_{\bf k}, {\rm out} ;
0_{- {\bf k}}, {\rm int} 
 \rangle |^2
 \nonumber \\
&& = 
| \langle n_{{\bf k} } , {\rm out} ;
 n_{ - {\bf k} } , {\rm int}| \left( 
 \frac{1}{{\rm cosh} ( \zeta({\bf k}) )} \sum_{\ell=0}^{+ \infty} ({\rm tanh} ( \zeta({\bf k}) ))^\ell  \right) 
 | \ell _{\bf k}, {\rm out} ;
\ell_{- {\bf k}}, {\rm int} 
 \rangle |^2
 \nonumber \\
 &&
 = \frac{1}{{\rm cosh}^2 ( \zeta({\bf k}) )}  ({\rm tanh} ( \zeta({\bf k}) ))^{2 n_{{\bf k}} }
= e^{- n_{\bf k} \frac{\omega( {\bf k})}{T}} \left( 1 - e^{-  \frac{\omega( {\bf k})}{T}} \right) 
\, ,
\label{Zeron}
 \ee
where for the final state we have taken:
\be \langle n_{{\bf k} } , {\rm out} 
; n_{- {\bf k} } , {\rm int} 
| 
= \langle {\rm out} , {\rm int} | \left( a^{ (\rm o) }_{\bf k}  \right)^n  \left( a^{  (\rm i)}_{ - {\bf k}} \right)^n.
\ee 
The following identity deserves to be noted, 
\be
\frac{ |N_{\bf k} |^{2 n} }{ \left( 1+|N_{\bf k}|^2 \right)^{n+1} } 
\equiv e^{- n \frac{\omega( {\bf k})}{T}} \left( 1 - e^{-  \frac{\omega( {\bf k})}{T}} \right) 
\, .
\label{ZeronB}
\ee
Finally, we evaluate the amplitude (\ref{Zeron}) including all modes in the squeezing operator according to (\ref{proofabio}) or (\ref{iSoi}) for a $2$-dimensional quantum field theory, namely
\be
&& | \langle n_{{\bf k} } , {\rm out} ;
 n_{ - {\bf k} } , {\rm int}| 
\prod_{{\bf k'}}  S_{\rm sq}({\bf k'}) 
| 0_{\bf k'}, {\rm out} ;
0_{- {\bf k'}}, {\rm int} 
 \rangle |^2 = \nonumber \\
&& =   \left( \prod_{{\bf k'}}  \frac{1}{{\rm cosh}^2 ( \zeta({\bf k'}) )}\right)   ({\rm tanh} ( \zeta({\bf k}) ))^{2 n_{{\bf k}} }
 =  e^{- n_{\bf k} \frac{\omega( {\bf k})}{T}} 
  \prod_{{\bf k'}}  \left( 1 - e^{-  \frac{\omega( {\bf k'})}{T}} \right) 
  =  e^{- n_{\bf k} \frac{\omega( {\bf k})}{T}} \, \left( e^{-\frac{\omega_0}{T}} ,  e^{-\frac{\omega_0}{T}}  \right)_{\infty}  
\, ,
\label{ZeronP}
 \ee
where $\omega({\bf k'}) \equiv \omega_0 \, n$ ($n \in \mathbb{N}$) and the last expression stands for the Euler function:
\be
\left( q,  q \right)_{\infty} \equiv \prod_{\bf k}^{\infty} \left(1 - q^{\bf k} \right)  . 
\ee
The product runs only over ${\bf k}_x\in \mathbb{N}$. 
The result is very similar for a $4$-dimensional quantum field theory, but the final result cannot be expressed in terms of the Euler function.

\begin{figure}
\begin{center}
\includegraphics[height=7cm]{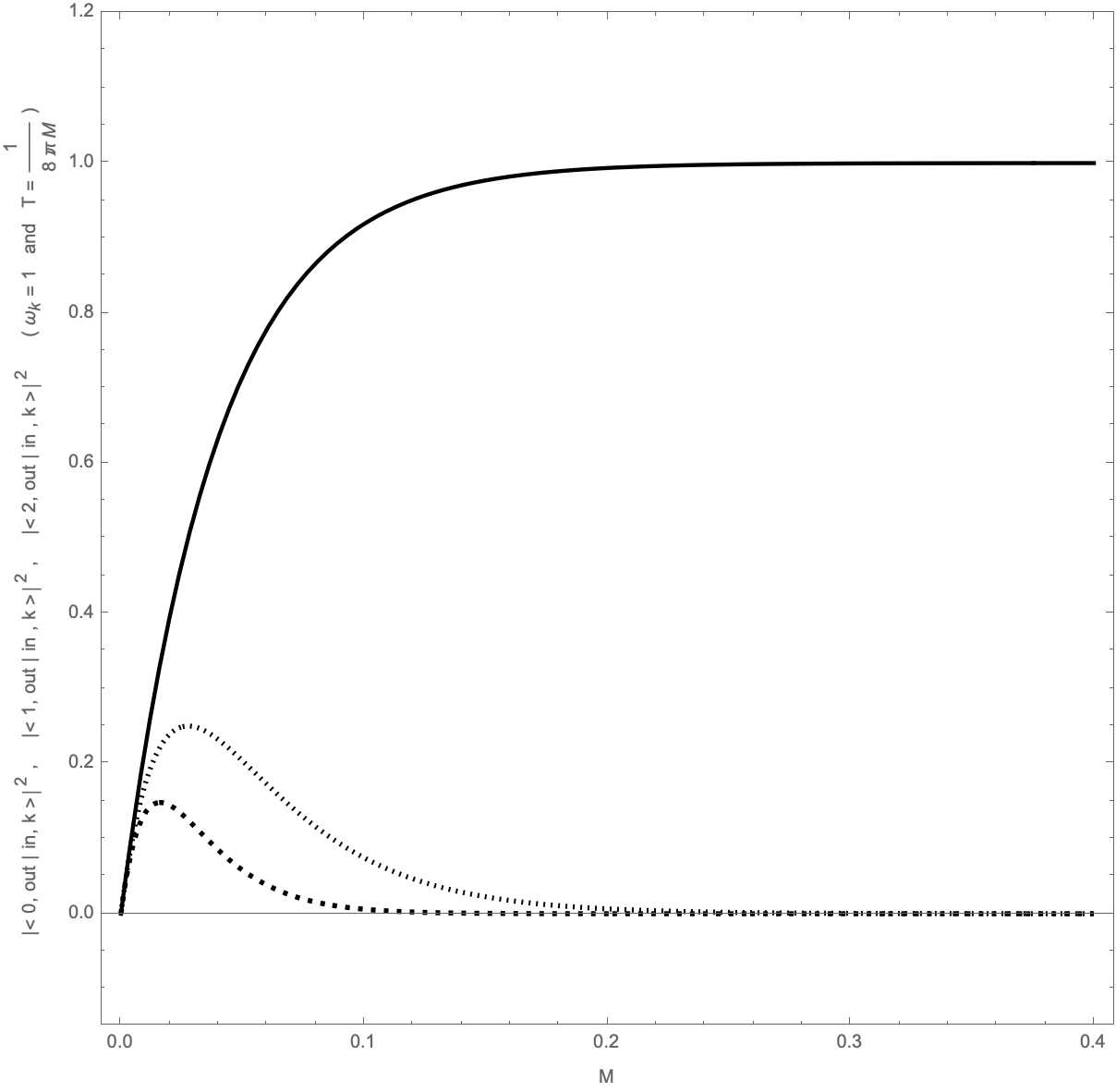} 
\\
\includegraphics[height=7cm]{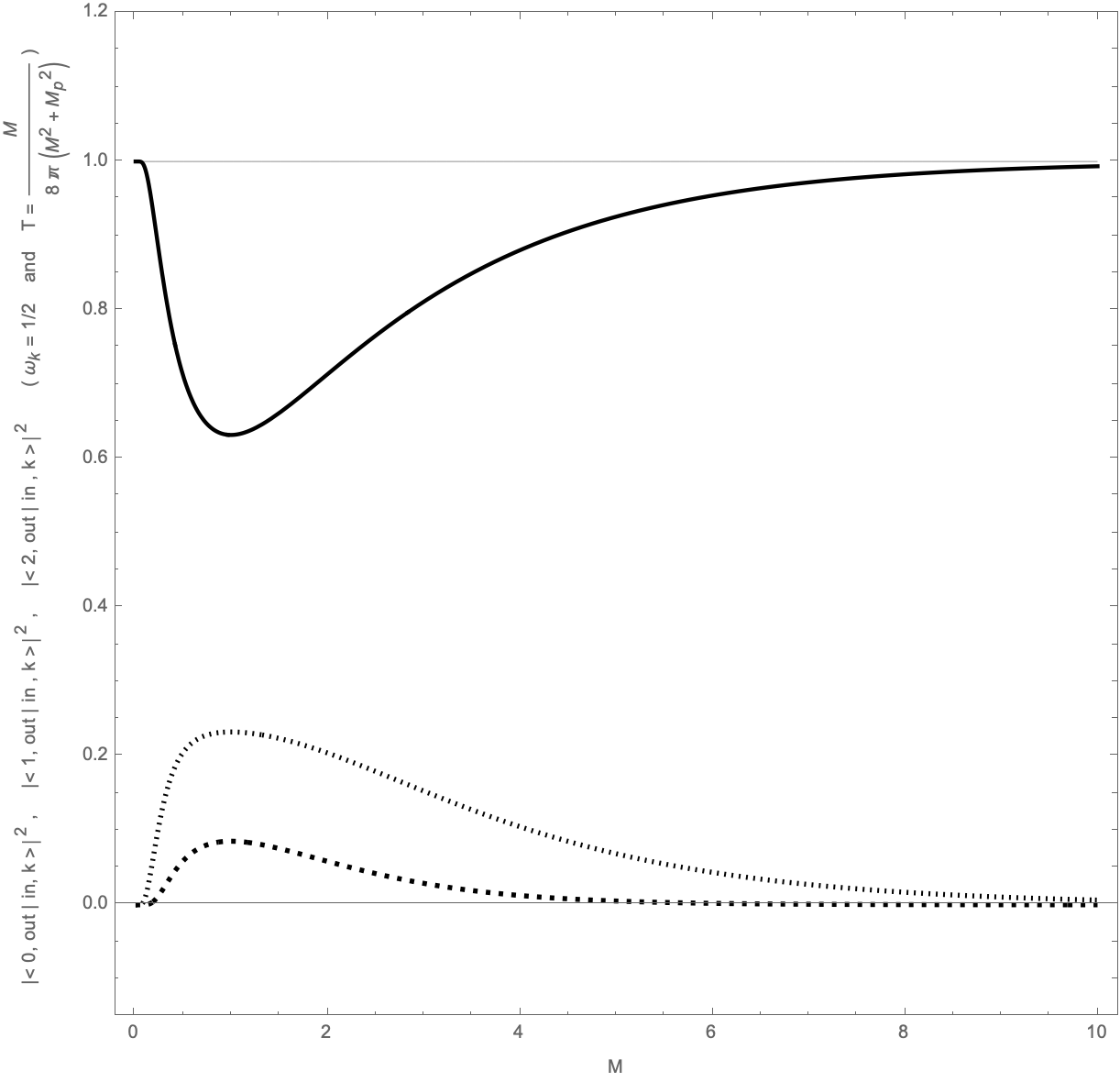} 
\hspace{1.5cm}
\includegraphics[height=7cm]{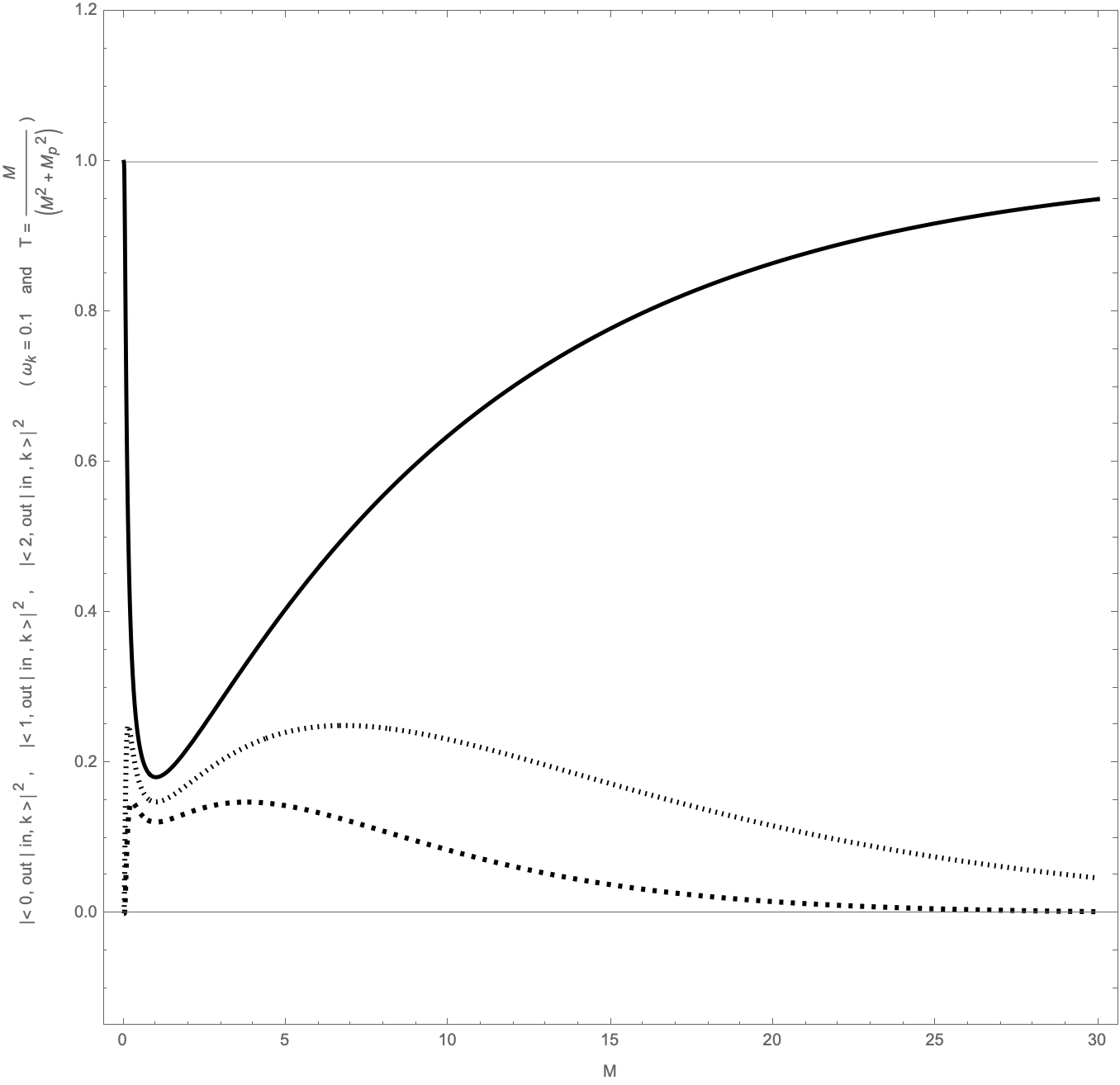} 
\includegraphics[height=7cm]{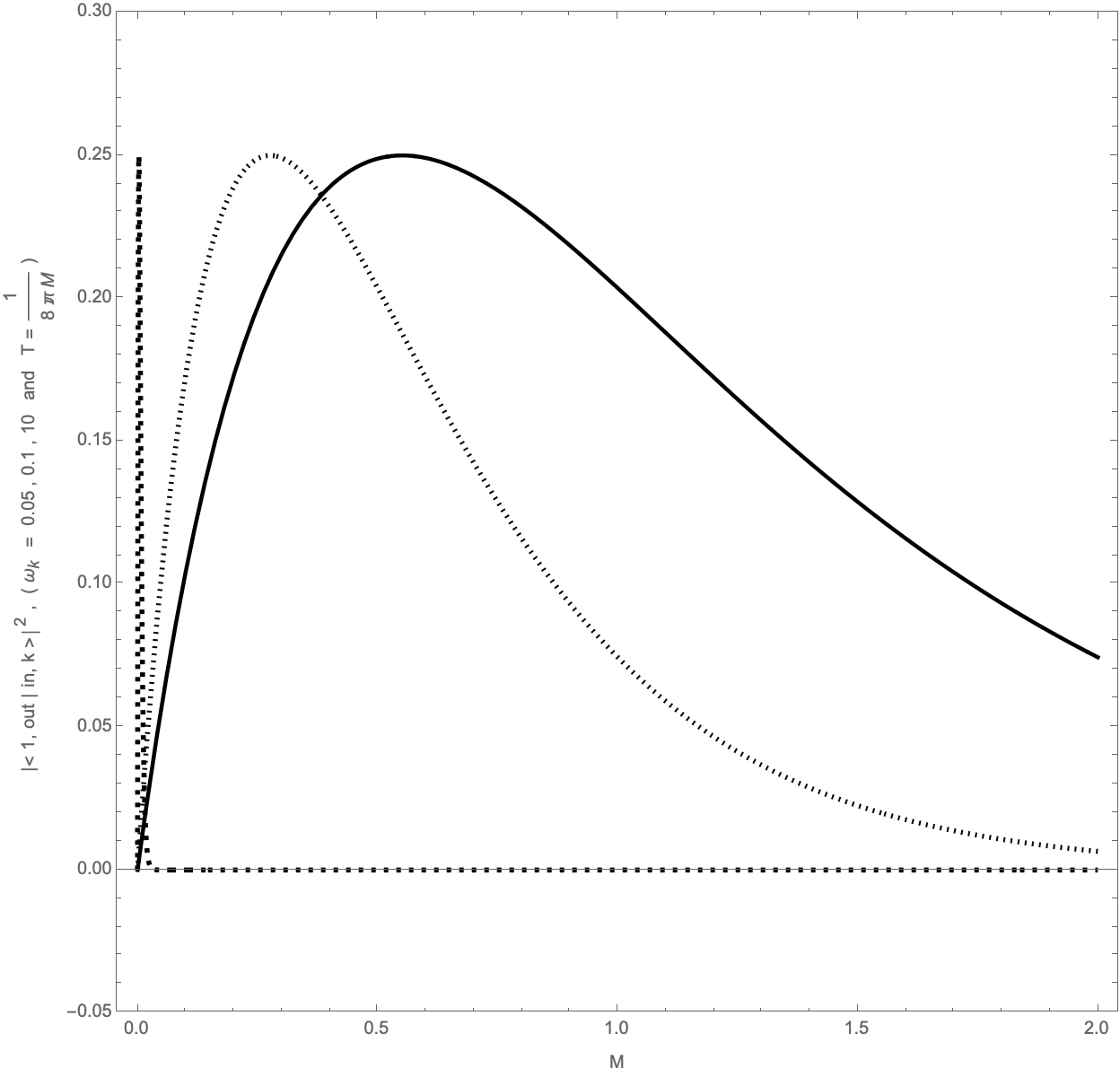} 
\hspace{1.5cm}
\includegraphics[height=7cm]{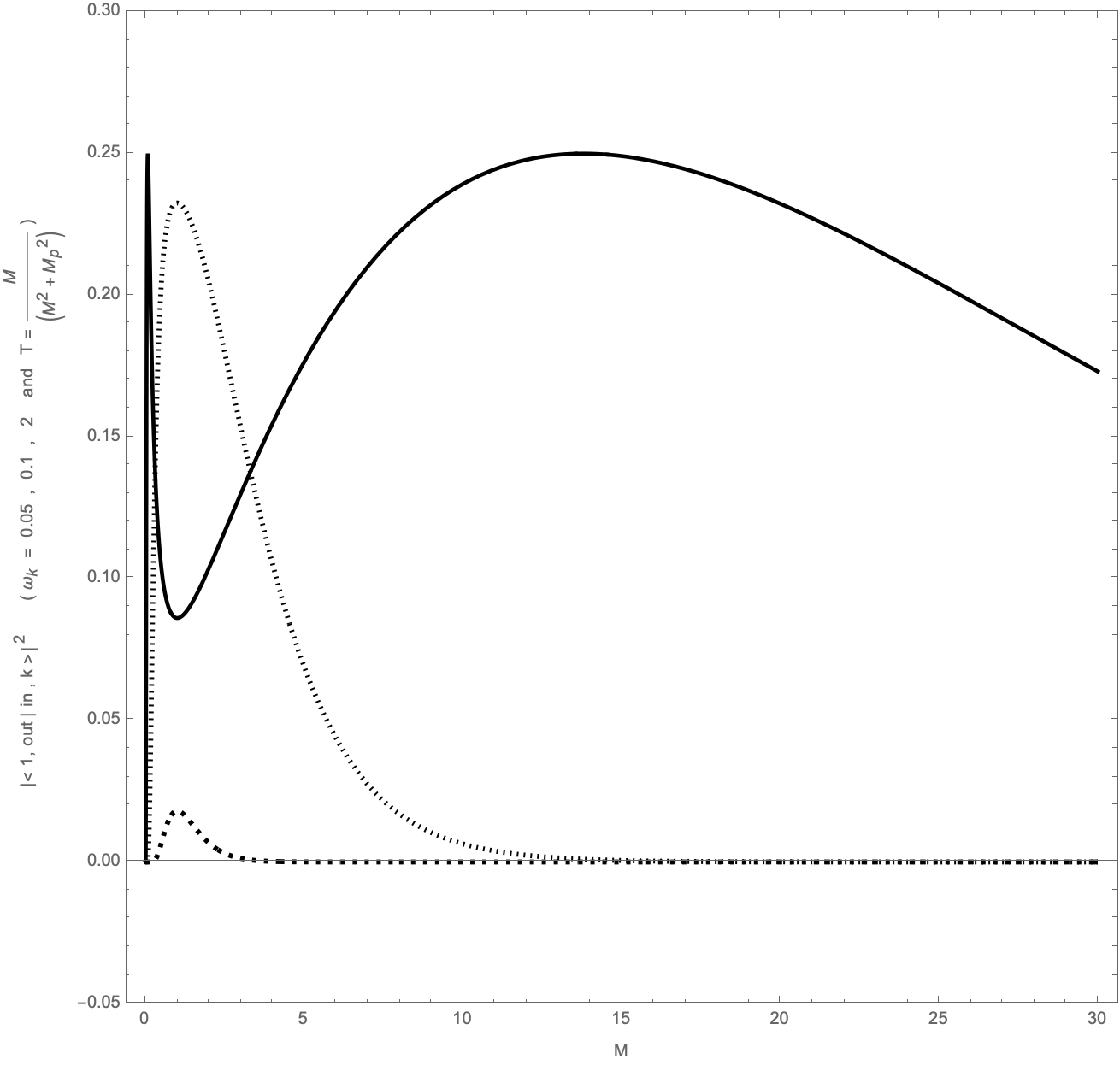} 
\caption{The first plot shows the probability density for the creation of zero, one, or two particles of energy 
$\omega({\bf k}) = 1$ in Planck units as a function of the black hole mass $M$, assuming the Hawking temperature $1/ 8 \pi M$. In the second and third plots, we consider the same probability density for $\omega({\bf k}) = 1/2$ and 
$\omega({\bf k}) = 0.1$, but with the modified temperature (\ref{TemperatureMod}), i.e.\ 
$T = M/(M^2+M_{\rm p}^2)$ (in order to enlarge the effect, we have omitted the factor $8 \pi$ in the denominator of the modified temperature). In the fourth and fifth plots, we show the transition amplitude for emitting one particle as a function of $M$, but for different values of the energy 
$\omega({\bf k})$ and for the Hawking or modified temperature.
}
\label{ZeronFigAmp}
\end{center}	
\end{figure}

\begin{figure}
\begin{center}
\includegraphics[height=7cm]{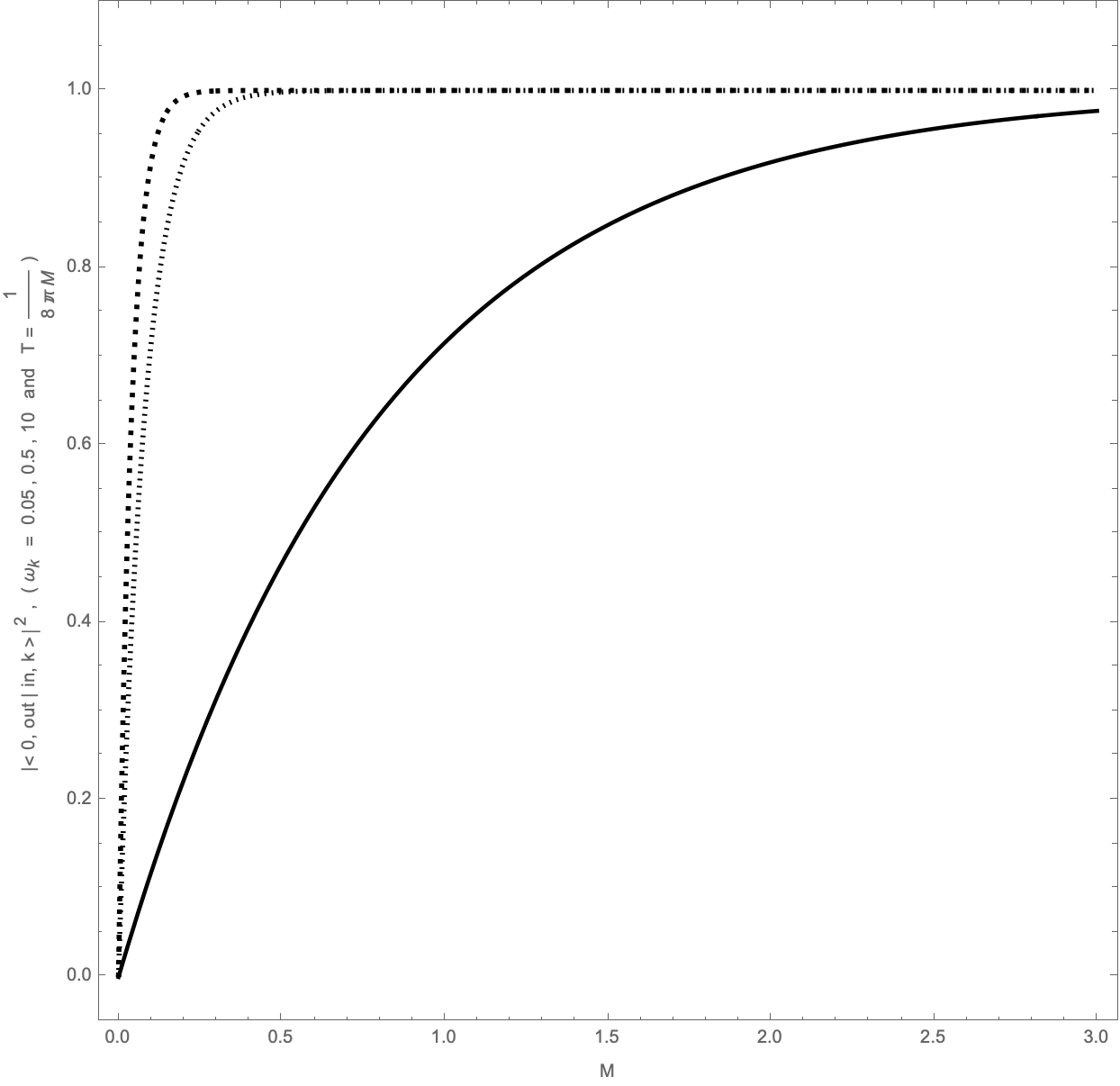} 
\hspace{1.5cm}
\includegraphics[height=7cm]{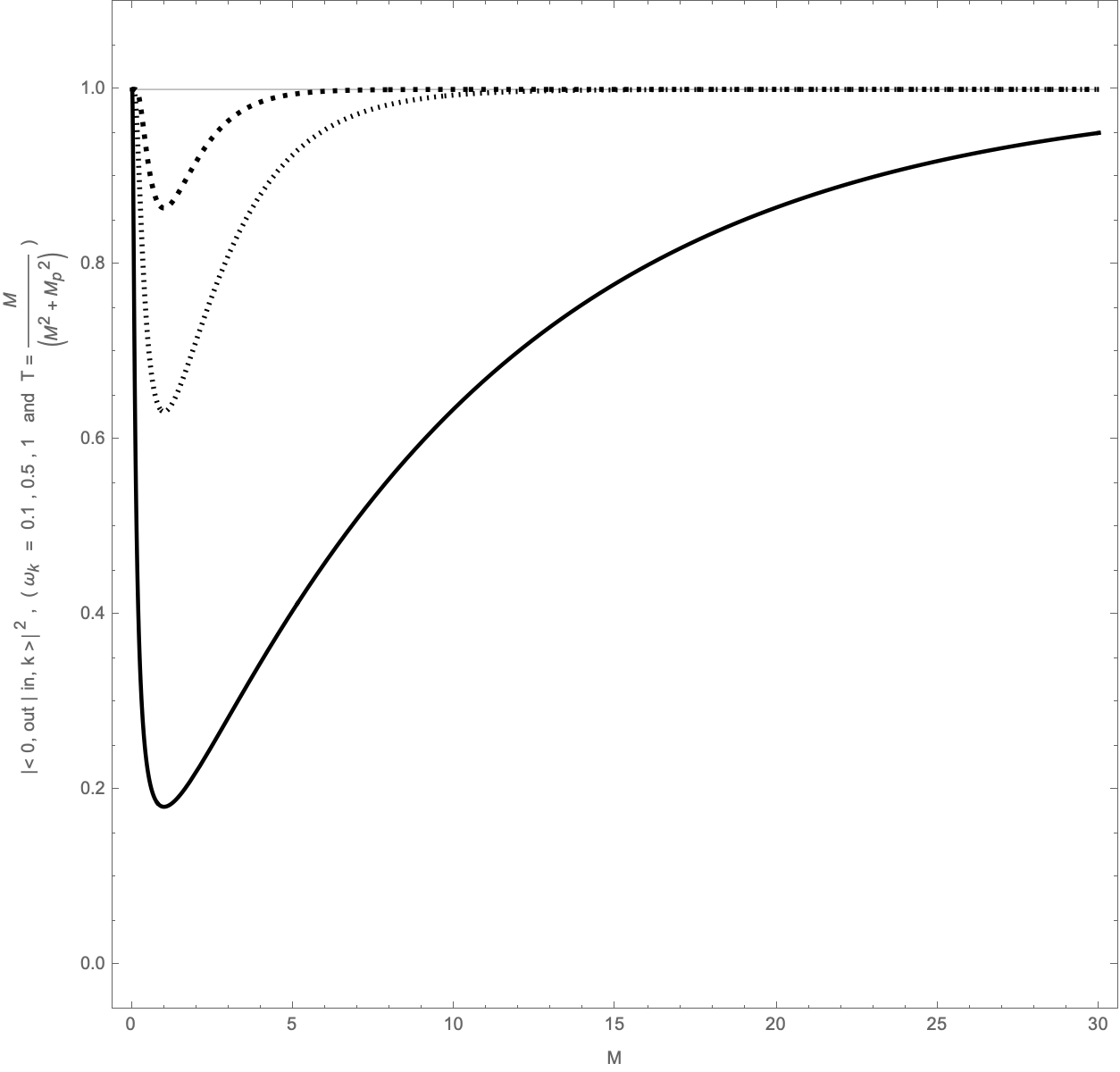} 
\caption{Plots of the probability density for the creation of zero particles for different values of the energy $\omega_{\bf k}$ in Planck units as a function of the black hole mass $M$. In the first plot we have assumed the Hawking temperature $1/ 8 \pi M$, while in the second $T = M/(M^2+M_{\rm p}^2)$.
}
\label{ZeronFig}
\end{center}	
\end{figure}
\hspace{-0.5cm}
 We provide several plots of the amplitudes (\ref{Zeron}) and (\ref{ZeronP}) in Fig.~\ref{ZeronFigAmp}. The first differs from the second and the third in the functional dependence of the temperature on the mass $M$: in the first we used the Hawking temperature, while in the second and third plots we implemented a modified temperature with the same features as (\ref{Temperatures}),
namely:
\be
T = \frac{G M}{8 \pi [ (GM)^2 + M_{\rm p}^2 ] } .
\label{TemperatureMod}
\ee
In the fourth and fifth plots, we show the transition amplitude for the emission of one particle as a function of $M$, for different values of the energy $\omega({\bf k})$, assuming the Hawking or modified temperature. 
In Fig.~\ref{ZeronFig}, we again consider the Hawking or the modified temperature and plot the amplitude for the creation of zero particles for different values of $\omega({\bf k})$. 
For $n>0$ the probability to produce particles approaches zero correctly for very large and very small black hole mass $M$. Indeed, for large $M$ the temperature goes to zero regardless of the modification to the temperature, while for small $M$ the modification of the temperature seems unavoidable, at least for $n=0$. Indeed, for $M=0$ there is no black hole and we should recover QFT in Minkowski spacetime. 
If the temperature goes to zero for $M\rightarrow 0$, then the number of produced particles also vanishes, while the normalization tends correctly to one, and the squeezing operator tends to the identity. Below we collect all the relevant limits for $T\rightarrow0$ according to (\ref{limitsB}), 
\be
&& \lim_{T \rightarrow 0} N_{\bf k} =
\lim_{T \rightarrow 0} 
\frac{1}{ {\rm e}^{ \frac{ \omega({\bf k}) }{T} }-1 } = 0 
\quad  \text{for} \quad
 \omega({\bf k}) > 0 \, , 
 \nonumber \\
&& 
\lim_{T \rightarrow 0} S_{\rm sq}(\omega({\bf k})) = 1\!\!1
\quad  \text{and} \quad 
\lim_{T \rightarrow 0} S_{\rm sq} = 1\!\!1  \, ,
\nonumber \\
&& 
\lim_{T \rightarrow 0}
| \langle n_{{\bf k} } , {\rm out} | 0_{\bf k}, {\rm in}  \rangle |^2 = 0 \quad  \forall \, n >0 \quad \text{but} \quad
\lim_{T \rightarrow 0}
| \langle n_{{\bf k} } , {\rm out} | 0_{\bf k}, {\rm in}  \rangle |^2 = 1 \quad \text{for} \quad  n=0 
\quad  \text{for} \quad
 \omega({\bf k}) > 0 \, 
 , \nonumber \\
&&
\lim_{T \rightarrow 0}
\langle {\rm out} | {\rm in} \rangle = 
\lim_{T \rightarrow 0}
\prod_{\omega({\bf k}) > 0 }  \sqrt{1 - {\rm e}^{- \frac{\omega ({\bf k}) }{T}} } = 1
\quad  \text{for} \quad
 \omega({\bf k}) > 0 \,  .
\ee
Most of the above limits hold if the scalar field has a mass term, meaning that in general one has an infrared divergence.

\section{The analogy 
with BCS theory and superconductivity}
\label{BCSBEC}
Well-established results and the mass gap for the low-temperature theory of superconductivity are derived in the mean-field approximation. In short, the theory describing electron-phonon interactions is first approximated with a quartic interaction between fermions, and afterwards replaced with a non-interacting effective model consisting of two operators: one quadratic in the creation operators and the other quadratic in the annihilation operators. The former creates, while the latter destroys, a Cooper pair.
%
Such a mean-field approximation consists in discarding quadratic fluctuations in the number operators themselves. 
We recall that phonons are not real particles but emerge as mediators of an effective interaction between electrons moving in a solid. This interaction is determined by the deformation of the crystal lattice as the electrons themselves pass through, and gives rise to the formation of Cooper pairs. 

In a curved spacetime, as for example in a gravitational collapse leading to the formation of a black hole, phonons are replaced by gravitons (fundamental particles), while the Cooper pairs are not made of electrons with opposite spin, but of electrons and positrons that experience an effective gravitational attraction (see later in this section). 
In the Rindler reference frame, the effective mean-field model takes into account the acceleration, namely the energy involved in the particle creation comes from the rocket fuel. 
On the other hand, in the presence of gravity the effective interaction between particles (scalars or fermions) is due to the presence of gravitons, while in an accelerating reference frame the energy stored in the fuel is partly converted into the creation of particles from the vacuum, but the effective Hamiltonian takes a very similar form.

%
As extensively shown in this paper, in a general spacetime the very same BCS state is obtained according to quantum field theory in curved spacetime \cite{Fabbri, BD, Townsend}. Hence, by reverse engineering, we can reconstruct a Hamiltonian for QFT in curved spacetime in the mean-field approximation \cite{NotesBCS}. 
The result will be a BCS Hamiltonian for the Bogoliubons, but with the mass gap function provided by the specific spacetime geometry. 
The BCS state is the vacuum for the Bogoliubons---the in-operators according to our definitions in curved spacetime---but an arbitrary combination of Cooper pairs for the out-operators; namely, the BCS state is an electron-positron condensate.
Moreover, since we already have the BCS state and the Bogoliubov coefficients, we can reconstruct the mass-gap function $\Delta_{\bf k}$ that will fully determine the effective Hamiltonian in the mean-field approximation. 

Let us very briefly recap the model and the main features of BCS theory in solid-state physics. 
The effective Hamiltonian of the BCS model in the mean-field approximation reads\footnote{The effective Hamiltonian before implementing the mean-field approximation includes four-fermion interactions and reads:
\be
H = \sum_{\bf{k}, {\rm s}} \xi_{\bf k} \, c_{{\bf k}, {\rm s}}^{ \dagger} c_{ {\bf k},{\rm s}} + \frac{1}{N}
 \sum_{ {\bf k}, {\bf k'} } V_{ {\bf k} {\bf k'} } c_{\bf{k} \uparrow}^{ \dagger} c_{- \bf{k} \downarrow}^{ \dagger}
  c_{- {\bf k'} \downarrow} c_{ {\bf k'} \uparrow} 
  . 
  \label{EffHBCS}
\ee
}:
\be
H_{\rm BCS} = 
\sum_{\bf{k}, {\rm s}} \xi_{\bf k} \, c_{{\bf k}, {\rm s}}^{ \dagger} c_{ {\bf k},{\rm s}} + 
\sum_{\bf k}  \left( \Delta_{\bf k} \, c_{\bf{k} \uparrow}^{ \dagger} c_{- \bf{k} \downarrow}^{ \dagger}
+ \Delta_{\bf{k}}^* \,  c_{- {\bf k} \downarrow} c_{\bf{k} \uparrow} 
\right)
+ \sum_{ \bf{k} } 
 \Delta_{\bf k} \, \left\langle c_{\bf{k} \uparrow}^{ \dagger} c_{- \bf{k} \downarrow}^{ \dagger} \right\rangle ,
 \label{H_BCS}
\ee
where $\Delta_{\bf k}$ is known as the mass gap function and is defined in terms of an effective interaction between electrons in the crystal and the correlation function of Hermitian conjugate operators with opposite spin and momentum, 
\be
\Delta_{\bf k} = - \frac{1}{N} \sum_{\bf k'} V_{{\bf k} {\bf k'} } 
\big\langle c_{ {\bf - k'} \downarrow } \, c_{ {\bf k'} \uparrow} \big\rangle
\label{consi1}
\ee
The BCS Hamiltonian can be diagonalized by a Bogoliubov transformation with coefficients $u_{\bf k}$ and $v_{\bf k}$, introducing new fermionic operators $\gamma_{ {\bf k}  s}$ (the Bogoliubons) defined by the BT,
\be
&& c_{{ \bf k} \uparrow} = u^*_{{ \bf k} } \gamma_{{\bf k} \uparrow} 
+ v_{{ \bf k}} \gamma_{- {\bf k} \downarrow}^\dagger \nonumber \\
&& c^\dagger_{ - { \bf k} \downarrow} = u_{{ \bf k}} \gamma^\dagger_{ - {\bf k} \downarrow}
- v^*_{{ \bf k}} \gamma_{{\bf k} \uparrow}^\dagger 
\ee
where the coefficients $u_{{ \bf k} }, v_{{ \bf k} }$ satisfy the normalization condition,
\be
| u_{{ \bf k} } |^2 + |v_{{ \bf k} }|^2 = 1 \, . 
\ee
The BCS Hamiltonian takes the following diagonal form, 
\be
H_{\rm BCS} = 
\sum_{\bf{k}, {\rm s}} E_{\bf k}  \gamma^\dagger_{{\bf k}, {\rm s}} \gamma_{{\bf k},{\rm s}} + E_0 
\, ,
\ee
provided the following relation between $u_{{ \bf k} }$, $v_{{ \bf k} }$, $\Delta_{\bf k}$, and $\xi_{ {\bf k} }$ 
is satisfied,
\be
\frac{v_{\bf k}}{u_{\bf k} }= \frac{\sqrt{\xi_{\bf k}^2 + |\Delta_{\bf k}|^2} - \xi_{\bf k}}{\Delta^*_{\bf k}} . 
\label{vonu}
\ee
Since the numerator is real, the phase of $\Delta_{\bf k}$ must be the same as the relative phase between 
$v_{\bf k}$ and $u_{\bf k}$. Moreover, we can set the phase of $u_{\bf k}$ to zero without loss of generality. 
Hence, $v_{\bf k}$ and $\Delta_{\bf k}$ have the same phase. 
The energy eigenvalue of the full BCS Hamiltonian and the ground-state energy are:
\be
E_{\bf k} = \sqrt{\xi_{\bf k}^2 + | \Delta_{\bf k}|^2} \, , \qquad 
E_0 = \sum_{\bf k} \left( \xi_{\bf k} - E_{\bf k} + \Delta_{\bf k} 
\langle c^\dagger_{ {\bf k} \uparrow} c^\dagger_{- {\bf k} \downarrow}
 \rangle \right) \, .
\label{BogoEn}
\ee
Finally, the BCS ground state corresponds to the vacuum of Bogoliubons,
\be
\gamma_{ {\bf k} s} | \Psi_{\rm BCS} \rangle = 0 \, ,
\ee
and reads:
\be
| \Psi_{\rm BCS} \rangle 
& = & \prod_{\bf k} u_{\bf k} \, {\rm e}^{ \frac{v_{\bf k}}{u_{\bf k}} c^\dagger_{{\bf k} \uparrow} 
c^\dagger_{ - {\bf k} \downarrow} } | 0 \rangle 
\nonumber \\
& = &  \prod_{\bf k} \left( u_{\bf k} + v_{\bf k} \, 
 c^\dagger_{{\bf k} \uparrow} 
c^\dagger_{ - {\bf k} \downarrow} 
\right)
| 0 \rangle \, . 
\label{BCSCM}
\ee 
Finally, since the Bogoliubons follow the Fermi-Dirac distribution and have energy $E_{\bf k}$, we can express the consistency condition (\ref{consi1}) as:
\be
\Delta_{\bf k} = - \frac{1}{N} \sum_{\bf k'} \frac{V_{{\bf k} {\bf k'} }  \, \Delta_{\bf k'}}{2 E_{\bf k'} } \tanh
\left( \frac{ E_{\bf k'}}{2 T}  \right)  .
\label{consi2}
\ee

To make contact with quantum field theory in the spacetime background of interest, the following identifications are in order. For the \underline{Rindler spacetime}:
\be
&& c_{{\bf k} \uparrow} \,\, \rightarrow \,\, c^{\rm I}_{\bf k} \, , \qquad  
c_{{\bf k} \uparrow}^{\dagger}  \,\, \rightarrow \,\, c^{\rm I \dagger}_{\bf k} \, ,
\nonumber \\
&& c_{-{\bf k} \downarrow}  \,\, \rightarrow \,\, d^{\rm II}_{ - {\bf k}} \, , \qquad 
c_{-{\bf k} \downarrow}^\dagger  \,\, \rightarrow \,\, d^{\rm II \dagger }_{ - {\bf k}} \, .
\nonumber \\
\ee
In short, this amounts to making the following identification: $\uparrow \equiv {\rm electron}$ and $\downarrow \equiv {\rm positron}$. 
Consistently, the Bogoliubons are identified with:
\be
&& \gamma_{{\bf k} \uparrow} \,\, \rightarrow \,\, a_{\bf k} \, , \nonumber \\
&& \gamma_{- {\bf k} \downarrow}^\dagger \,\, \rightarrow  \,\, b^\dagger_{ - {\bf k}} \, . 
\ee
In the \underline{black hole background} the identifications read:
\be
&& c_{{\bf k} \uparrow} \,\, \rightarrow \,\, c^{({\rm out})}_{\bf k} \, , \qquad  
c_{{\bf k} \uparrow}^{\dagger}  \,\, \rightarrow \,\, c^{( {\rm out}) \dagger}_{\bf k} \, ,
\nonumber \\
&& c_{-{\bf k} \downarrow}  \,\, \rightarrow \,\, d^{({\rm int})}_{ - {\bf k}} \, , \qquad 
c_{-{\bf k} \downarrow}^\dagger  \,\, \rightarrow \,\, d^{({\rm int}) \dagger }_{ - {\bf k}} \, .
\ee
Consistently, the Bogoliubons are identified with:
\be
&& \gamma_{{\bf k} \uparrow} \,\, \rightarrow \,\, a^{({\rm in})}_{\bf k} \, , \nonumber \\
&& \gamma_{- {\bf k} \downarrow}^\dagger \,\, \rightarrow \,\, b^{({\rm in})\dagger}_{ - {\bf k}} \, .
\label{BogoBH}
\ee

According to the derivation of BCS theory in condensed matter physics, the Hamiltonian in the mean-field approximation, taking into account the above correspondence between operators, reads:
\be
\hspace{-0.5cm}
H_{\rm eff} = 
\sum_{ {\bf k}, s} \xi_{ {\bf k}} \, c_{\bf k}^{( {\rm out}) \dagger} c_{\bf k}^{( {\rm out})} + 
\sum_{{ \bf k}, s} \xi_{\bf k} \, d_{- \bf k}^{( {\rm int}) \dagger} d_{- \bf k}^{( {\rm int})} + 
\sum_{{\bf k}, s}  \left( \Delta_{\bf k} \, c_{ \bf k}^{( {\rm out}) \dagger} d_{-{\bf k}}^{( {\rm int}) \dagger}
+ \Delta_{\bf k}^* \,  d_{- {\bf k}}^{( {\rm int})} c_{ \bf k}^{( {\rm out})} 
\right)
+ \sum_{{\bf k}, s} 
 \Delta_{\bf k} \, \left\langle c_{{\bf k}}^{( {\rm out}) \dagger} d_{- {\bf k}}^{( {\rm int}) \dagger} \right\rangle \!  .
 \label{EffBH-R}
\ee
Contrary to the condensed matter system, the sum over the spin $s$ appears in every term of the Hamiltonian.
In terms of Bogoliubons the Hamiltonian reads:
\be
H_{\rm eff} = 
\sum_{\bf{k}, {\rm s}} E_{\bf k}  \left( a^{({\rm in})\dagger}_{{\bf k}, {\rm s}} \, 
a^{({\rm in})}_{{\bf k}, {\rm s}} 
+
 b^{({\rm in})\dagger}_{ - {\bf k}, {\rm s}} \, 
b^{({\rm in})}_{ - {\bf k}, {\rm s}} \right) 
+ E_0
\, . 
\ee
According to (\ref{vacM2}), 
the BCS state for fermions and anti-fermions in the black hole background reads:
\be
\label{BCSBH} 
|0_{\bf k}\rangle^+ |0_{- {\bf k} }\rangle^- 
= \left(
\cos r_{\bf k}  \, + e^{-i\phi}\,\sin r_{\bf k} \,c^{ ({\rm out}) \dagger}_{ \bf k}\,d^{ ({\rm int}) \dagger}_{-{\bf k}}
\right) \,|0_{\bf k}\rangle^+_{\rm out }\,|0_{-{\bf k}}\rangle^-_{\rm int } 
= S_{\rm sq} \,|0_{\bf k}\rangle^+_{\rm out }\,|0_{- {\bf k}}\rangle^-_{\rm int} \equiv | \Psi_{{\rm BCS}} \rangle
\, , 
\ee
where the $\{+,-\}$ ket labels stand for a
particle or an anti-particle state, and 
\be
\tan r_{\bf k} = {\rm e}^{- \pi \Omega_{\bf k}} \, , \quad \Omega_{\bf k} = \frac{\sqrt{ {\bf k}^2 + m^2}}{a} 
= \frac{\sqrt{ {\bf k}^2 + m^2}}{2 \pi  T } 
= \frac{ \xi_{\bf k} }{2  \pi T } = \frac{\xi_{\bf k}}{\kappa_+} 
\, .
\label{tanrk}
\ee
where we have considered massive fermions in order to avoid infrared divergences. However, all the results are correct for massive as well as massless particles. 
By comparing (\ref{BCSBH}) and (\ref{BCSCM}), we infer that:
\be
\cos r_{\bf k} \equiv u_{\bf k} \, , \quad \sin r_{\bf k} \, e^{- i \phi}  \equiv v_{\bf k} \, ,
\label{cossin}
\ee
where we have introduced the temperature according to (\ref{T_U}). However, all the above formulas from (\ref{EffBH-R}) to (\ref{tanrk}) are valid for both Rindler spacetime and the black hole background. In the latter case we have to replace in (\ref{tanrk}) the Hawking temperature (\ref{T_H}).

The BCS state is annihilated by the operators $a_{\bf k}^{(\rm in)}$, $b_{- {\bf k}}^{(\rm in)}$ (see (\ref{BogoBH})). 
Therefore, the latter operators, together with their Hermitian conjugates, 
destroy and create what are usually called Bogoliubons. 
The ground state of the system is not simply an empty Fermi sea, but a state where all particle levels are unoccupied, i.e., 
\be
a_{\bf k}^{(\rm in)} | \Psi_{\rm BCS} \rangle = 0 \, ,  \quad  b_{- {\bf k}}^{(\rm in)} | \Psi_{\rm BCS} \rangle = 0 \,\, \text{for all} \,\, {\bf k} \, .
\ee
Such a state, namely the BCS state in the context of superconductivity, is a coherent superposition of states with different particle numbers and represents the macroscopic condensate.
Therefore, a general background metric (in this case the black hole metric or the Rindler metric) 
can turn the QFT vacuum into a superconducting condensate.

\begin{figure}
	\begin{center}
		\includegraphics[height=7cm]{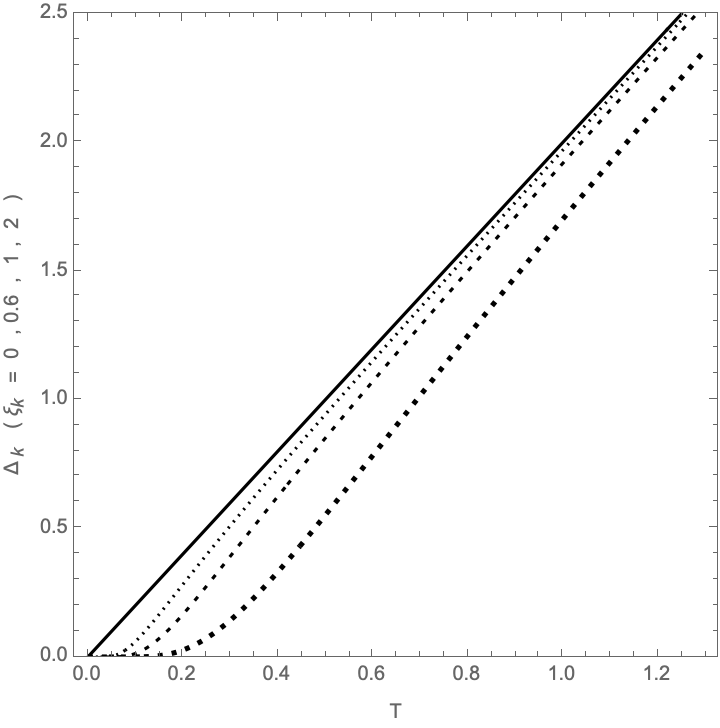}
		\hspace{1.5cm} 
		\includegraphics[height=7cm]{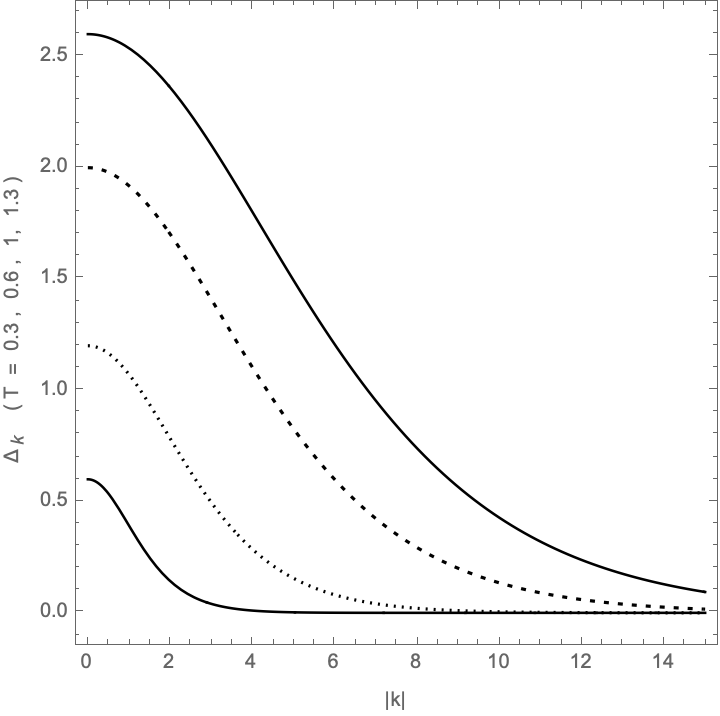}
		\caption{Plot of $|\Delta_{\bf k}|$ as a function of $T$ in the left panel and as a function of 
$\xi_{\bf k} = |{\bf k}|$ in the right panel (we have considered the massless case). In the left panel the solid line corresponds to $|\Delta_{\bf k}| = 2 T$, 
i.e.\ $\xi_{\bf k} \rightarrow 0$ (or large $T$), and the other curves, from left to right, are for increasing values of 
$|{\bf k}|$. 
Notice that $|\Delta_{\bf k}| \simeq 2 T$ for large $\xi_{\bf k}$. 
In the right panel, from bottom to top the curves refer to increasing values of the temperature.
}
		\label{DeltaGap}
	\end{center}
\end{figure}

\begin{figure}
	\begin{center}
		\includegraphics[height=7cm]{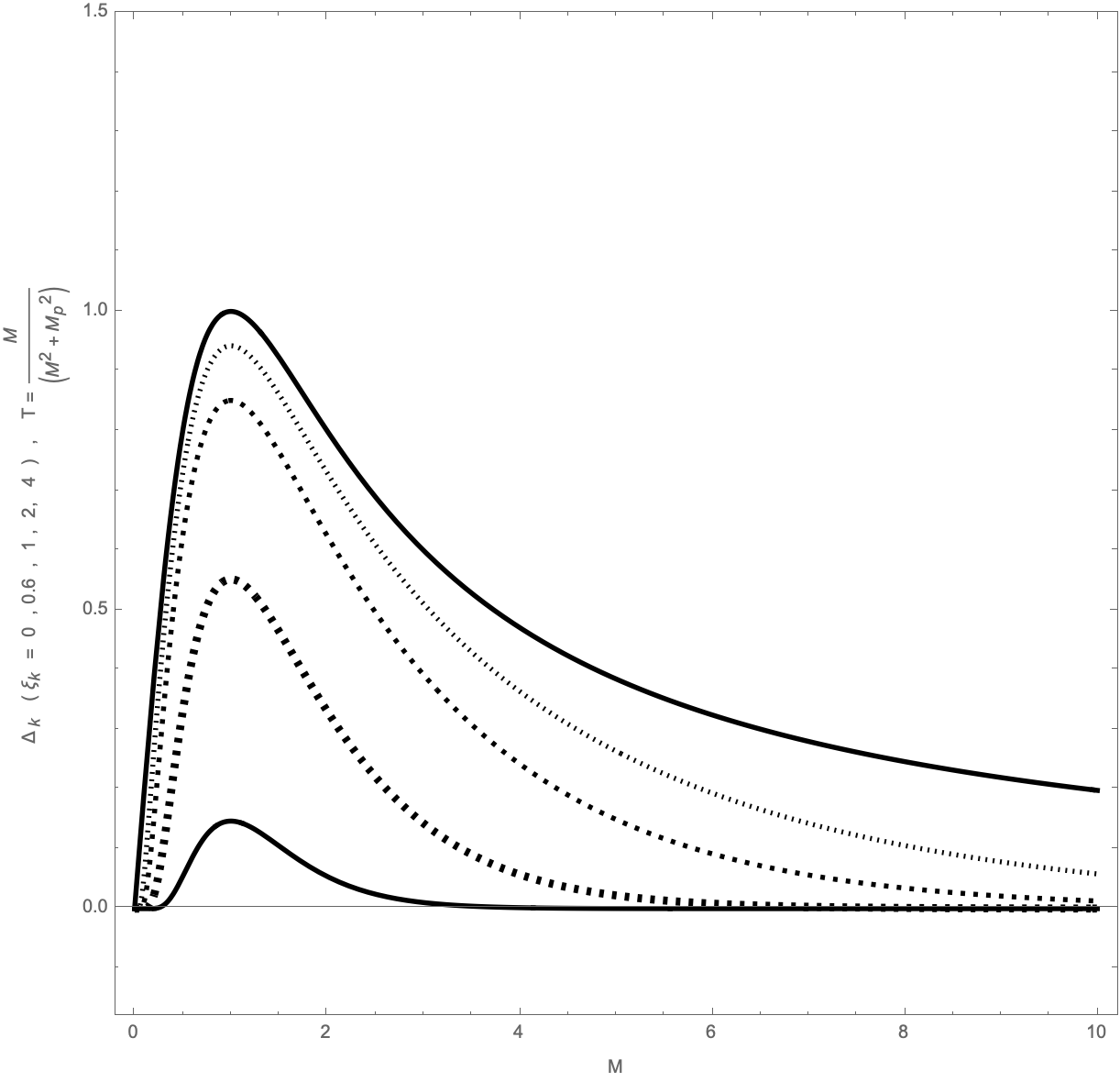}
		\caption{Plot of the gap function $|\Delta_{\bf k}|$ as a function of $M$ for five different values of $\xi_{\bf k}$ and temperature $T = M/(M^2 + M_{\rm p}^2)$. Clearly, for large and small $M$, the mass gap goes to zero and we recover quantum field theory in Minkowski spacetime. 
}
		\label{DeltaGap2}
	\end{center}
\end{figure}

In the Rindler, or the black hole, metric, $\Delta_{\bf k}$ is fully specified for $\xi_{\bf k} > 0$ because the Bogoliubov transformation is known, having been evaluated by Unruh and Hawking. In particular, substituting (\ref{cossin}) into (\ref{vonu}), the latter becomes:
\be
e^{- i \phi} \frac{ \sin r_{\bf k} }{  \cos r_{\bf k}  }= \frac{\sqrt{\xi_{\bf k}^2 + | \Delta_{\bf k}|^2 } - \xi_{\bf k}}{\Delta^*_{\bf k}} . 
\label{vonuBH0}
\ee
Hence, $ \Delta_{\bf k} =  |\Delta_{\bf k}| \exp ( - i \phi)$ and the modulus $| \Delta_{\bf k} |$ is obtained by solving the following equation,
\be
e^{- i \phi} \frac{ \sin r_{\bf k} }{  \cos r_{\bf k}  }= \frac{\sqrt{\xi_{\bf k}^2 + | \Delta_{\bf k}|^2 } - \xi_{\bf k}}{ e^{i \phi} | \Delta_{\bf k} |} 
\quad 
\Longrightarrow 
\quad 
\boxed{
| \Delta_{\bf k} | = \frac{2 e^{\pi  \Omega_{\bf k} } \xi_{\bf k}  }{e^{2 \pi  \Omega_{\bf k} } - 1} 
= \xi_{\bf k}   \,\text{csch}(\pi  \Omega_{\bf k} )}
\, , 
\label{vonuBH}
\ee
which has been derived for $\xi_{\bf k} > 0$. However, the result for $| \Delta_{\bf k} |$ is analytic and can be extended to $\xi_{\bf k} = 0$ as well. In particular, near $\xi_{\bf k} = 0$ the mass gap is constant and proportional to the temperature,
\be
| \Delta_{\bf k} | = 2 T + O\left(\xi_{\bf k}^2 \right)  \, .
\label{Dxi}
\ee
In other words, the mass gap is non-zero even at the Fermi level, where $\xi_{\bf k} = 0$. 
Moreover, for large values of the temperature $| \Delta_{\bf k} |$ is independent of $\xi_{\bf k}$, 
\be
| \Delta_{\bf k} | = 2 T - \frac{\xi_{\bf k}}{12 T} +  O \left( T^{-3} \right) \, .
\label{DT}
\ee
In Fig.~\ref{DeltaGap}, we provide a plot of the mass gap function for different values of the temperature or different values of the energy $\xi_{\bf k} = |{\bf k} |$. 
In Fig.~\ref{DeltaGap2}, we provide a plot of $|\Delta_{\bf k}|$ as a function of the mass $M$ and the temperature given in (\ref{TemperatureMod}). Also:
\be
\lim_{T \rightarrow 0} | \Delta_{\bf k} | = 0 \quad \text{for} \quad \xi_{\bf k} >   0 \, .
\ee

\begin{figure}
	\begin{center}
		\includegraphics[height=6cm]{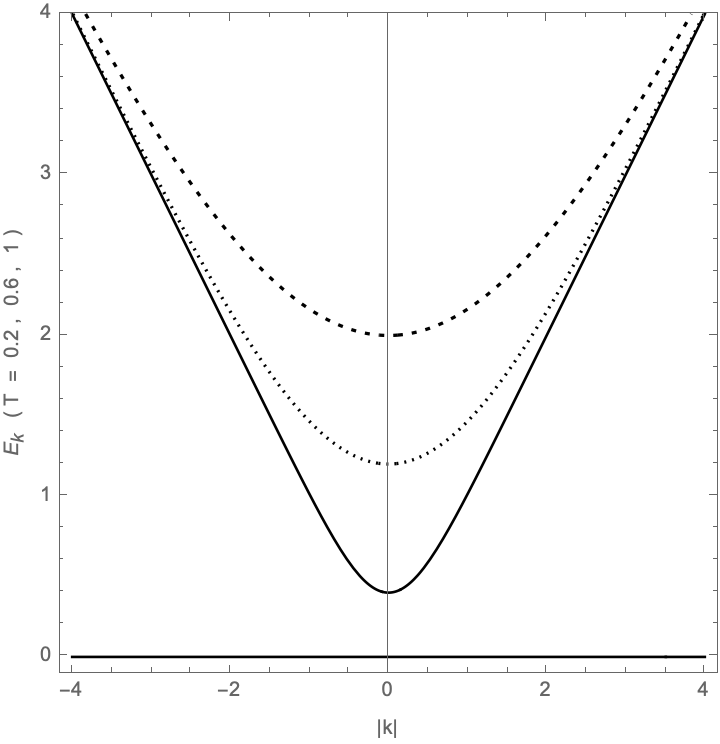}
		\caption{Plot of the energy $E_{\bf k}$ as a function of $|{\bf k}|$ for three different values of $T$; clearly the energy exhibits a mass gap for any positive, non-vanishing value of the temperature. 
}
		\label{Energy}
	\end{center}
\end{figure}

Substituting (\ref{vonuBH}) into (\ref{BogoEn}), we obtain the energy eigenvalues, i.e.\ 
\be 
E_{\bf k} = \xi_{\bf k} \coth (\pi \, \Omega_{\bf k} ) \, , 
\label{EnBH}
\ee
whose plot is given in Fig.~\ref{Energy}.

Finally, we have to consider the consistency condition (\ref{consi2}). 
However, we would like to stress that all the results, including the effective Hamiltonian (\ref{EffBH-R}), the gap function (\ref{vonuBH}), the energy (\ref{EnBH}), etc., are correct regardless of the consistency condition. Indeed, we do not derive $V_{{\bf k} {\bf k'}}$ here from first principles starting from a scalar or a Dirac field in curved spacetime. We believe this project is already rich enough in novel results and self-consistent enough to leave further direct derivations to the future. 
Nevertheless, we are going to show that the usual approximation of a constant interaction potential gives a very good approximation at relatively high temperature. 
Substituting (\ref{vonuBH}) and (\ref{EnBH}) into (\ref{consi2}) we get:
\be
&& \xi_{\bf k} \,\text{csch}(\pi  \Omega_{\bf k} ) = - \frac{1}{N} \sum_{\bf k'} \frac{V_{ {\bf k} {\bf k'} }}{2} 
\,\text{sech}(\pi  \Omega_{\bf k'} )
\tanh
\left[ ( \pi  \Omega_{\bf k'} ) \coth ( \pi   \Omega_{\bf k'} ) 
   \right] \, , 
    \nonumber \\
   &&  \Omega_{\bf k} = \frac{\sqrt{ {\bf k}^2 + m^2}}{2 \pi T } \equiv  \frac{\sqrt{ {\bf k}^2 + m^2}}{\kappa_+ }
   \, , \quad \xi_{\bf k} = \sqrt{{\bf k}^2 + m^2} \, .
\label{consi3}
\ee

\begin{figure}
	\begin{center}
		\includegraphics[height=7cm]{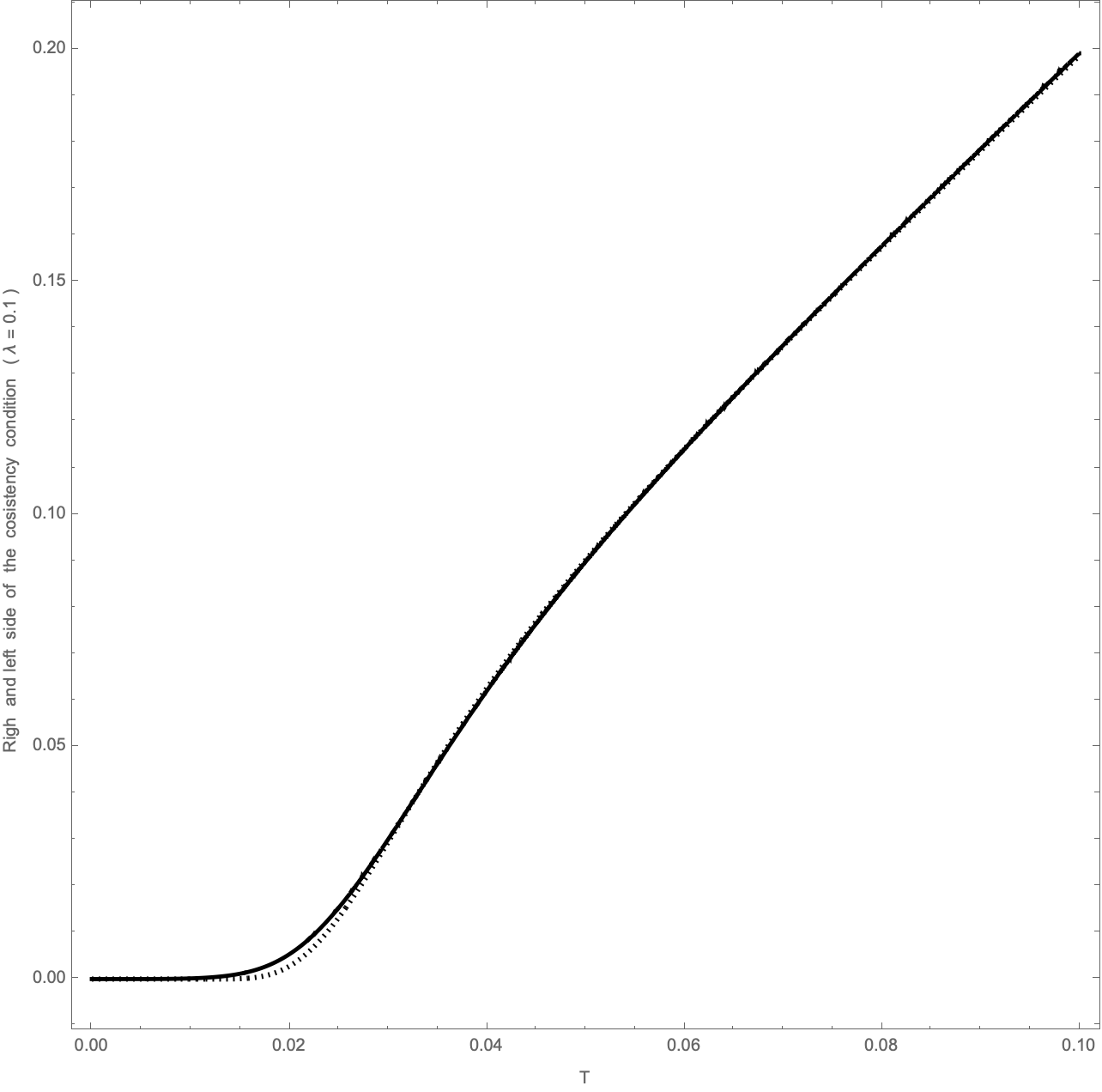}
		\hspace{1cm} 
		\includegraphics[height=7cm]{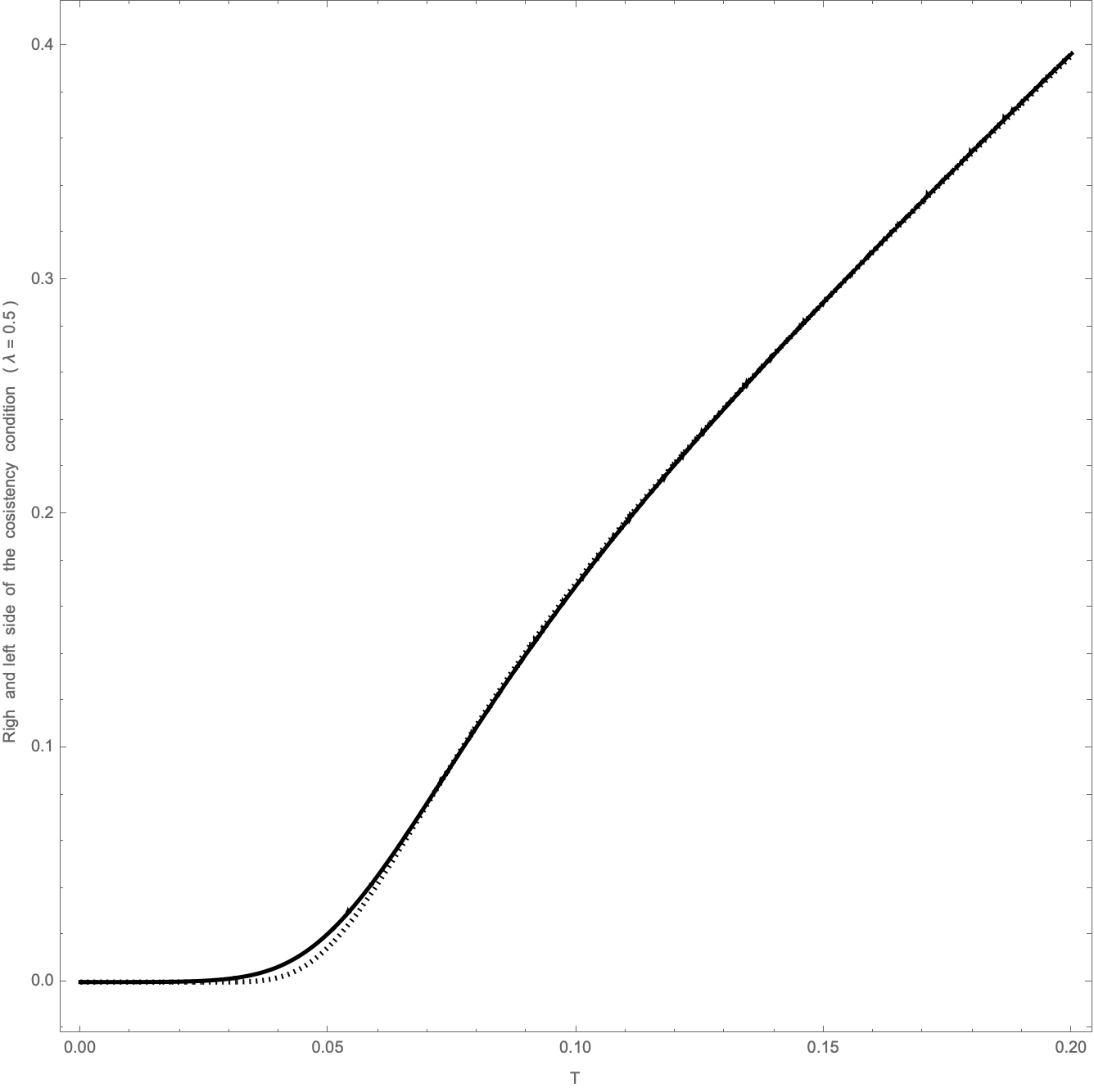}
		\includegraphics[height=7cm]{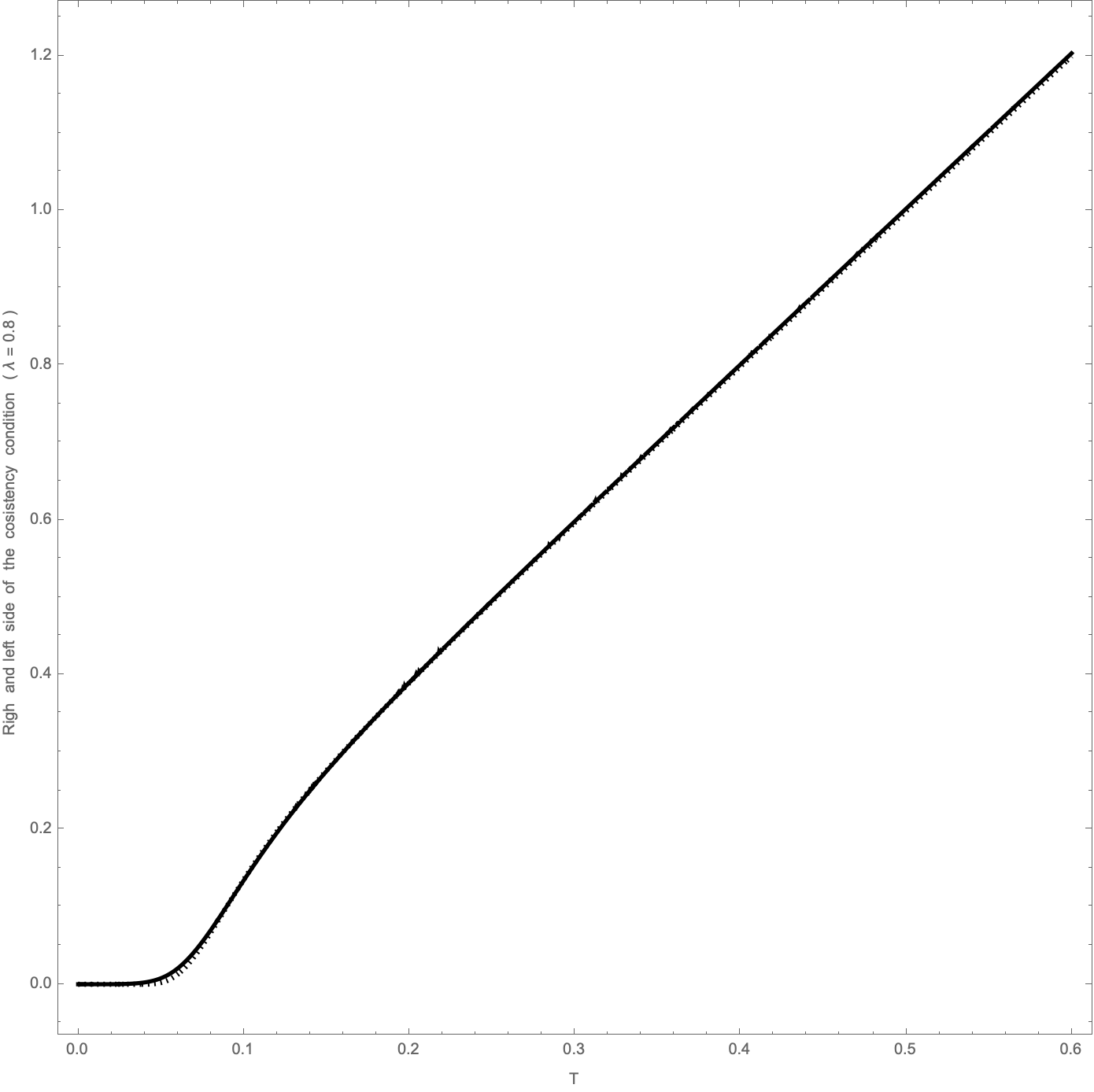}
		\hspace{1cm}
		\includegraphics[height=7cm]{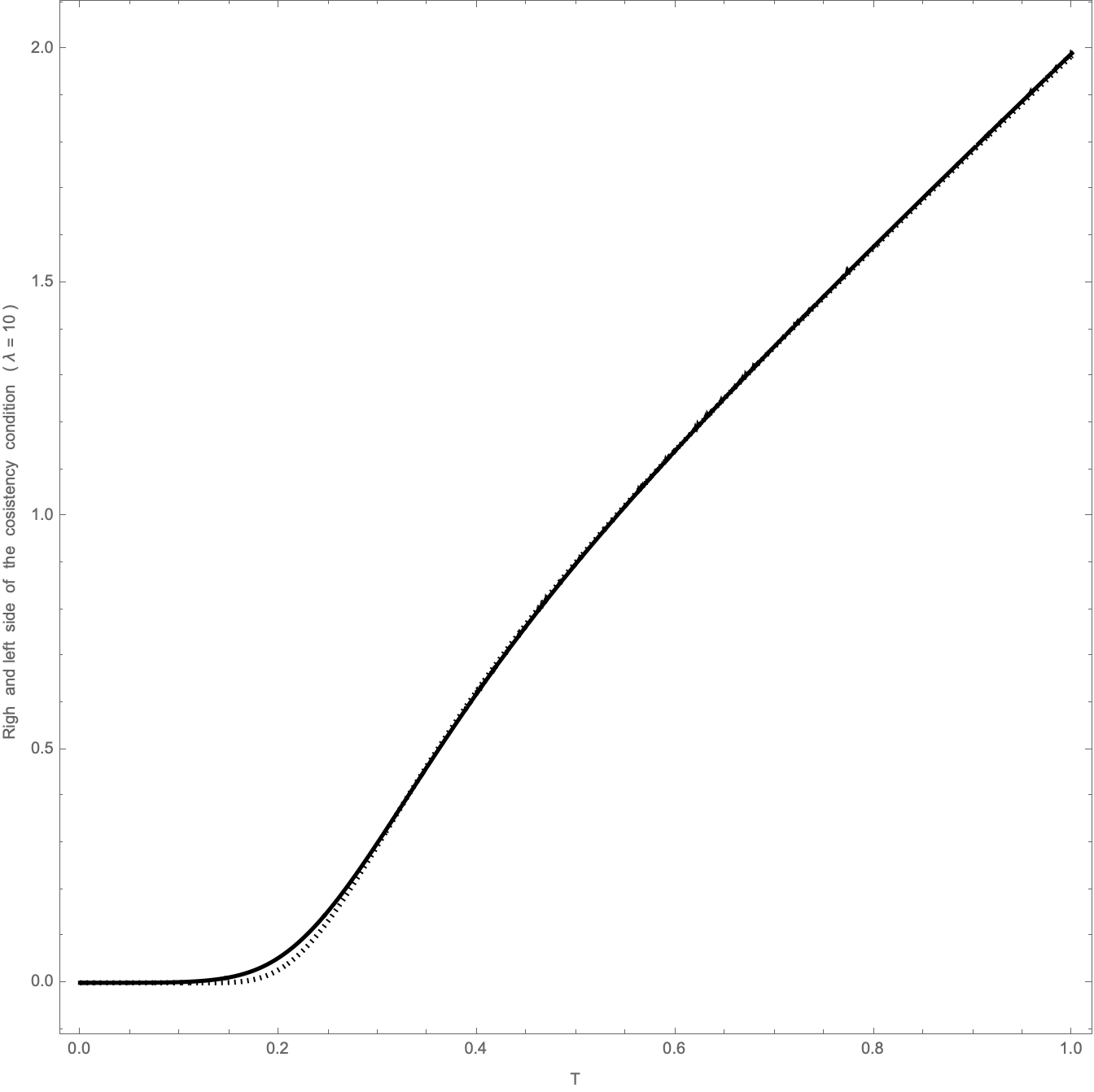}
		\caption{Plots of the right-hand side (solid line) and the left-hand side (dashed line) of (\ref{consistencyC}) as a function of the temperature. Clearly, for large temperature the mass gap function $|\Delta_{\bf k}| \simeq 2 T$ satisfies the consistency condition very well $\forall \, |{\bf k}|$. We have assumed the realistic gravitational potential (\ref{gravipoteff}) and massless particles, namely $m=0$.  
}
		\label{consistencyC}
	\end{center}
\end{figure}

We can study the consistency equation (\ref{consi3}) in the continuum by replacing the sum with a three-dimensional integral and assuming a Newtonian gravitational interaction proportional to a constant mass squared (in natural units) that we will fix later, 
\be
V_{{\bf k} {\bf k'} } = - \frac{ {\rm Mass}^2 }{ ( \bf{k' - k} )^2 } .
\label{gravipot}
\ee
 Therefore, (\ref{consi3}) becomes:
\be
- \frac{1}{N} \sum_{\bf k'} V_{ {\bf k} {\bf k'} } & = & - \frac{{\rm Vol}}{N} \int \frac{d^3 k' }{(2 \pi)^3} \left[ - \frac{ {\rm Mass}^2 }{ ( \bf{k' - k} )^2 } \right] 
=  
- 2 \pi \frac{{\rm Vol}}{N}  \int_0^{+ \infty} \int_0^{\pi}  \frac{{\bf k'}^2 d |{\bf k'}|   \sin \theta \, d \theta}{(2 \pi)^3} \left[ - \frac{ {\rm Mass}^2 }{ ( \bf{k' - k} )^2 } \right] \nonumber \\
& = &
- 2 \pi \frac{{\rm Vol}}{N}  \int_0^{+ \infty} \int_0^{\pi}  \frac{{\bf k'}^2 d |{\bf k'}|   \sin \theta \, d \theta}{(2 \pi)^3} \left[ - \frac{ {\rm Mass}^2 }{ {\bf k'}^2 + {\bf k}^2 - 2 |{\bf k'}| |{\bf k}| \cos \theta }\right] \nonumber 
\ee
\be
& = &
-  \frac{{\rm Vol} }{(2 \pi )^2 N}  \int_0^{+ \infty} \int_0^{\pi}  {\bf k'}^2 d |{\bf k'}|   \sin \theta \, d \theta 
\left[ - \frac{  {\rm Mass}^2  }{ {\bf k'}^2 + {\bf k}^2 - 2 |{\bf k'}| |{\bf k}| \cos \theta }\right] \nonumber \\
& = &
 \frac{{\rm Vol} \,  {\rm Mass}^2 }{(2 \pi )^2 N}  \int_0^{+ \infty}  {\bf k'}^2 d |{\bf k'}|   \int_0^{\pi} \sin \theta \, d \theta \left[ \frac{1 }{ {\bf k'}^2 + {\bf k}^2 - 2 |{\bf k'}| |{\bf k}| \cos \theta }\right] \nonumber \\
& =  &
 \frac{{\rm Vol}\,  {\rm Mass}^2 }{(2 \pi )^2 N}  \int_0^{+ \infty}  {\bf k'}^2 d |{\bf k'}|   
\, \frac{1}{2 {\bf k'} {\bf k}}  \, \log \frac{( \bf{ |k'| + |k|} )^2}{( \bf{ |k'| - |k|} )^2} \nonumber \\
& = &
 \frac{{\rm Vol} \, {\rm Mass}^2 }{(2 \pi )^2 N}  \int_0^{+ \infty}   {\bf | k' |} d |{\bf k'}| 
\, \frac{1}{2  {\bf k}}  \, \log \frac{( \bf{|k'| + |k|} )^2}{( \bf{|k'| - |k|} )^2} \, .
 \ee
Therefore, the consistency condition (\ref{consi3}) turns into the following integral form, 
\be
&&
\boxed{ \xi_{\bf k} \,\text{csch}(\pi  \Omega_{\bf k} ) = 
{\rm \ell}  
  \int_0^{+ \infty}   {\bf | k' |} d |{\bf k'}| 
\, \frac{1}{2  | {\bf k} |}  \, \log \frac{( \bf{|k'| + |k|} )^2}{( \bf{|k'| - |k|} )^2} 
\, 
\frac{1}{2}
\,\text{sech}(\pi  \Omega_{\bf k'} )
\tanh
\left[ ( \pi  \Omega_{\bf k'} ) \coth ( \pi   \Omega_{\bf k'} ) 
   \right]
   }  \, , 
    \nonumber \\
   &&  \Omega_{\bf k} = \frac{\sqrt{ {\bf k}^2 + m^2}}{2 \pi T } \, , \quad \xi_{\bf k} = \sqrt{{\bf k}^2 + m^2} \, ,
   \quad 
 {\rm \ell } \equiv   \frac{{\rm Vol} \, {\rm Mass}^2  }{(2 \pi )^2 N} 
 =  \frac{{\rm Vol} \, {\rm Mass}^2 }{(2 \pi )^2 \, {\rm Vol}/\ell_{\rm P}^3 }
 = \frac{\ell_{\rm P}^3 \, {\rm Mass}^2 }{( 2 \pi )^2 }
 \, ,
\label{consi3Con}
\ee
where we have assumed $N = {\rm Vol}/\ell_{\rm P}^3$.
Since the energy $\xi_{\bf k} = |{\bf k}|$ (massless case) cannot exceed the mass $M$, and $T=1/8 \pi M$, we have to respect the following inequality,
\be
|{\bf k}| < M = \frac{1}{ 8 \pi T} \, .
\ee
In Fig.~\ref{consistencyC}, we make several plots of the right- and left-hand sides of the consistency condition (\ref{consi3Con}) assuming 
\be
|{\bf k}| = \frac{\lambda}{ 8 \pi T} \, , \quad \lambda < 1\, .
\ee
We are able to achieve a good agreement between the left- and right-hand sides of (\ref{consi3Con}) for:
\be
{\rm \ell } = 0.74 \, \ell_{\rm P}. 
\label{good0}
\ee
Hence, comparing (\ref{good0}) with (\ref{consi3Con}), we get:
\be
{\rm Mass}^2 = (2 \pi)^2 \frac{0.74}{\ell_{\rm P}^2} =  (2 \pi)^2 \, 0.74 \, M_{\rm P}^2 \, .
\label{Msq}
\ee

According to the numerical analysis and the above result, the interaction potential depends neither on the black hole mass $M$ nor on the energy of the matter particles. This is consistent with the condensed matter analogue, where the potential $V_0$ does not depend on the temperature. If the mass $M$ were present in the potential
(\ref{gravipot}), the temperature would also be present there because of the relation $T =1/8 \pi M$. In other words, an interaction potential independent of the temperature requires the absence of $M$ in (\ref{gravipot}). Finally, according to the numerical result (\ref{Msq}), the effective gravitational potential reads:
\be
V_{{\bf k} {\bf k'} } \approx  - \frac{ {5.4^2 \, {\rm M_{P}}^2 } }{ ( \bf{k' - k} )^2 } .
\label{gravipoteff}
\ee
The effective Hamiltonian is given by (\ref{EffHBCS}) with the replacement (\ref{gravipoteff}) for the potential $V_{\bf k k'}$.

Let us try to give an interpretation of the result. 
According to the potential obtained from the scattering amplitude in the Regge limit \cite{Graviballs} for two massless 
particles of energy $E_1$ and $E_2$, we get:
\be
V(r) = - \frac{ G_{\rm N}  E_1 E_2}{r} \, .
\label{regge}
\ee
Comparing (\ref{gravipoteff}) with (\ref{regge}), 
two particles moving in the black hole background interact with each other as if having approximately Planck mass, exchanging a massless particle (likely the graviton, according to (\ref{gravipoteff})),    
and behaving like a quantum liquid at the temperature $1/8 \pi M$. 
In analogy with a solid, we could say that the presence of a black hole heats up the spacetime and the matter within. 
At the temperature $1/8 \pi M$, the {\em atoms of space} (analogues of the ions in a crystal), having about Planck mass, interact gravitationally with matter, which is captured by such excited chunks of space. If we consider two particles and two chunks of space, each atom of space captures a particle and the two new atoms, made of matter and space, interact gravitationally with each other with, to a good approximation, Planck mass---actually about five times the Planck mass.
Hence, we get a Cooper pair made of an electron and a positron because of the mediation of the atoms of space as described above.
Conversely, if the black hole is not present, such a capture mechanism is not triggered.
The above interpretation equally follows by replacing the chunks of space with bound states of gravitons, or graviballs \cite{Graviballs}.

Regardless of the above interpretation,
we can safely conclude that the presence of a black hole turns the Minkowski vacuum into a superconducting condensate at high temperature. Indeed, contrary to superconductivity in a solid as described by BCS theory, in the presence of a black hole the mass gap function goes to zero at low temperature (large black hole mass) and grows with the temperature, quickly reaching the asymptotic value $2T$. The last stage of the evaporation process consists of a superconducting condensate at extremely high temperature. 
Conversely, when $a \rightarrow 0$ or $T\rightarrow 0$ ($M \rightarrow + \infty$), $\Delta_{\bf k} \rightarrow 0$ and the vacuum goes back to the Minkowski one, i.e.\ the theory describes free electrons and positrons in flat spacetime. When $a$ and $M$ take finite values, the interactions between electrons and positrons take place due to the coupling with gravity and the Minkowski vacuum turns into a BCS state, namely into a superconducting condensate in Minkowski spacetime. 
This is an extremely elegant and clean description of the evaporation mechanism, consistent with a unitary QFT at any stage of the evaporation.  
Indeed, in the Rindler metric we can claim unitarity for any value of the acceleration because the spacetime is singularity-free, 
but what about the last stage of the black hole evaporation process?
According to the paragraph above, the whole process seems to be well defined and unitary for any value of the mass $M$. However, the main difference between the Rindler and a black hole spacetime is the presence of a curvature singularity at $r=0$, i.e.\ the spacetime is geodesically incomplete. 
The pair production is well defined as long as the dynamics is well defined, namely as long as the collapsing matter has not reached the singular point; but afterwards the mechanism of pair creation breaks down. The effective BCS Hamiltonian $H_{\rm eff}$ can describe our QFT only for the matter that has not yet reached $r=0$. 
All the remaining matter that reaches the core of the black hole is gone, and we cannot infer anything about its future. Conversely, if there were no singularity, or if the time to reach $r=0$ were infinite, then the whole process would be under control, as in the Rindler case. In other words, the only real issue is the singularity.
In the next section, we will expand on the black hole singularity and show how it is harmless in Einstein's gravity, which is secretly conformally invariant. 
However, if we do not concern ourselves with the singularity problem, as in most of the literature of the last almost $60$ years, then we have a perfectly well-defined and unitary quantum field theory in curved spacetime. Moreover, the similarities with BCS theory provide a clear physical mechanism at the foundation of particle creation in curved spacetime, i.e.\ the concept of particle in a curved spacetime is no longer so mysterious, but actually very well defined because it can be traced back to Minkowski spacetime, where particle physics takes place, but in a BCS superconducting condensate. 

Only a slight forcing is needed in order to push the model to the very last stage of the evaporation: 
we can replace the Hawking temperature with a proposal similar to (\ref{Temperatures2}). In this way, QFT is unitary up to the last stage of the evaporation process, regardless of the singularity issue. 
In other words, the model described by the effective Hamiltonian is universal and insensitive to the presence of the singularity at $r=0$. Indeed, the Hawking temperature is singular only for $M=0$, but regular at any other stage of the evaporation with $M>0$, regardless of the amount of mass that has already ended up in the singularity.

For the bosonic case, the analogy is very strict with BEC superfluids \cite{Guo_2017} or the He~II theory of superfluidity \cite{Superfluidity}. The reader can simply compare (\ref{Sq1}) with formula (II.25) in \cite{Superfluidity}, or formula (45) (and the text below) in \cite{Guo_2017}.

In the following table, we summarize the similarities and the differences between a superconductor and QFT in the black hole background.
\begin{center}
    \begin{tabular}{ | l | l | l | }
    \hline
     & Superconductor  & Black Hole   \\ \hline
    State  & 
  $  
  | \Psi_{\rm BCS} \rangle  = 
 S^{\rm BCS}_{\rm sq}
 | 0 \rangle 
 =   \prod_{\bf k} \left( u_{\bf k} + v_{\bf k} \, 
 c^\dagger_{{\bf k} \uparrow} 
c^\dagger_{ - {\bf k} \downarrow} 
\right)
| 0 \rangle 
$
     & 
     $  | \Psi_{\rm H} \rangle = 
  S^{\rm H}_{\rm sq} \,|0 \rangle 
=  \prod_{\bf k} \left(
u_{\bf k}  + v_{\bf k} \, c^{ ({\rm out}) \dagger}_{ \bf k} \, d^{ ({\rm int}) \dagger}_{ - {\bf k}}
\right)  | 0 \rangle 
 $ 
      \\ \hline
 Vacuum      & 
 $| 0 \rangle  \equiv |0_{ {\bf k} \uparrow }\rangle  |0_{- {\bf k} \downarrow}\rangle \equiv | {\rm out } \rangle$
  and $| {\rm in} \rangle \equiv | \Psi_{\rm BCS} \rangle$
  & $ | 0 \rangle  \equiv |0_{\bf k}\rangle^+_{\rm out }\,|0_{-{\bf k}}\rangle^-_{\rm int } \equiv | {\rm out} \rangle$ and 
  $| {\rm in} \rangle \equiv | \Psi_{\rm H} \rangle$
      \\ \hline 
    Bogoliubov coef.   & $u_{{ \bf k} }, v_{{ \bf k} }$ & 
    $u_{{ \bf k} } = \cos r_{\bf k}$, $v_{{ \bf k} } = \sin r_{\bf k}$ and $\tan r_{\bf k} = {\rm e}^{- \frac{\xi_{\bf k}}{2T}}$
      \\ \hline
   Pairing & 
  $  c^\dagger_{{\bf k} \uparrow} c^\dagger_{ - {\bf k} \downarrow}$
    & 
     $c^{ ({\rm out}) \dagger}_{ \bf k}\,d^{ ({\rm int}) \dagger}_{-{\bf k}}$
     \\ \hline
     Bogoliubons & $\gamma_{ {\bf k} s}$ s.t.\  $\gamma_{ {\bf k} s}| \Psi_{\rm BCS} \rangle = 0$ & 
     $a_{\bf k}^{(\rm in)}, b_{- {\bf k}}^{(\rm in)}$ s.t.\ 
   $a_{\bf k}^{(\rm in)} | \Psi_{\rm H} \rangle = 0$, $b_{- {\bf k}}^{(\rm in)} | \Psi_{\rm H} \rangle = 0$
     \\ \hline
    Order parameter  & $\Delta_{\bf k} (T)$  & 
    $ \Delta_{\bf k}(T)  =  \xi_{\bf k}   \,\text{csch} \left( \frac{\xi_{\bf k}}{2 T} \right)$  \\
    \hline
  Condensate   &  $\langle c^\dagger_{{\bf k} \uparrow} 
c^\dagger_{ - {\bf k} \downarrow}  \rangle_{\rm BCS} \neq 0$ &
$\langle c^{ ({\rm out}) \dagger}_{ \bf k} \, d^{ ({\rm int}) \dagger}_{ - {\bf k}} \rangle_{\rm H} \neq 0$
    \\
    \hline
   Interaction  & Electromagnetic & Gravitational
    \\
    \hline
  $U(1)$ symmetry   & Broken & Unbroken 
    \\
    \hline
    \end{tabular}
\end{center}
Notice that the BCS state, or the H-state, is the vacuum for the Bogoliubons for all ${\bf k}$, i.e.\ we have introduced the product over ${\bf k}$ according to (\ref{vacM2BCS}). 
For $T \rightarrow 0$,
\be
\lim_{T \rightarrow 0} \tan r_{\bf k} = \lim_{T \rightarrow 0} {\rm e}^{- \frac{\xi_{\bf k}}{2T}} = 0 
\quad \Longrightarrow 
\quad r_{\bf k} \rightarrow 0 \quad \Longrightarrow 
\quad u_{\bf k} \rightarrow 1 \, , \quad  v_{\bf k} \rightarrow 0  \quad \Longrightarrow \quad |\Psi_{\rm H} \rangle = | 0 \rangle . 
\ee

\begin{figure}
	\begin{center}
		\includegraphics[height=7cm]{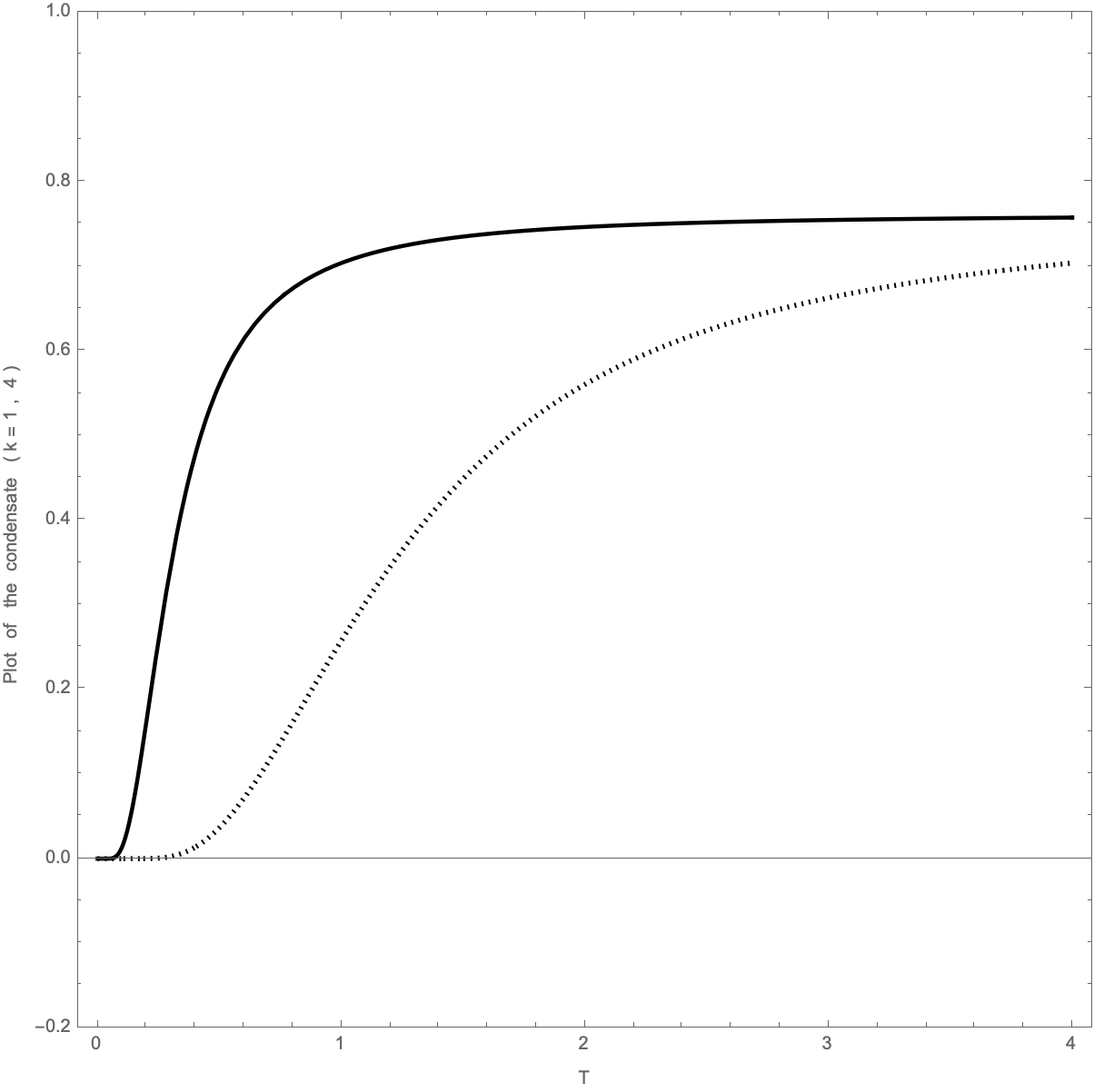}
		\hspace{1cm} 
		\includegraphics[height=7cm]{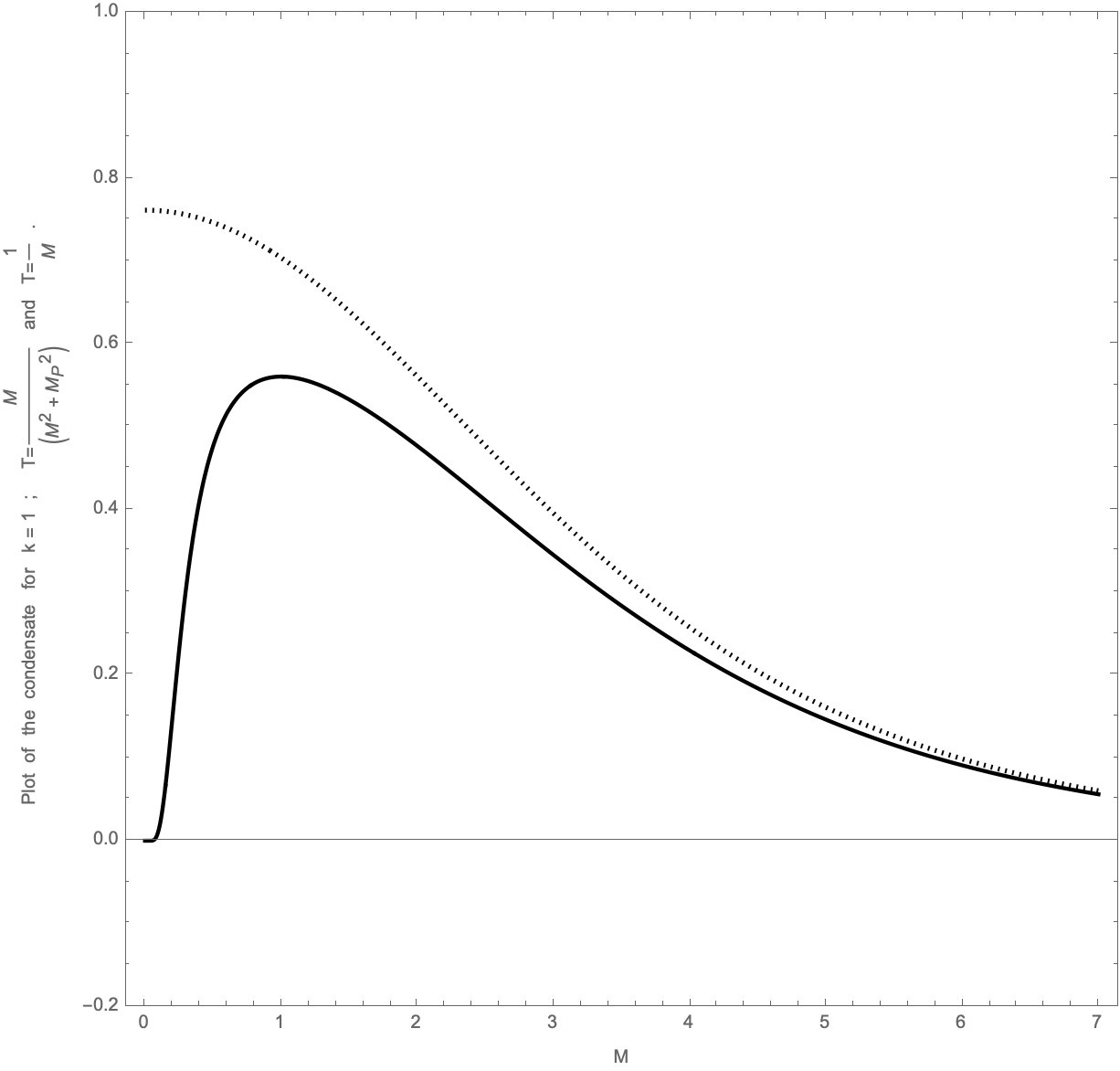}
		%
		\caption{Left panel: plot of twice the condensate according to (\ref{condensateF}) as a function of the temperature for $k=1,4$. Right panel:
		plot of the condensate as a function of the black hole mass $M$ for $k=1$. We have considered the Hawking temperature $1/M$ (dashed line) and the modified temperature $M/(M^2+M_{\rm p}^2)$ (solid line). 
}
		\label{condensate}
	\end{center}
\end{figure}

Comparing (\ref{consi1}) with (\ref{consi3}) and taking the Hermitian conjugate of 
$\big\langle c_{ {\bf - k} \downarrow } \, c_{ {\bf k} \uparrow} \big\rangle$, namely
$\big\langle c^\dagger_{ {\bf k} \downarrow } \, c^\dagger_{ {\bf - k} \uparrow} \big\rangle$, 
the condensate reads:
\be
2 \, \big\langle c^{ ({\rm out}) \dagger}_{ \bf k}\,d^{ ({\rm int}) \dagger}_{-{\bf k}} \big\rangle = \text{sech}(\pi  \Omega_{\bf k} )
\tanh
 \left[ ( \pi  \Omega_{\bf k} ) \coth ( \pi   \Omega_{\bf k} ) 
   \right] ,
   \label{condensateF}
\ee
whose plot is given in Fig.~\ref{condensate}. 

\section{Entanglement Entropy} 
Last but not least, in this section we evaluate the entanglement entropy of the radiated particle state by tracing over the int-particles that cross the horizon. In Rindler spacetime the very same result consists in tracing over the particles in the causally disconnected wedge. 
The reader can find all the steps of the derivation in \cite{Guo_2017}. 

Since the BCS or Hawking state cannot be factorized into a product of single-mode states, it possesses an entanglement between the out and int quanta. 
In order to measure the entanglement of a bipartite system with components A and B, we have to introduce a generalization of the von Neumann entropy of a subsystem. 
Given a general state $\Psi$, the associated density matrix is:
\be
\rho = | \Psi \rangle \langle \Psi | \, ,
\ee
and, taking the partial trace over the basis of subsystem B, we obtain the reduced density matrix 
\be
\rho_{\rm A} = {\rm Tr}_{\rm B} ( \rho) \, .
\ee
The von Neumann entropy of subsystem A provides the entanglement entropy, 
\be
S_{\rm EE} = - {\rm Tr} \left( \rho_{\rm A} \ln \rho_{\rm A} \right) \, .
\ee
In our model ${\rm A}$ stands for ${\rm out}$ and ${\rm B}$ corresponds to ${\rm int}$. 

Therefore, according to the derivation in \cite{Guo_2017}, it can be simply expressed in terms of the Bogoliubov coefficients $u_{\bf k}$ and $v_{\bf k}$. In the \underline{fermionic} case, using (\ref{cossin}) and the first relation in (\ref{tanrk}), i.e.\ $\tan r_{\bf k} = \exp[- \pi \Omega_{\bf k}]$, the entanglement entropy reads:
\be
S^{(\rm F)}_{\rm EE}  & = & - \sum_{\bf k} \left(   | u_{\bf k}|^2 \ln | u_{\bf k}|^2 +  | v_{\bf k}|^2 \ln | v_{\bf k}|^2 
\right)
\\
& = & - \sum_{\bf k} \left(   | \cos r_{\bf k}|^2 \ln | \cos r_{\bf k}|^2 +  | \sin r_{\bf k}|^2 \ln | \sin r_{\bf k}|^2 
\right)
\\
& = &  \sum_{\bf k} \left[ \log \left( 1+ e^{2 \pi  \Omega_{\bf k} }\right)-\frac{2 \pi  \Omega_{\bf k} }{e^{-2 \pi  \Omega_{\bf k} }+1}
\right] .
\label{EEFermi}
\ee
Instead of evaluating the sum over ${\bf k}$, we replace the sum with an integral over the same variable, 
\be
S^{\rm (F)}_{\rm EE}   =   {\rm Vol} \int \frac{d^3 {\bf k} }{(2 \pi)^3} \left[ \log \left( 1+  e^{2 \pi  \Omega_{\bf k} } \right)-\frac{2 \pi  \Omega_{\bf k} }{e^{-2 \pi  \Omega_{\bf k} }+1} 
\right] \, .
\label{SEEint}
\ee
For massless fermions $\Omega_{\bf k} = |{\bf k}|/\kappa_{+}$ (see (\ref{tanrk})), thus we can evaluate (\ref{SEEint}) analytically. The result reads:
\be
S^{\rm (F)}_{\rm EE}   =  \frac{7 \, {\rm Vol} }{180} \pi ^2 \kappa_+^3 \, .
\ee

\begin{figure}
	\begin{center}
		\includegraphics[height=6.9cm]{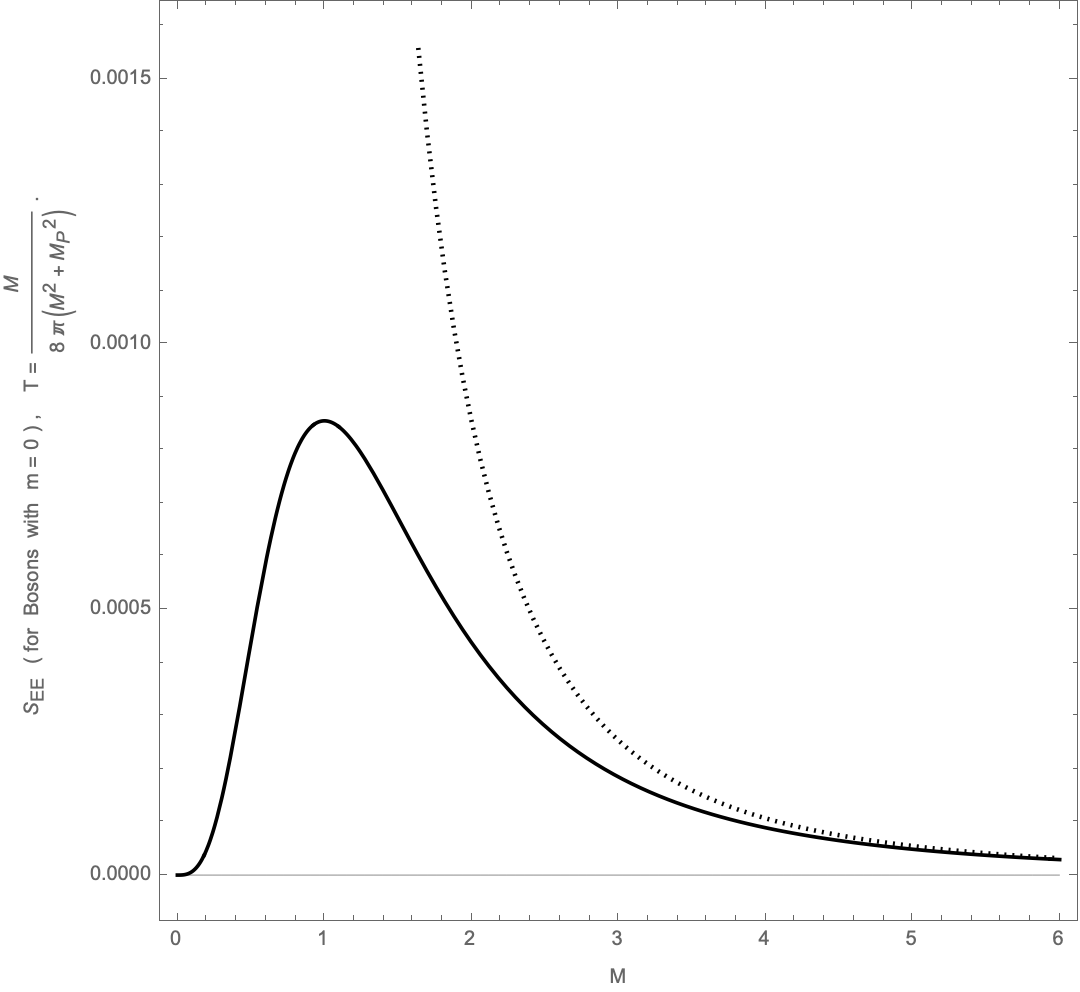}
		\hspace{1cm} 
		\includegraphics[height=7cm]{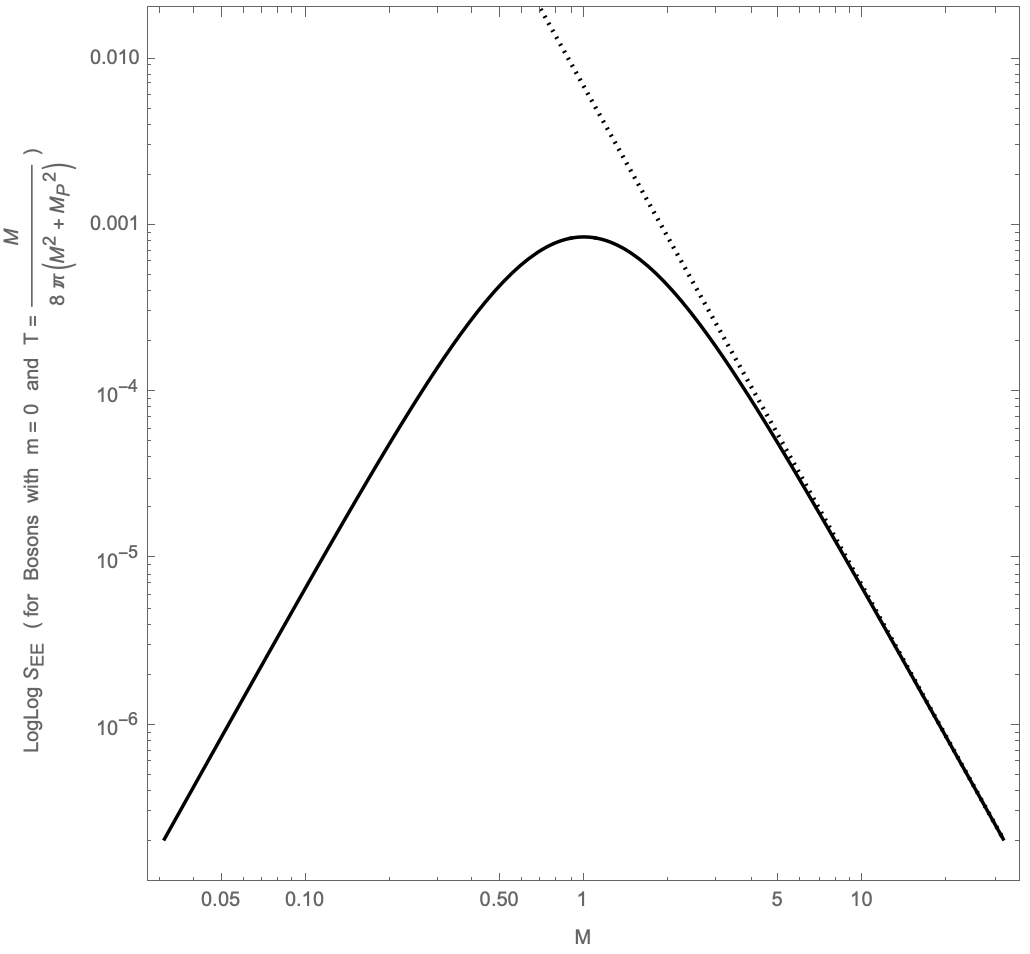}
		\caption{Left panel: plot of the entanglement entropy (\ref{SEEintBose}) for a massless scalar in the black hole background as a function of the modified temperature (solid line) or the Hawking temperature (dotted line).
		A very similar plot is obtained in the fermionic case.  
}
		\label{EEBF}
	\end{center}
\end{figure}

Similarly, in the \underline{bosonic} case, 
\be
S^{\rm (B)}_{\rm EE}  & = &  \sum_{\bf k} \left(   | u_{\bf k}|^2 \ln | u_{\bf k}|^2 -  | v_{\bf k}|^2 \ln | v_{\bf k}|^2 
\right)
\\
& = & \sum_{\bf k} \left( | \cosh r_{\bf k}|^2 \ln | \cosh r_{\bf k}|^2 +  | \sinh r_{\bf k}|^2 \ln | \sinh r_{\bf k}|^2 
\right)
\\
& = &  \sum_{\bf k} \left[ -  \log \left( 1- e^{ - 2 \pi  \Omega_{\bf k} } \right) + \frac{2 \pi  \Omega_{\bf k} }{e^{2 \pi  \Omega_{\bf k} }-1} 
\right]  . 
\label{EEBose}
\ee
Performing the integral in place of the sum, we get:
\be
S^{\rm (B)}_{\rm EE}   =   {\rm Vol} \int \frac{d^3 {\bf k} }{(2 \pi)^3} \left[  -  \log \left( 1- e^{ - 2 \pi  \Omega_{\bf k} } \right) + \frac{2 \pi  \Omega_{\bf k} }{e^{2 \pi  \Omega_{\bf k} }-1} 
\right] = \frac{2 \,  {\rm Vol} }{45} \pi ^2 \kappa_+^3 
\, .
\label{SEEintBose}
\ee
In Fig.~\ref{EEBF} we provide a plot of the entanglement entropy for bosons (left panel) and for fermions (right panel) as a function of the black hole mass $M$ and constant volume $V$. The dashed line corresponds to the Hawking relation between mass and temperature, while the solid line corresponds to the modified temperature (\ref{TemperatureMod}).

\section{The Area Law} 
\label{sec:area}

In this subsection we derive the area law $S \propto A$ from the full entanglement entropy calculation when the local Hawking temperature and the proper volume element are correctly identified, and the integration is performed over the \emph{entire} space outside the event horizon \cite{Bombelli:1986rw}. 

We work here with the Schwarzschild metric in Planck units: 
\begin{equation}
ds^2 = -\left(1-\frac{2M}{r}\right)dt^2 + \frac{dr^2}{1-\frac{2M}{r}} + r^2 d\Omega^2 \, , 
\end{equation}
while the Hawking temperature at infinity and the local blueshifted temperature for a general value of the radial coordinate are:
\begin{equation}
T_{\rm H} = \frac{1}{8\pi M}, \qquad T_{\rm loc}(r) = \frac{T_{\rm H}}{\sqrt{1-\frac{2M}{r}}}.
\label{Tempr}
\end{equation}
For a scalar field, the entanglement entropy has been derived in the previous section and reads:
\begin{equation}
S^{\rm (B)}_{\rm EE} = \frac{2 \, {\rm Vol}}{45} \pi^2 \kappa_+^3 \, , \qquad \kappa_+ = 2\pi T \, .
\label{SEEglobal}
\end{equation}
Therefore, the entropy density is:
\begin{equation}
s(r) = \frac{2}{45} \pi^2 (2\pi T_{\rm loc}(r))^3 = \frac{16\pi^5}{45} \, T_{\rm loc}^3(r) \, .
\label{eden}
\end{equation}
Substituting (\ref{Tempr}) into (\ref{eden}), we get:
\begin{equation}
s(r) = \frac{16\pi^5}{45} \frac{T_{\rm H}^3}{\left(1-\frac{2M}{r}\right)^{3/2}} = \frac{1}{1440\pi^2 M^3} \cdot \frac{1}{\left(1-\frac{2M}{r}\right)^{3/2}} \, .
\end{equation}
The other main quantity is the physical volume element of a spherical shell, i.e.\ 
\begin{equation}
d{\rm V}(r) = \frac{4\pi r^2}{\sqrt{1-\frac{2M}{r}}} \, dr \, .
\label{propVol}
\end{equation}
Therefore, the differential entropy of a shell at radius $r$ is:
\begin{equation}
dS = s(r) \, d {\rm V}_{\rm prop}(r) = \frac{1}{360\pi M^3} \cdot \frac{r^2}{\left(1-\frac{2M}{r}\right)^2} \, dr \, .
\end{equation}
Changing the radial coordinate to the dimensionless variable $x = r/(2M)$ ($dr = 2M\,dx$), 
\begin{equation}
\boxed{dS = \frac{1}{45\pi} \cdot \frac{x^4}{(x-1)^2} \, dx \, , }
\end{equation}
which shows how the entropy is distributed in coordinate space. 

Integrating over all space from the horizon ($x=1+\delta$, where $\delta = \delta r/(2M)$) to a large radius $X = R/(2M)$, we get:
\begin{equation}
S = \frac{1}{45\pi} \int_{1+\delta}^{X} \frac{x^4}{(x-1)^2} \, dx \, .
\end{equation}
The exact integral can be evaluated using the partial fraction decomposition, 
\begin{equation}
\int \frac{x^4}{(x-1)^2} dx = \frac{x^3}{3} + x^2 + 3x + 4\ln|x-1| - \frac{1}{x-1}.
\label{Integrale}
\end{equation}
Evaluating the above expression between $1+\delta$ and $X$, and expanding the lower limit for $\delta \ll 1$, the dominant terms contributing to the entropy are:
\begin{equation}
S = \frac{1}{45\pi} \left[ \frac{1}{\delta} + 4\ln\delta + \frac{13}{3} + \text{(terms in }X\text{)} + O(\delta) \right] \, , 
\label{Ediv}
\end{equation}
which is divergent at the horizon, where $\delta = 0$.
In order to regularize the result, we introduce a physical cut-off defined as follows. 
We evaluate the proper distance $\alpha$ from the horizon to $r = 2M + \delta r$, i.e.\ 
\begin{equation}
\alpha = \int_{2M}^{2M+\delta r} \frac{dr}{\sqrt{1-\frac{2M}{r}}} \approx 4M\sqrt{2\delta} \quad \Longrightarrow \quad
\delta = \frac{\alpha^2}{32 M^2} \, .
\label{deltaCO}
\end{equation}
Now substituting $\delta$ back into (\ref{Ediv}), using the area in terms of the mass, $A = 16\pi M^2$, and choosing 
the cut-off to be proportional to the Planck length, $\alpha^2 = \ell_{\rm p}^2/720$, 
\begin{equation}
\boxed{S = \frac{1}{45\pi} \cdot \frac{2A}{\pi\alpha^2} - \frac{4}{45\pi} \ln\left(\frac{2A}{\pi\alpha^2}\right) + O(1)
= 
 \frac{A}{4 \ell_{\rm p}^2} - \frac{4}{45\pi} \ln\left(\frac{A}{4 \ell_{\rm p}^2}\right) + O(1)
} \, .
\end{equation}

\subsection{Area law with a smooth Compton wavelength cutoff}
\label{sec:area_compton}
In this subsection we incorporate the natural cutoff provided by the Compton wavelength of the emitted quanta \ref{Temperatures}. The physical condition is that a black hole can exist only if its Schwarzschild radius is at least of the order of the Compton wavelength of the quanta it emits. When the Compton wavelength exceeds the Schwarzschild radius, the notion of a thermal black hole ceases to be well-defined, and the effective temperature must vanish.
For the sake of simplicity, we will use the smooth temperature (\ref{TemperatureMod}) in place of the first expression in (\ref{Temperatures}), namely 
\begin{equation}
\boxed{
T_{\rm eff}(M) = \frac{1}{8\pi M} \cdot \frac{1}{1 + \frac{1}{2M^2}} , 
}
\label{eq:Teff}
\end{equation}
which coincides with (\ref{TemperatureMod}) for $M_{\rm p} = 1/2$. 

The function (\ref{eq:Teff}) has the following properties:
\begin{itemize}
    \item For $M \gg 1$: $T_{\rm eff} \approx \frac{1}{8\pi M} = T_{\rm H}$ (Hawking temperature).
    \item For $M = 1/\sqrt{2}$: $T_{\rm eff} = \frac{\sqrt{2}}{16\pi}$ (maximum).
    \item For $M \to 0$: $T_{\rm eff} \approx \frac{M}{4\pi} \to 0$.
\end{itemize}

The maximum occurs at $M_{\rm max} = 1/\sqrt{2}$, with value $T_{\rm max} = \sqrt{2}/16\pi$.
Now the entropy density reads: 
\begin{equation}
s(r) = \frac{16\pi^5}{45} \frac{T_{\rm eff}^3(M)}{\left(1 - \frac{2M}{r}\right)^{3/2}}.
\end{equation}
Taking into account the blueshifted temperature (\ref{Tempr}) and the proper volume element (\ref{propVol}),  
the differential entropy of a shell at radius $r$ is:
\begin{equation}
dS = s(r) \, dV_{\rm prop}(r)
= \frac{64\pi^6}{45} T_{\rm eff}^3(M) \frac{r^2}{\left(1 - \frac{2M}{r}\right)^2} \, dr.
\end{equation}
Changing again to the dimensionless variable $x = r/(2M)$ ($dr = 2M\,dx$):
\begin{equation}
dS = \frac{512\pi^6}{45} M^3 T_{\rm eff}^3(M) \frac{x^4}{(x-1)^2} \, dx.
\end{equation}
Using (\ref{eq:Teff}), we have the following identity, 
\begin{equation}
M^3 T_{\rm eff}^3(M)
= \frac{1}{512\pi^3} \frac{1}{\left(1 + \frac{1}{2M^2}\right)^3} \, , 
\end{equation}
which, substituted into the differential entropy, gives:
\begin{equation}
dS = \frac{1}{45\pi} \cdot \frac{1}{\left(1 + \frac{1}{2M^2}\right)^3} \cdot \frac{x^4}{(x-1)^2} \, dx.
\label{eq:dS_compton}
\end{equation}
Integrating from the horizon ($x=1+\delta$, where $\delta = \delta r/(2M)$) to a large radius $X = R/(2M)$:
\begin{equation}
S = \frac{1}{45\pi} \cdot \frac{1}{\left(1 + \frac{1}{2M^2}\right)^3}
\int_{1+\delta}^{X} \frac{x^4}{(x-1)^2} \, dx.
\end{equation}
Evaluated between $1+\delta$ and $X$, and expanding for $\delta \ll 1$, the leading terms are:
\begin{equation}
S = \frac{1}{45\pi} \cdot \frac{1}{\left(1 + \frac{1}{2M^2}\right)^3}
\left[
\frac{1}{\delta} + 4\ln\delta + \frac{13}{3} + \text{(terms in }X\text{)} + O(\delta)
\right].
\end{equation}
According to (\ref{deltaCO}) and using $A = 16\pi M^2$, we get:
\begin{equation}
\frac{1}{\delta} = \frac{32M^2}{\alpha^2} = \frac{2A}{\pi\alpha^2} \, .
\end{equation}
Therefore, 
\be
\ln\delta = \ln\left(\frac{\alpha^2}{32M^2}\right) = \ln\left(\frac{\pi \alpha^2}{2A}\right)
= -\ln\left(\frac{2A}{\pi \alpha^2}\right) 
\quad \text{and} \quad 
1 + \frac{1}{2M^2} = 1 + \frac{8\pi}{A}.
\ee
The final expression for the entanglement entropy is:
\begin{equation}
S(A) = \frac{1}{45\pi} \cdot \frac{1}{\left(1 + \frac{8\pi}{A}\right)^3}
\left[
\frac{2A}{\pi\alpha^2} - 4\ln\left(\frac{2A}{\pi\alpha^2}\right) + \frac{13}{3} + \mathcal{O}(X)
\right] , 
\label{eq:S_final}
\end{equation}
where the terms depending on the large-distance cutoff $X$ are:
\begin{equation}
\mathcal{O}(X) = \frac{X^3}{3} + X^2 + 3X + 4\ln X,
\end{equation}
which are finite and do not affect the divergent structure near the horizon.
In order to recover the Bekenstein-Hawking entropy $S_{\rm BH} = A/(4\ell_{\rm P}^2)$ in the large-area limit, we fix the proper-distance cutoff $\alpha$ as:
\begin{equation}
\alpha^2 = \frac{1}{720} \, \ell_{\rm P}^2.
\label{eq:alpha_cutoff}
\end{equation}
This choice is natural because it relates the cutoff to the Planck scale through the numerical coefficient that comes from the field content (here a single massless scalar field).
Moreover, restoring the Planck length in the first round bracket that comes from the temperature, we get:
\begin{equation}
\boxed{
S(A) = \frac{1}{\left(1 + \frac{8\pi \ell_{\rm p}^2 }{A}\right)^3}
\left[
\frac{A}{4 \ell_{\rm p}^2} - \frac{4}{45\pi} \ln\left(\frac{A}{4 \ell_{\rm p}^ 2}\right) + {\rm const.} + \mathcal{O}(X)
\right] .
}
\label{eq:S_finalfinal}
\end{equation}

For a large black hole ($A \gg \ell_{\rm P}^2$), $\left(1 + \frac{8\pi \ell_{\rm P}^2}{A}\right)^3 \approx 1$, 
and the entropy reduces to: 
\begin{equation}
S(A) = \frac{A}{4\ell_{\rm P}^2} - \frac{4}{45\pi} \ln\left(\frac{A}{\ell_{\rm P}^2}\right) + O(1).
\label{eq:area_law}
\end{equation}
This matches exactly the Bekenstein-Hawking entropy as the leading term, with a logarithmic correction.

For sub-Planckian black holes ($A \to 0$), $\left(1 + \frac{8\pi \ell_{\rm P}^2}{A}\right)^3 \approx \frac{512\pi^3 \ell_{\rm P}^6}{A^3}$, and 
\begin{equation}
S(A) \to 0 \quad \text{as} \quad A \to 0.
\label{eq:S_zero}
\end{equation}

\subsection{Short discussion of the result}
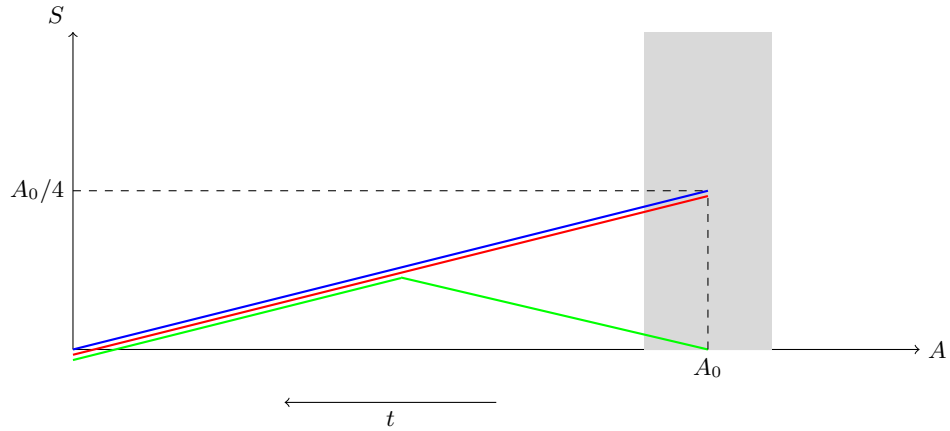
\begin{figure}[htbp]
\centering
\begin{tikzpicture}[scale=1.4]
\draw[->] (0,0) -- (8,0) node[right] {$A$};
\draw[->] (0,0) -- (0,3) node[above left] {$S$};

\fill[gray!30] (5.4,0) rectangle (6,3);

\fill[gray!30] (6,0) rectangle (6.6,3);

\draw[blue, thick] (0,0) -- (6,1.5);

\draw[red, thick] (0,-0.05) -- (6,1.45);

\draw[green, thick] (6,0) -- (3.1,0.68);

\draw[green, thick] (0,-0.10) -- (3.1,0.675);

\draw[black, dashed] (6,0) -- (6,1.5);

\draw[black, dashed] (0,1.5) -- (6,1.5);

\node[left] at (0,1.5) {$A_0/4$};

\node[below] at (6,0) {$A_0$};

\draw[->] (4,-0.5) -- (2,-0.5);
\node[below] at (3,-0.5) {$t$};
\end{tikzpicture}
\caption{Diagram of the entropy with time direction. The blue solid line indicates the Bekenstein-Hawking entropy $A/4$, while the red solid line stands for the entanglement entropy. Clearly they overlap perfectly, but the issue is at the formation of the black hole, when the area is $A_0$. Indeed, at the beginning the temperature is very low and one would expect the entanglement entropy to be zero or very small. However, according to our derivation it is exactly equal to $A_0/4$. What one would expect would be more similar to the green solid line. }
\label{SBHEE}
\end{figure}

In Fig.~\ref{SBHEE} we compare the Bekenstein-Hawking entropy and the entanglement entropy. They perfectly overlap at every stage of the evaporation process. However, one would expect the entanglement entropy to be very small for $A=A_0$, consistently with the very low temperature. In order to solve this minor issue we need a dynamical model in which we include the back-reaction, i.e.\ we need a microscopic model of the black hole matter as well. In other words, quantum field theory in curved spacetime is defined for a fixed value of the black hole mass, without taking into account the dynamical mechanism by which the matter that constitutes the black hole is converted into particles that escape towards infinity.
We expect that a proper interaction Hamiltonian will solve this issue. However, this program is beyond the scope of this paper, which is already rich in new ideas and results, and 
will be investigated in a subsequent work focusing on a dynamical toy model describing the entire evaporation process. 

\section{The singularity issue and black hole evaporation} 

Contrary to the previous sections, this one is more speculative, but useful for understanding various aspects that will likely be developed in future projects.

As mentioned in Section~(\ref{BCSBEC}), quantum field theory in curved spacetime and the associated black hole evaporation could break down only when the int-particles and/or the collapsing matter reach the timelike singularity located at $r=0$. From that moment on we would no longer be able to predict the future and we would lose our predictive power. In order to solve this problem, following the previous literature \cite{Bambi:2016wdn, Bambi:2016yne}, we can invoke a symmetry principle, namely Weyl conformal invariance \cite{narlikar:1977nf}. Let us briefly recap the main idea. It has been known for a long time that Einstein's gravity is actually a scalar-tensor theory invariant under Weyl conformal transformations, but in the broken phase of the conformal symmetry \cite{Bambi:2016wdn, Kubo:2022jwu}. According to this symmetry, singular and singularity-free solutions are actually gauge equivalent, meaning that geodesically complete and incomplete spacetimes are related by a Weyl conformal rescaling. 
In practice, we can make the Schwarzschild metric, and any other spacetime, singularity-free by a conformal rescaling: all curvature invariants are regular, and, most importantly, 
the spacetime is geodesically complete \cite{Bambi:2016wdn, Bambi:2016yne}. 
Regarding this latter crucial property, nothing can reach the core of the Schwarzschild spacetime at $r=0$ in a finite amount of time. Therefore, the collapse time is actually infinite for both the collapsing black hole particles and the created int-particles, which will never reach the core at $r=0$. Hence, the Hawking process can be very schematically described as follows.

%
%

We investigate the black hole evaporation process 
by simply considering energy conservation, and without requiring any nonlocal interaction. 
Let us consider a simple toy model consisting of $N$ massive particles (constituting the black hole of mass $M$) turning into $N^\prime$ massive particles and one Hawking pair. Regarding the latter two particles, one escapes towards infinity while the other crosses the horizon, reuniting with the black hole.
The process reads:
\be
\underbrace{N \omega_\bfk}_{M^{\rm initial}_{\rm BH}} = N^\prime \omega_{\bfk^{\prime}} +\omega_0 + \omega_{\infty} 
\quad 
\Longrightarrow 
\quad 
\underbrace{N^\prime \omega_{\bfk^{\prime}} + \omega_0}_{M_{\rm BH}^{\rm final}} 
= N \omega_\bfk -  \omega_{\infty} < M_{\rm BH}^{\rm initial} \, .
\label{evapF}
\ee
The particle of energy $\omega_0$ contributes again to the mass of the black hole after the first stage of the evaporation process because it falls back into the black hole. In other words, the black hole takes back one of the two particles it created. 
Therefore, the mass of the black hole after the creation of a Hawking pair is smaller than the initial one, as observed from infinity.

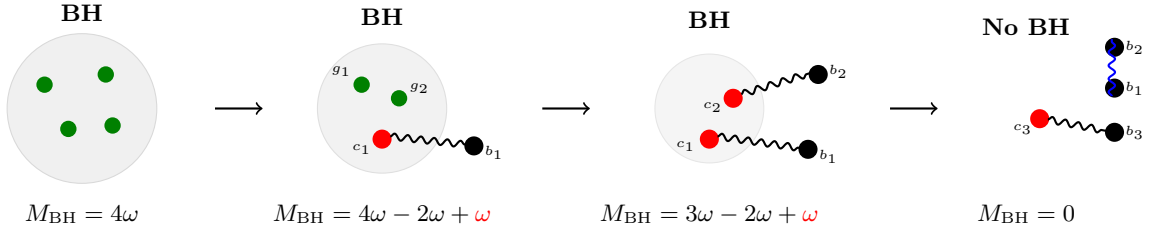
\begin{figure}[h]
\centering
\hspace{-2.5cm} 
\begin{tikzpicture}[scale=0.9, every node/.style={font=\small}]

\draw[gray!30, fill=gray!12] (0,0) circle (1.1);
\node at (0,1.4) {\textbf{BH}};
\node at (0,-1.55) {$M_{\rm BH}  = 4\omega$};
\fill[green!50!black] (-0.55,0.35) circle (0.12);
\fill[green!50!black] (0.35,0.5) circle (0.12);
\fill[green!50!black] (-0.2,-0.3) circle (0.12);
\fill[green!50!black] (0.45,-0.25) circle (0.12);

\hspace{0.5cm} 
\draw[->, thick] (1.4,0) -- (2.1,0);
\hspace{0.5cm} 

\draw[gray!25, fill=gray!10] (3.3,0) circle (0.95);
\node at (3.3,1.35) {\textbf{BH}};
\node at (3.3,-1.55) {$M_{\rm BH} =  4 \omega - 2 \omega + {\color{red}\omega}
$};
\fill[green!50!black] (3.0,0.35) circle (0.12);
\fill[green!50!black] (3.55,0.15) circle (0.12);
\node[font=\tiny] at (2.7,0.55) {$g_1$};
\node[font=\tiny] at (3.85,0.3) {$g_2$};
\fill[red] (3.3,-0.45) circle (0.14);
\node[font=\tiny] at (3.0,-0.6) {$c_1$};
\fill[black] (4.65,-0.55) circle (0.14);
\node[font=\tiny] at (4.95,-0.65) {$b_1$};
\draw[decorate, decoration={snake, amplitude=1.2pt, segment length=5pt}, thick] 
    ($(3.3,-0.45)+(0.12,0.04)$) -- ($(4.65,-0.55)+(-0.12,0.02)$);

\hspace{0.5cm} 
\draw[->, thick] (5.1,0) -- (5.8,0);
\hspace{0.5cm} 

\draw[gray!20, fill=gray!8] (7.0,0) circle (0.8);
\node at (7.0,1.3) {\textbf{BH}};
\node at (7.0,-1.55) 
{$M_{\rm BH} = 3 \omega - 2 \omega + {\color{red}\omega}$};
\fill[red] (7.0,-0.45) circle (0.14);
\fill[red] (7.35,0.15) circle (0.14);
\node[font=\tiny] at (6.65,-0.55) {$c_1$};
\node[font=\tiny] at (7.05,0.02) {$c_2$};
\fill[black] (8.45,-0.6) circle (0.14);
\fill[black] (8.6,0.5) circle (0.14);
\node[font=\tiny] at (8.75,-0.7) {$b_1$};
\node[font=\tiny] at (8.9,0.55) {$b_2$};
\draw[decorate, decoration={snake, amplitude=1.2pt, segment length=5pt}, thick] 
    ($(7.0,-0.45)+(0.12,0.04)$) -- ($(8.45,-0.6)+(-0.12,0.02)$);
\draw[decorate, decoration={snake, amplitude=1.2pt, segment length=5pt}, thick] 
    ($(7.35,0.15)+(0.12,0.06)$) -- ($(8.6,0.5)+(-0.12,0.02)$);

\hspace{0.5cm} 
 \draw[->, thick] (9.1,0) -- (9.8,0);

\node at (11.1,1.2) {\textbf{No BH}};
\node at (11.1,-1.55) {$M_{\rm BH} = 0
$};
\fill[red] (11.3,-0.15) circle (0.14);
\node[font=\tiny] at (11.05,-0.3) {$c_3$};
\fill[black] (12.4,0.9) circle (0.14);
\fill[black] (12.4,0.3) circle (0.14);
\fill[black] (12.4,-0.35) circle (0.14);
\node[font=\tiny] at (12.7,0.9) {$b_2$};
\node[font=\tiny] at (12.7,0.3) {$b_1$};
\node[font=\tiny] at (12.7,-0.35) {$b_3$};
\draw[decorate, decoration={snake, amplitude=1.2pt, segment length=5pt}, thick] 
    ($(11.3,-0.22)+(0.12,0.06)$) -- ($(12.4,-0.35)+(-0.12,0.02)$);
\draw[decorate, decoration={snake, amplitude=1.2pt, segment length=5pt}, thick, blue] 
    ($(12.4,0.3)+(-0.04,-0.12)$) -- ($(12.4,0.9)+(-0.04,0.12)$);

\end{tikzpicture}
\caption{
A cartoon of particle production based on energy conservation. When a black hole, here simply made of four green particles, creates a Hawking pair, its mass decreases by two units, but right afterwards the Hawking int-particle falls into the event horizon and the total mass of the black hole, at the end of the first stage of the evaporation, consists of three particles. The process repeats until the black hole consists of only two green particles (Page time). At that stage, the creation of another Hawking pair makes the black hole evaporate completely because of energy conservation. Notice that after full evaporation the created red int-particle travels towards $r=0$, but since there is no horizon anymore, it will move towards infinity in the antipodal direction with respect to the Hawking out-particle. 
For the sake of simplicity, we have assumed here that all the particles involved have the same energy.
Notice that we have not taken into account the relation between energy and temperature of the Hawking pairs. Indeed, as explained in the text, the limit $M\rightarrow 0$ is not analytic, signalling the presence of a black hole remnant instead of full evaporation.
}
\label{cartoon} 
\end{figure}

Indeed, gravity extends everywhere in space and causes the production of Hawking pairs mainly near the horizon. In other words, we could think of the Hawking mechanism as pair production in an external field \cite{Dobado:1998mr, Wondrak:2023zdi, Ferreiro:2023jfs, Wondrak:2023hcz}. Such production causes the mass of the black hole to decrease from $M$ to $M - \omega_0 - \omega_\infty$ because of energy conservation. Afterwards, the particle of energy $\omega_0$ is captured again by the black hole, which 
ends up with energy $M - \omega_\infty$. Indeed, the new black hole of mass $M - \omega_0 - \omega_\infty$ will absorb the particle of energy $\omega_0$ to produce a black hole of mass $M - \omega_\infty$ (we have assumed local energy conservation here). 
Notice that, contrary to the argument in \cite{Hawking}, here all the particles have positive energy.
The reader can find a very simple example of evaporation consisting of only three stages in 
Fig.~\ref{cartoon}. 
Notice that everyone assumes the emitted particles not to be black holes. This is the very same assumption made by us to cut off the temperature to a finite value. In other words, everyone introduces the Compton wavelength unconsciously and assumes it to be larger than the Schwarzschild radius.

So far so good, and one could think of pushing this mechanism all the way to the full evaporation of the black hole.
However, this process is under control only until all the collapsing matter has reached the singularity. Afterwards, we have no grasp of the dynamics of the system: we completely lose predictability. 
Therefore, in order to solve the information loss problem we need to solve the singularity problem. This was done in \cite{Akil:2021hhr} in a conformally invariant theory \cite{Bambi:2016wdn, Bambi:2016yne}. Most of the arguments in \cite{Akil:2021hhr} work for the scenario presented in this paper, as long as the energy of the int-particles in \cite{Akil:2021hhr} is flipped to a positive value. 
The solution in \cite{Bambi:2016wdn, Bambi:2016yne} is geodesically complete, meaning that any massive or massless particle never reaches $r=0$. Hence, the collapsing matter and the Hawking int-particles have enough time to interact, causing the black hole mass to decrease.


Let us make the energy conservation argument more quantitative. We assume the initial state of the black hole to be described by the following $N$-particle state, 
\be
| N \omega, {\rm in}  \rangle = \frac{1}{\sqrt{N !}} \left(a^{(\rm in) \dagger}_{\omega} \right)^N | {\rm in} \rangle.
\ee
Hence, the initial energy is simply:
\be
E_{\rm initial} = \langle N \omega, {\rm in}  | H^{(\rm in)} | N \omega, {\rm in}  \rangle = N \omega 
= M^{\rm initial}_{\rm BH} \equiv M \,  .
\ee
The final energy reads:
\be
E_{\rm final} & = & \langle N \omega', {\rm in}  | H^{( \rm in) } | N \omega', {\rm in}  \rangle  
+
\langle {\rm in}  | H^{( \rm i) }  + H^{(\rm o)} | {\rm in} \rangle  \label{strange} \\
& = & \langle N \omega', {\rm out}  | H^{( \rm out) } | N \omega', {\rm out}  \rangle  
+
\langle {\rm in}  | H^{( \rm i) }  + H^{(\rm o)} | {\rm in} \rangle  \nonumber \\
& = &  N \omega' 
+ \langle {\rm out}  | \, S_{\rm sq}^\dagger \left( H^{( \rm i) }  + H^{(\rm o)} \right)  S_{\rm sq} | {\rm out} \rangle 
 \nonumber \\
& = & N \omega' 
+ \sum_{{q}}  \langle {\rm out}  | \, S_{\rm sq}^\dagger(\omega_q) 
\left( a_q^{(\rm i)^\dagger} a_q^{(\rm i)}  +   a_q^{(\rm o)^\dagger} a_q^{(\rm o)}  \right)  \omega_q \, S_{\rm sq} (\omega_q) | {\rm out} \rangle
 \nonumber \\
& = & N \omega' 
+ 2 \sum_{{q}} \frac{\omega_q }{{\rm e}^{8 \pi M \, \omega_q} -1 } 
 \nonumber \\
& \rightarrow & N \omega' 
+ 2 \, {\rm Vol}  \int_0^{+ \infty} 4 \pi \omega^2 d \omega \, \frac{\omega }{{\rm e}^{8 \pi M \, \omega} -1 } 
= \underbrace{N \omega' 
+ \frac{ \pi  \, {\rm Vol}}{15360 M^4}}_{M^{\rm final}_{\rm BH}}  +  \frac{ \pi  \, {\rm Vol}}{15360 M^4} \, .
\label{energyF2}
\ee
At the fourth step only $S_{\rm sq}(\omega_q)$ survives, while the other squeezing 
operators in the product give the identity. 
The final energy (\ref{energyF2}) in terms of the Hawking temperature $T= 1/8 \pi M$ takes the following form,
\be
E_{\rm final}  = 
\underbrace{N \omega' 
+ \frac{ 4 \pi^5 \, T^4  \, {\rm Vol}}{15}}_{M^{\rm final}_{\rm BH}}  +  \frac{4 \pi^5 \, T^4  \, {\rm Vol}}{15} \, .
\ee
Therefore, the variation of the black hole mass is:
\be
\Delta E  = 
-  \frac{4 \pi^5 \, T^4  \, {\rm Vol}}{15}  = 
- 
 \frac{ \pi  \, {\rm Vol}}{15360 M^4} 
\, .
\label{DE}
\ee
Since ${\rm Vol}$ has been introduced to take care of the infrared divergence, we first divide by the volume of the Universe to obtain the energy density and, afterwards, we integrate over the volume of the black hole to get:
\be
\Delta E  = 
-  \frac{4 \pi^5 \, T^4  \, {\rm V_{\rm BH}}}{15}  = 
- 
 \frac{ \pi  \, {\rm V_{\rm BH}}}{15360 M^4} 
\, .
\label{DEVBH}
\ee
Finally, the power output per unit area is obtained simply by dividing (\ref{DE}) again by the volume (in natural units) of the black hole:
\be
\frac{P}{A} = \frac{4 \pi^5 \, T^4 }{15}, 
\ee
which is the Stefan-Boltzmann law.

Regarding the first equality (\ref{strange}), a remark is needed about the expectation values.
It may seem strange that the final energy of the black hole, i.e.\ after pair production, has been evaluated as the expectation value on the Hilbert space $\mathcal{H}_{\rm in}$. However, this quantity can be rewritten as:
\be
 \langle N \omega', {\rm in}  | H^{ ( \rm in) } | N \omega', {\rm in}  \rangle 
 & = & \langle N \omega', {\rm out}  |  S_{\rm sq}^\dagger \, H^{( \rm in) }  S_{\rm sq} |  N \omega', {\rm out}  \rangle \nonumber \\
 & = & 
  \langle N \omega', {\rm out}  |  S_{\rm sq}^\dagger \left(   S_{\rm sq}  H^{( \rm out) \dagger} S_{\rm sq}^\dagger \right)  S_{\rm sq} |  N \omega', {\rm out}  \rangle
  \nonumber \\
  & = & 
  \langle N \omega', {\rm out}  |  H^{( \rm out)}  |  N \omega', {\rm out}  \rangle \, , 
  \label{remark}
\ee
which is the expectation value of the energy after pair production. 
On the other hand, the second expectation value in (\ref{strange}) is correct because the particle pair is created on the initial hypersurface, similarly to (\ref{Ninfinity}).

A more speculative implication concerns the final stage of the black hole evaporation. Does the black hole evaporate completely, or is a kind of remnant left behind?
According to the energy balance, 
\be
M^{\rm final}_{\rm BH} = M -  \frac{ \pi  \, {\rm Vol} \, M_{\rm Pl}^8}{15360 M^4} < M \, .
\ee
Assuming 
\be
{\rm Vol} \equiv \frac{4}{3} \pi (A_{\rm EH})^{\frac{3}{2}} = \frac{4}{3} (4 \pi (2 G M)^2)^{\frac{3}{2}} = 
  \frac{4}{3} \pi (4 \pi)^{\frac{3}{2}} (2 G M)^3 = 
   \frac{4}{3} \pi (4 \pi)^{\frac{3}{2}} 
  (2M/M^2_{\rm Pl})^3,
\ee
 and imposing the mass of the black hole to be positive, we obtain a minimum value for the black hole mass,
\be
 M -  \frac{4}{3} \pi (4 \pi)^{\frac{3}{2}} \frac{ \pi  \, { 2^ 3 M_{\rm Pl}^2}}{15360 M}  \geqslant 0 \quad \Longrightarrow 
 \quad M  \geqslant  M_{\rm Pl} \,  
 \sqrt{\frac{4}{3} \pi (4 \pi)^{\frac{3}{2}} }
 \left( \frac{\pi}{1920} \right)^{1/2} 
 = 
\frac{\pi ^{7/4}}{6 \sqrt{5}} M_{\rm Pl}
 \approx 0.5 \, M_{\rm Pl} \, .
 \label{Mmin}
\ee
The result (\ref{Mmin}) is very close to the constraint we obtained by requiring the Schwarzschild radius to be larger than the Compton wavelength, namely $0.7 \, M_{\rm Pl}$ (\ref{Temperatures}).
Since the minimal mass for having a black hole (\ref{Mmin}) is smaller than the constraint obtained by comparing the Schwarzschild horizon and the Compton wavelength, 
the two independent constraints seem to suggest that the final stage of the evaporation process is actually not a remnant, but a horizonless particle-like object.

\section*{Conclusions}

On the basis of solid results in the condensed matter literature, we showed that any Bogoliubov transformation is realized by unitary operators, namely the Squeezing operator. Therefore, we proved that quantum field theory in a general spacetime is described by the product of two unitary operators: the S-matrix, defined in the out- or the in-Hilbert space, times the Squeezing operator. The outcome is a unitary realization of
the $ \$ = S S_{\rm sq}$ operator introduced long ago by S. Hawking.
Hence, we explicitly evaluated the $n$-particle production in presence of a black hole (or in the Rindler spacetime) as a function of the black hole mass and assuming Hawking temperature or a modified temperature consistent in zero black hole mass. 

Therefore, we figured out that a general spacetime turns the vacuum state into a condensate described by: the BCS state of super-conduction, for Fermions, or the Bose-Einstein condensation (BEC) ground state, for Bosons.
Accordingly, there is no issue about the particle interpretation in curved spacetime, which stays the same as in Minkowski spacetime (if at least one between the in- or the out-Hilbert space is defined in Minkowski), with the dynamics taking place in a condense rather then the vacuum for the Bogoliubons. In other words, the definition of particle is inferred according to the out- or the in-Hilbert space while all the difficulties due to the general or curved spacetime are moved into the BCS or BEC ground state.

Following the main steps in the construction of the BCS theory, 
by reverse engineering, we derived the effective Hamiltonian for electrons and positrons in the black hole (or the Rindler) background in the mean field approximation. We also reconstruct the analog of the four-fermions interaction emerging in a solid due to the exchanging of phonons in between fermions. In curved spacetime, we are tented to think that the phonons are replaced by gravitons, while the crystal's ions by chunks or atoms of the space itself. 
In support of the above statement, according to the consistency condition for the BCS theory, it turns out that the Fourier's transform of the interaction potential corresponds to a massless particle (likely the graviton), while the effective mass of the interacting Fermions is close to the Planck mass. 

More quantitatively, we derived a simple analytic form of the mass gap function $\Delta(T)$, which contrary to the solid physics case, is here uniquely fixed by the Bogoliubov transformation. In particular, $\Delta(T)$ goes to zero at zero temperature and grows linearly like $2T$ for large temperature. In other words, the quantum field theory in a general spacetime is described by an high temperature superconductive condensate.  

Regarding unitarity in curved spacetime, for $T \rightarrow 0$, the gap function goes to zero too, $\Delta(T) \rightarrow0$, and the effective Hamiltonian turns into the Hamiltonian for free Fermions in Minkowski spacetime. 
Therefore, simply assuming that the temperature tends to zero when the black hole mass vanishes, we have a unitary evolution described by the squeezing operator at any stage of the evaporation, including the very end when the black hole disappearances. 

We explicitly evaluated the entanglement entropy of the emitted particles (for fermions and bosons) by tracing over the int-particles that cross the horizon, and found a very simple result cubic in the temperature. 
Furthermore, starting from the entanglement entropy, we derived the area law and discussed the limit of the result.
A natural quantum modification of the temperature, consistent with the Compton wavelength of matter, yields zero entanglement entropy after complete evaporation.

Finally, we comment on the singularity problem and its crucial role in the unitarity issue. 
We have provided a simplified model for the evaporation process based on energy conservation and involving only positive-energy particles, contrary to the original argument by Hawking, which invoked negative-energy particles. 
In particular, we derived the Stefan-Boltzmann law in terms of the Hawking temperature by only assuming energy conservation.
The singularity resolution is here based on Weyl conformal invariance \cite{Bambi:2016wdn}, but the very same discussion applies to any other regular black hole metric.

\section*{Acknowledgments}
We would like to thank Ali Akil for the numerous discussions and for providing the relevant literature, which has been indispensable and crucial in the preparation of this article.
We are grateful to Alessio Filippetti for the important teachings in solid-state physics pertaining to our work.

\end{document}